\documentclass[11pt, leqno]{article}
 \usepackage[toc,page]{appendix}
\usepackage{amsmath, setspace, multirow, lineno}
\usepackage[font=small,labelfont=bf]{caption}
\usepackage[top=1.25in, bottom=1.25in, left=1.2in, right=1.2in]{geometry}
\usepackage[nolists]{endfloat}

\usepackage{natbib}
\usepackage{color,soul}

\DeclareCaptionStyle{italic}[justification=centering]{labelfont={bf},textfont={it},labelsep=colon}
\usepackage{graphicx,psfrag,epsf,textcomp,epstopdf, amsthm}
\usepackage{enumerate, dsfont}

\usepackage{url,xcolor}
\usepackage{booktabs, subfig, bm, paralist,mathpazo,tikz,todonotes,longtable,microtype}
\usepackage[linesnumbered,ruled,vlined]{algorithm2e}

\usepackage[pdftex,colorlinks=true]{hyperref}
\definecolor{darkblue}{rgb}{0,0,.6}
\hypersetup{citecolor=darkblue,linkcolor=darkblue,urlcolor=darkblue}
\definecolor{DarkRed}{rgb}{.7,0,.4}

\newcommand{\argmax}{\operatornamewithlimits{argmax}}
\newcommand\Tau{\mathcal{T}}
\usepackage{comment,orcidlink}

\newcommand{\blind}{0}

\newcommand{\X}{\mathcal{X}}
\newcommand{\Y}{\mathcal{Y}}

\graphicspath{{plots/}}
\DeclareMathOperator*{\argmin}{\arg\!\min}

\newsavebox\CBox

 \newtheorem{@definition}{\sc Definition}[section]

\newcommand{\Rlogo}{\protect\includegraphics[height=1.8ex,keepaspectratio]{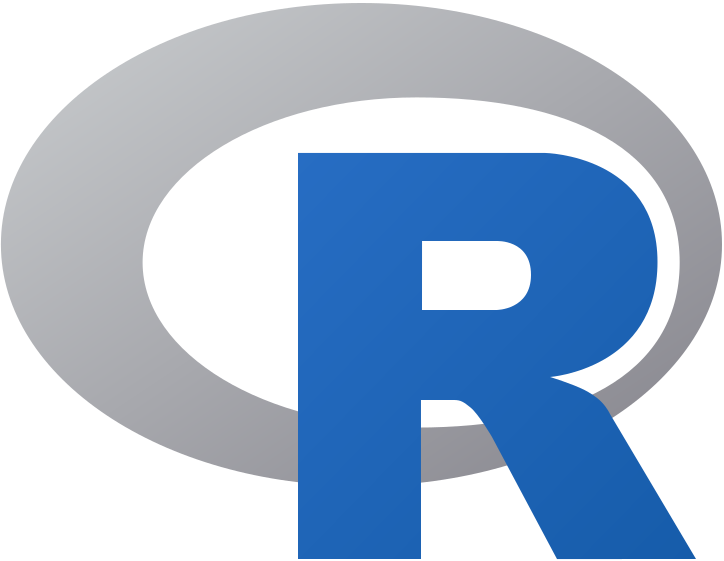}}

  \renewcommand\X{\mathcal{X}}

\begin{document}

\def\spacingset#1{\renewcommand{\baselinestretch}{#1}\small\normalsize} \spacingset{1}

\if0\blind
{
\title{\bf On function-on-function linear quantile regression}}
\author{
Muge Mutis \orcidlink{0000-0002-9801-4835} \footnote{Corresponding address: Department of Statistics, Yildiz Technical University, 34220 Esenler-Istanbul, Turkiye; Email: muge.mutis@yildiz.edu.tr} \\
Department of Statistics \textendash Yildiz Technical University \\
\\
Ufuk Beyaztas \orcidlink{0000-0002-5208-4950} \\
Department of Statistics \textendash Marmara University \\
\\
Filiz Karaman \orcidlink{0000-0002-8491-674X} \\
Department of Statistics \textendash Yildiz Technical University \\
\\
Han Lin Shang \orcidlink{0000-0003-1769-6430} \\
Department of Actuarial Studies and Business Analytics \textendash \\
Macquarie University 
}
\maketitle
\fi

\if1\blind
{
\title{\bf On function-on-function linear quantile regression}
} \fi

\maketitle

\begin{abstract}
We present two innovative functional partial quantile regression algorithms designed to accurately and efficiently estimate the regression coefficient function within the function-on-function linear quantile regression model. Our algorithms utilize functional partial quantile regression decomposition to effectively project the infinite-dimensional response and predictor variables onto a finite-dimensional space. Within this framework, the partial quantile regression components are approximated using a basis expansion approach. Consequently, we approximate the infinite-dimensional function-on-function linear quantile regression model using a multivariate quantile regression model constructed from these partial quantile regression components. To evaluate the efficacy of our proposed techniques, we conduct a series of Monte Carlo experiments and analyze an empirical dataset, demonstrating superior performance compared to existing methods in finite-sample scenarios. Our techniques have been implemented in the \texttt{ffpqr} package in \Rlogo.
\end{abstract}


\newpage
\spacingset{1.6} 

\section{Introduction} \label{sec:intro}

Conditional mean regression, while informative, offers only a glimpse into the relationship between a response variable and a set of predictor variables. However, in contexts like natural disasters, environmental pollution, and health risks, where understanding various conditional percentiles of the response variable is vital for modeling events and predicting future risks accurately (see, e.g., \cite{Magzamen2015, Abbas2019, Lara2019, Vasseur2021}), relying solely on conditional mean regression might prove insufficient. This deficiency arises because results from conditional mean regression may not effectively generalize to alternative locations of the response variable away from the center.

In addressing this limitation, quantile regression, initially proposed by \cite{Koenker1978}, emerges as a valuable alternative. It facilitates the estimation of percentiles of the response variable for a given set of predictors, offering a comprehensive understanding of the potential relationship between predictor and response variables. Unlike conditional mean regression, quantile regression boasts several advantages. Notably, it exhibits reduced sensitivity to heavy-tailed errors, heteroskedasticity, and the presence of outliers (for a deeper exploration of quantile regression, see \cite{Koenker2005}, for instance).

The classical quantile regression method is traditionally employed for analyzing data structured in a discrete matrix format. However, technological advancements, particularly in data collection tools, have significantly enhanced the availability of observations that can be continuously recorded across grids, a format known as functional data. The infinite-dimensional nature of functional data poses considerable challenges, often rendering classical techniques, including quantile regression inadequate. Consequently, a growing focus is on developing robust methodologies tailored to functional data analysis. Noteworthy contributions in this area include the works of \cite{ramsay2002}, \cite{ferraty2006}, \cite{RamsaySilverman2005}, \cite{horvath2012}, \cite{cuevas2014}, \cite{Hsing2015}, and \cite{Kokoszka2017}, which offer comprehensive overviews, theoretical advancements, and case studies in functional data analysis.

Several functional linear quantile regression models have been proposed, differing according to whether the response and/or predictor variables are scalar or functional. These models include:
\begin{inparaenum}
\item[(i)] scalar response-functional predictor models (see, for instance, \cite{cardot2005, ferraty2005, cardot2007, chenmuller2012, Kato2012, Tang2014, Yu2016, yao2017, Ma2019, Sang2020, Chaouch2020, Zhang2021, Li2022, Zhu2022, Zhou2022});
\item[(ii)] functional response-scalar predictor models (see, for example, \cite{Kim2007, Wang2009, yang2020, Liu2020}); and
\item[(iii)] functional response-functional predictor models (see, for instance, \cite{BSA2022, BSMMA}).
\end{inparaenum}
This study focuses on case (iii), function-on-function linear quantile regression (FFLQR).

In the realm of FFLQR, the function-on-function partial quantile regression (FPQR) approach introduced by \cite{BSMMA} utilizes a functional partial least squares (FPLS) based strategy for parameter estimation. Conversely, the approach proposed by \cite{BSA2022} (referred to as ``FLQR'') employs a functional principal component (FPC) based methodology for the same objective. In both FPLS and FPC methodologies, the infinite-dimensional observations of the functional response and predictors undergo projection onto finite-dimensional orthonormal FPLS and FPC bases. This process yields orthogonal latent components in both cases. The developed regression model utilizes latent components derived from both the functional response and predictor to approximate the regression task concerning them. The FPC technique aims to maximize the covariance among the functional predictors to extract these components. However, only a few components typically contain the most relevant information about this covariance, which might not necessarily be essential for representation. Some crucial terms explaining the interaction between the basis functions and functional predictors may likely come from later FPCs \citep{Delaigle2012}. In contrast, FPLS extracts components by maximizing the covariance between the functional response and predictors. Consequently, compared to FPC, FPLS produces more informative latent components with fewer terms, making it more favorable (see, for example, \cite{Reiss2007, Delaigle2012, Bande2017, BSMMA}). Thus, we opt for the FPLS method to estimate the FFLQR in this study.

In FPQR, as proposed by \cite{BSMMA}, the framework of partial quantile regression initially introduced by \cite{Dodge2009} is adapted, which acts as a counterpart to conventional partial least squares regression, for use within the FPLS framework. To address the estimation of FFLQR, FPQR employs a mechanism of partial quantile covariance among functional predictor variables to derive the FPQR bases. These bases are then utilized to compute FPQR components and estimate the final model. Due to the infinite-dimensional nature of response and predictor variables, directly computing FPQR bases is infeasible. Therefore, in FPQR, basis functions are initially approximated using a finite-dimensional expansion technique. Subsequently, the FPQR model, constructed with functional predictor variables, is approximated using the multivariate extension of the partial quantile regression method proposed by \cite{Dodge2009}, utilizing basis expansion coefficients for both response and predictors. Despite its practical utility, FPQR has limitations, mainly because the partial quantile regression approach of \cite{Dodge2009} is designed solely for univariate response models, making its extension to multiple response models challenging. This challenge arises from the need to compute quantile covariance individually for each response variable, leading to significant increases in computational time. FPQR may become computationally burdensome when dealing with a large number of basis expansion functions in the model and/or a substantial number of functional predictor variables.

Drawing inspiration from \cite{Alvaro2022}, we introduce two enhanced versions of FPQR to improve the computational efficiency and numerical accuracy of FPQR when estimating FFLQR. These modified FPQR methods incorporate alterations to the original FPQR framework proposed by \cite{BSMMA}, utilizing two distinct quantile covariance techniques. In the first approach, we integrate the quantile covariance method introduced by \cite{Choi2018} into the FPLS framework. Empirical evidence suggests that the quantile covariance proposed by \cite{Choi2018} yields outcomes comparable to those of \cite{Dodge2009}'s quantile covariance for multi-response models. This method involves solving quantile regression models for both the response variables concerning the predictors and the predictors concerning the response. Thus, it requires more computing time than \cite{Dodge2009}'s method. However, our numerical results, presented in detail in Section~\ref{sec:montecarlo}, demonstrate that the FPQR constructed via \cite{Choi2018}'s quantile covariance produces better estimates for the FFLQR model than those constructed via \cite{Dodge2009}'s quantile covariance. The second approach modifies FPQR by extending the quantile covariance method proposed by \cite{Li2015} to accommodate functional data. \cite{Li2015}'s quantile covariance method relies on traditional covariance and readily extends to multiple response models; thus, it requires considerably less computing time than its competitors.

In the existing FPQR method, the partial quantile regression algorithm from \cite{Dodge2009} is utilized to derive the FPQR components. However, in our proposed methods, we extend the partial quantile regression algorithm introduced by \cite{Alvaro2022} to handle functional data. This updated algorithm supports multiple responses in the model, reducing computational time compared to FPQR. Similar to the FPQR approach by \cite{BSMMA}, our proposed methods approximate FPQR components using an appropriate basis expansion technique. Following the basis expansion, we employ singular value decomposition on the quantile covariance obtained through either \cite{Choi2018}'s or \cite{Li2015}'s approach to extract partial quantile components. These components are then utilized in quantile regression to approximate the regression coefficient function of the FFLQR model.

The subsequent sections of this paper are structured as follows. Section~\ref{sec:FFLQR} outlines the FFLQR model. Section~\ref{sec:FPQR} provides an overview of the FPQR method. The technical details of the proposed methods are elaborated in Section~\ref{sec:props}. We perform a series of Monte Carlo experiments in Section~\ref{sec:montecarlo} to evaluate the finite-sample performance of the proposed methods. The results obtained by applying the proposed methods to empirical data, specifically the North Dakota weather dataset, are presented in Section~\ref{sec:empiricaldata}. Finally, Section~\ref{sec:conc} concludes the paper and discusses potential extensions of the methodology.

\section{Function-on-function linear quantile regression model} \label{sec:FFLQR}

Consider a dataset comprising pairs of functional response $\Y(u)$ and functional predictor $\X(v)$, denoted by $\left\lbrace \Y_{i}(u), \X_{i}(v): i=1,\ldots,n \right\rbrace $, drawn independently and identically from a pair $(\Y(u), \X(v))$. These functions are real-valued and square-integrable, defined on closed intervals $u \in \mathcal{I}_{\Y}$ and $v \in \mathcal{I}_{\X}$, respectively, and they belong to the separable Hilbert space $\mathcal{L}_{2}$, denoted by $\mathcal{H}$. Assuming mean-zero processes for both the functional response and predictor, that is, $\text{E}[\Y(u)] = \text{E}[\X(v)] = 0$, we aim to estimate the entire conditional distribution of $\Y$ for a given quantile level $\tau \in (0,1)$. This involves computing the $\tau^{th}$ conditional quantile $Q_{\tau}\left [ \Y_{i} \vert \X_{i} \right ]$ of $\Y$ conditioned on the entire trajectory $\X_{i}$. Then, the FFLQR aims to characterize the entire conditional distribution of $\Y$, expressed as follows:
\begin{equation}\label{denk2.2}
Q_{\tau}\left [ \Y_{i} \vert \X_{i} \right ]= \int_{\mathcal{I}_{\X}} \X_{i}(v) \beta_{\tau}(v,u)dv,
\end{equation}
where $\beta_{\tau}(v,u)$ signifies the two-dimensional regression coefficient function, evaluating the impact of $\X_{i}$ on the $\tau$\textsuperscript{th} quantile of $\Y_{i}$.

The FFLQR in~\eqref{denk2.2} is estimated by minimizing the check-loss function $\rho_{\tau}(x) = x \left\lbrace \tau- \mathds{1} (x< 0)\right\rbrace$ (cf., e.g., \cite{Koenker1978}), incorporating a binary indicator function $\mathds{1}\left\lbrace \cdot \right\rbrace $, as follows:
\begin{equation*}\label{denk2.3}
\underset{\begin{subarray}{c}
  \beta_{\tau}(v,u) \in \mathcal{H}^2 
  \end{subarray}}{\argmin}~ \sum_{i=1}^n \rho_{\tau} \left[ \Y_i(u) - \int_{\mathcal{I}_x} \X_i (v) \beta_{\tau}(v,u) dv \right].
\end{equation*}

\section{The FPQR for the estimation of FFLQR}\label{sec:FPQR}

This section provides an overview of the FPQR method proposed by \cite{BSMMA} for estimating the FFLQR model presented in~\eqref{denk2.2}. In FPLS, the orthonormal eigenfunctions are derived by maximizing the squared covariance between the functional response and functional predictor. This is achieved by optimizing the covariance operator $\text{Cov} \left( \cdot, \cdot \right)$ using the least squares loss function as follows:
\begin{align}\label{eq:cov_cv}
&\underset{\begin{subarray}{c}
  \Vert p \Vert_{\mathcal{L}_2[\mathcal{I}_{\Y}]} = 1 \\
  \Vert w \Vert_{\mathcal{L}_2[\mathcal{I}_{\X}]} = 1
  \end{subarray}}{\argmax} \text{Cov}^2 \left( \int_{\mathcal{I}_{\Y}} \Y(u) p(u) du, ~ \int_{\mathcal{I}_{\X}} \X(v) w(v) dv \right), \\
 \Longleftrightarrow &  \underset{\begin{subarray}{c}
  \Vert p \Vert_{\mathcal{L}_2[\mathcal{I}_{\Y}]} = 1 \\
  \Vert w \Vert_{\mathcal{L}_2[\mathcal{I}_{\X}]} = 1
  \end{subarray}}{\argmin} \text{E}^2 \left[ \rho \left( \int_{\mathcal{I}_{\Y}} \Y(u) p(u) du - \int_{\mathcal{I}_{\X}} \X(v) w(v) dv \right) \right], \label{eq:cov_ls}
\end{align}
where $\rho(x) = x^2$. The conventional covariance in~\eqref{eq:cov_cv} is employed to predict the mean value of the response variable based on given predictors. However, to predict the quantiles of the response, the covariance operator must be redefined. For this purpose, we introduce the quantile covariance, denoted by $\text{Cov}_{\tau} \left( \cdot, \cdot \right) $, which modifies $\text{Cov} \left( \cdot, \cdot \right)$ by replacing the least squares loss function in~\eqref{eq:cov_ls} with the quantile loss function $\rho_{\tau}(x) = x \left\lbrace \tau - \mathbb{1} (x < 0) \right\rbrace$. Since optimizations using the quantile loss function at different $\tau$ levels yield different results, the quantile covariance depends on the $\tau$ level proposed by \cite{BSMMA}.

Let $\mathcal{C}_{\Y \X}^{(\tau)}$ and $\mathcal{C}_{\X \Y}^{(\tau)}$ denote the cross-quantile-covariance operators, as follows:
\begin{align*}
\mathcal{C}_{\Y \X}^{\tau} &= \mathcal{L}_2[\mathcal{I}_{\Y}] \rightarrow \mathcal{L}_2[\mathcal{I}_{\X}], \qquad f \rightarrow g = \int_{\mathcal{I}_{\Y}} \text{Cov}_{\tau} \left[ \Y(u), \X(v) \right] f(u) du,\\
\mathcal{C}_{\X \Y}^{\tau} &= \mathcal{L}_2[\mathcal{I}_{\X}] \rightarrow \mathcal{L}_2[\mathcal{I}_{\Y}], \qquad g \rightarrow f = \int_{\mathcal{I}_{\X}} \text{Cov}_{\tau} \left[ \Y(u), \X(v) \right]  g(v) dv.
\end{align*} 
We define operators $\mathcal{U} = \mathcal{C}_{\X \Y}^{\tau} \circ \mathcal{C}_{\Y \X}^{\tau}$ and $\mathcal{V} = \mathcal{C}_{\Y \X}^{\tau} \circ \mathcal{C}_{\X \Y}^{\tau}$ as self-adjoint, positive, and compact, respectively. Their spectral analyses provide a countable collection of positive eigenvalues $\lambda_{\tau}$ coupled with orthonormal eigenfunctions $w_{\tau} \in \mathcal{L}_2[\mathcal{I}_{\X}]$. These eigenfunctions satisfy $\mathcal{U} w_{\tau} = \lambda_{\tau} w_{\tau}$, while adhering to the constraint $\int_{\mathcal{I}_{\X}} w_{\tau}(v) w_{\tau}^\top(v) dv = 1$ (refer to, for instance, \cite{BSMMA}, for further elucidation).

Subsequently, the FPQR components of~\eqref{denk2.2} are derived by maximizing the squared quantile covariance between the response and predictor, expressed as:
\begin{align*}
& \underset{\begin{subarray}{c}
  p_{\tau} \in \mathcal{L}_2[\mathcal{I}_{\Y}],~ \Vert p_{\tau} \Vert_{\mathcal{L}_2[\mathcal{I}_{\Y}]} = 1 \\
  w_{\tau} \in \mathcal{L}_2[\mathcal{I}_{\X}],~ \Vert w_{\tau} \Vert_{\mathcal{L}_2[\mathcal{I}_{\X}]} = 1
  \end{subarray}}{\argmax} \text{Cov}_{\tau}^2 \left( \int_{\mathcal{I}_{\Y}} \Y(u) p_{\tau}(u) du, ~ \int_{\mathcal{I}_{\X}} \X(v) w_{\tau}(v) dv \right), \\
  \Longleftrightarrow &  \underset{\begin{subarray}{c}
  p_{\tau} \in \mathcal{L}_2[\mathcal{I}_{\Y}],~ \Vert p_{\tau} \Vert_{\mathcal{L}_2[\mathcal{I}_{\Y}]} = 1 \\
  w_{\tau} \in \mathcal{L}_2[\mathcal{I}_{\X}],~ \Vert w_{\tau} \Vert_{\mathcal{L}_2[\mathcal{I}_{\X}]} = 1
  \end{subarray}}{\argmin} \text{E}^2 \left[\rho_{\tau} \left( \int_{\mathcal{I}_{\Y}} \Y(u) p_{\tau}(u) du - \int_{\mathcal{I}_{\X}} \X(v) w_{\tau}(v) dv \right) \right].
\end{align*}
In this context, $p_{\tau}(u)$ and $w_{\tau}(v)$ stand for the eigenfunctions linked to the largest eigenvalue of $\mathcal{V}$ and $\mathcal{U}$, respectively, for quantile level $\tau$.

The initial FPQR element, designated as $T_{\tau}^{(1)}$, is computed through functional linear regression, where $T_{\tau}^{(1)} = \int_{\mathcal{I}_{\X}} w_{\tau}^{(1)}(v) \X(v) dv$, and $w_{\tau}^{(1)}(v)$ represents the eigenfunction of $\mathcal{U}$ associated with the largest eigenvalue for quantile level $\tau$. Subsequent components are obtained iteratively. Let $h$ range from $1$ to $H$, serving as the iteration index. At each step $h$, the eigenfunction, $w_{\tau}^{(h)}(v)$, is determined as the solution of
\begin{align*}
w_{\tau}^{(h)} = \underset{\begin{subarray}{c}
  p_{\tau} \in \mathcal{L}_2[\mathcal{I}_{\Y}],~ \Vert p_{\tau} \Vert_{\mathcal{L}_2[\mathcal{I}_{\Y}]} = 1 \\
  w_{\tau} \in \mathcal{L}_2[\mathcal{I}_{\X}],~ \Vert w_{\tau} \Vert_{\mathcal{L}_2[\mathcal{I}_{\X}]} = 1
  \end{subarray}}{\argmax} \text{Cov}_{\tau}^2 \left( \int_{\mathcal{I}_{\Y}} \Y(u) p_{\tau}(u) du, ~ \int_{\mathcal{I}_{\X}} \X^{(h-1)}(v) w_{\tau}(v) dv \right),
\end{align*}
and the $h$\textsuperscript{th} FPQR component, $T_{\tau}^{(h)}$, is then obtained as $T_{\tau}^{(h)} = \int_{\mathcal{I}_{\X}} w_{\tau}^{(h)}(v) \X^{(h-1)}(v) dv$. Here, $\X^{(h)}(v) = \X^{(h-1)}(v) - \text{E}\left[ \X^{(h-1)}(v) \big\vert T_{\tau}^{(h)}\right]$ represents the residuals of the functional linear regression of $\X^{(h-1)}(v)$ on $T_{\tau}^{(h)}$.

The FPQR components kept after $H$ iterations are represented as $\left\lbrace T_{\tau}^{(1)}, \ldots, T_{\tau}^{(H)} \right\rbrace$. Ultimately, in the final step of FPQR, a quantile regression model of the response based on the retained FPQR components, $\text{E} \left[\rho_{\tau} \left( \Y(u) \big\vert T_{\tau}^{(1)}, \ldots, T_{\tau}^{(h)} \right) \right] $, is utilized to derive the estimation of the regression coefficient function as well as the estimation of the quantile of the response.

Although the functional random variables $\left[ \Y(u), \X(v) \right]$ inherently exist within an infinite-dimensional domain, the observed sample curves are constrained to finite sets of time points. Consequently, directly computing the FPQR basis and its associated components becomes infeasible. To address this challenge, a common practical strategy involves approximating the functional representations of the random variables and FPQR components using basis expansion techniques. These techniques encompass a variety of methods, including Fourier, B-spline, radial, and wavelet basis functions \citep{RamsaySilverman2005}. Notably, this study employs the B-spline basis expansion method.

Let $K_{\Y}$ and $K_{\X}$ denote the number of basis expansion functions utilized for projecting the functional response and predictor into a finite-dimensional space. The functional forms are then expressed in a basis expansion form, given by:
\begin{equation*}\label{denk2.4}
\Y(u) \approx \sum_{k=1}^{K_{\Y}} \zeta_{k} \phi_{k}(u) = \bm{\zeta}^\top \bm{\phi}(u), \quad \X(v) \approx \sum_{k=1}^{K_{\X}} \delta _{k} \psi _{k}(v) = \bm{\delta}^\top \bm{\psi}(v),
\end{equation*}
where $\bm{\phi}(u)=\left [ \phi_{1}(u),\ldots,\phi_{K_{\Y}}(u) \right ]^\top$ and $\bm{\psi}(v)=\left [ \psi_{1}(v),\ldots,\psi_{K_{\X}}(v) \right ]^\top$ denote the basis expansion functions and $\bm{\zeta}=\left [ \zeta_{1},\ldots,\zeta_{K_{\Y}} \right ]^\top$ and $\bm{\delta}=\left [ \delta_{1},\ldots,\delta_{K_{\X}} \right ]^\top$ are the vectors of basis expansion coefficients. 

Consider the symmetric matrices of inner products of the basis expansion functions $\bm{\Phi} = \int_{\mathcal{I}_{\Y}}\bm{\phi}(u) \bm{\phi}^\top(u)du$ and $\bm{\Psi} = \int_{\mathcal{I}_{\X}} \bm{\psi}(v) \bm{\psi}^\top(v)dv$ with dimensions $K_{\Y} \times K_{\Y}$ and $K_{\X} \times K_{\X}$, respectively. Additionally, denote the square roots of $\bm{\Phi}$ and $\bm{\Psi}$ as $\bm{\Phi}^{1/2}$ and $\bm{\Psi}^{1/2}$, respectively. It has been demonstrated by \cite{BSMMA} that the FPQR of $\Y(u)$ with respect to $\X(v)$ is analogous to conducting multivariate partial quantile regressions of $\bm{\Lambda} = \bm{\zeta}^\top \bm{\Phi}^{1/2}$ on $\bm{\Pi} = \bm{\delta}^\top \bm{\Psi}^{1/2}$, ensuring that both methods yield identical partial quantile regression components at each iteration step $h$.

\section{FPQR methods to estimate the FFLQR}\label{sec:props}

In the FPQR framework outlined in Section~\ref{sec:FPQR}, the determination of the eigenfunctions $w_{\tau}(v)$, crucial for defining the FPQR components $T_{\tau}$, relies on the quantile covariance method introduced by \cite{Dodge2009}. While this approach effectively approximates the FPQR of $\Y(u)$ with respect to $\X(v)$ using basis expansion coefficients, it is subject to limitations when applied to multivariate scenarios. Specifically, the original algorithm proposed by \cite{Dodge2009} is tailored for univariate response models, necessitating iterative application for multiple response models, that is, this approach requires $K_{\Y} \times K_{\X}$ independent univariate quantile regression models to be conducted, thereby substantially increasing computational overhead. To address this challenge, this study presents two adapted FPQR algorithms leveraging the quantile covariance methodologies proposed by \cite{Choi2018} or \cite{Li2015}, specifically designed to accommodate multiple response models.

Let us investigate the partial quantile regression of $\bm{\Lambda} = \bm{\zeta}^\top \bm{\Phi}^{1/2}$ with respect to $\bm{\Pi} = \bm{\delta}^\top \bm{\Psi}^{1/2}$, offering an approximate representation of the FFLQR model outlined in~\eqref{denk2.2}. Essentially, we seek to approximate the $\tau^\textsuperscript{th}$ conditional quantile of $\Y$ conditioned on $\X_i$, denoted by $Q_{\tau}\left [ \Y_{i} \vert \X_{i} \right ]$, within the finite-dimensional space defined by basis expansion coefficients. This approximation is expressed as follows:
\begin{equation*}
Q_{\tau}[\bm{\Lambda}_i \vert \bm{\Pi}_i] = \bm{\Pi}_i \bm{\Omega}_{\tau},
\end{equation*}
where $\bm{\Omega}_{\tau}$ represents the coefficient matrix, which can be estimated through the minimization of the check loss function.
\begin{equation*}
\widehat{\bm{\Omega}}_{\tau} = \underset{
\begin{subarray}
{c}\bm{\Omega}_{\tau}
\end{subarray}}{\argmin} \left[ \sum_{i=1}^n \rho_{\tau} (\bm{\Lambda}_i - \bm{\Pi}_i \bm{\Omega}_{\tau}) \right].
\end{equation*}

We employ the partial quantile regression methodology to derive an estimate $\bm{\Omega}_{\tau}$. However, unlike the utilization of the quantile covariance proposed by \cite{Dodge2009} in the FPQR framework introduced by \cite{BSMMA}, we explore the quantile covariances introduced by \cite{Choi2018} and \cite{Li2015}. These are outlined in Algorithms~\ref{alg:Choi} and~\ref{alg:Li}. 
\begin{algorithm}
\caption{Computation of quantile covariance of \cite{Choi2018}.}\label{alg:Choi}
For a specified quantile level $\tau$, perform quantile regression modeling of $\bm{\Lambda}$ with respect to $\bm{\Pi}$, repeating this procedure for each column of $\bm{\Pi}$, $k=1, \ldots, K_{\X}$. Record the estimated coefficient matrix as $\widehat{\bm{\Omega}}_{\tau}^{\bm{\Lambda},\bm{\Pi}} = [\widehat{\Omega}^{\bm{\Lambda},\bm{\Pi}}_{\tau,1}, \ldots, \widehat{\Omega}^{\bm{\Lambda},\bm{\Pi}}_{\tau,K_{\X}}]$.

Conduct quantile regression modeling of $\bm{\Pi}$ conditioned on $\bm{\Lambda}$ for the given $\tau$, iterating this process for each column of $\bm{\Pi}$, $k=1, \ldots, K_{\X}$. Capture the estimated coefficient matrix as $\widehat{\bm{\Omega}}_{\tau}^{\bm{\Pi}, \bm{\Lambda}} = [\widehat{\Omega}^{\bm{\Pi}, \bm{\Lambda}}_{\tau,1}, \ldots, \widehat{\Omega}^{\bm{\Pi}, \bm{\Lambda}}_{\tau,K_{\X}}]$.

Compute the $\tau$\textsuperscript{th} quantile covariance $\rho_{\text{cov}}^{(k)} = \text{sign}\left(\widehat{\bm{\Omega}}_{\tau}^{\bm{\Pi}, \bm{\Lambda}}\right) \sqrt{\widehat{\bm{\Omega}}_{\tau}^{\bm{\Pi}, \bm{\Lambda}} \widehat{\bm{\Omega}}_{\tau}^{\bm{\Lambda},\bm{\Pi}}}\sqrt{\text{var}(\bm{\Pi})\text{var}(\bm{\Lambda)}}$, where $\text{var}(\bm{\Pi})$ and $\text{var}(\bm{\Lambda})$ are the variances of $\bm{\Pi}$ and $\bm{\Lambda}$, respectively.

Repeat steps~one to~three for each column of $\bm{\Lambda}$, $k = 1,\ldots, K_{\Y}$.

Compute the quantile covariance $Q_{\text{cov}, \tau}$ with column entries $\rho_{\text{cov}}^{(k)}$.
\end{algorithm}

\begin{algorithm}
\caption{Computation of quantile covariance of \cite{Li2015}.}\label{alg:Li}
Compute the $n \times K_{\Y}$ dimensional matrix $\Gamma  = \bm{\Tau}-\mathds{1} ((\bm{\Lambda}-Q_{\bm{\Lambda},\tau})< 0)$ where $\bm{\Tau}$ is the $n \times K_{\Y}$ dimensional matrix with elements representing the quantile level $\tau \in (0,1)$ and $Q_{\bm{\Lambda},\tau}$ is the $\tau$\textsuperscript{th} quantile of $\bm{\Lambda}$.

Compute the centered and scaled version of $\bm{\Pi}$, denoted by $\bm{\Pi}^*$. 

Subsequently, compute the quantile covariance $Q_{\text{cov}, \tau} = \frac{1}{K_{\X}} \Gamma^\top \bm{\Pi}^*$.
\end{algorithm}

In contrast to the FPQR method developed by \cite{BSMMA}, which relies on the quantile covariance of \cite{Dodge2009}, we adapt the partial quantile regression technique introduced by \cite{Alvaro2022} to the domain of functional data and use the quantile covariances outlined in Algorithms~\ref{alg:Choi} and~\ref{alg:Li} to estimate the regression coefficient function $\widehat{\beta}_{\tau}(v,u)$. The fundamental steps of our proposed approach are outlined below:

\begin{enumerate}
\item[1)] \underline{Computation of the FPQR components.} \\
For a given functional response $\Y(u)$ and predictor $\X(v)$, derive the basis expansion coefficients $\bm{\Lambda} = \bm{\zeta}^\top \bm{\Phi}^{1/2}$ and $\bm{\Pi} = \bm{\delta}^\top \bm{\Psi}^{1/2}$ using predetermined constants $K_{\Y}$ and $K_{\X}$. Let $\widetilde{\bm{\Lambda}}$ and $\widetilde{\bm{\Pi}}$ represent the centered versions of $\bm{\Lambda}$ and $\bm{\Pi}$, respectively, and set $\widetilde{\bm{\Lambda}}_{0} = \widetilde{\bm{\Lambda}}$ and $\widetilde{\bm{\Pi}}_{0} = \widetilde{\bm{\Pi}}$. Then, for a given quantile level $\tau$ and the desired number of extracted components $h$, iterate over $h=1,2,\ldots$ as follows.
 \begin{enumerate}
    \item Compute the quantile covariance $Q_{\text{cov}, \tau}$ between $\widetilde{\bm{\Lambda}}_{h-1}$ and $\widetilde{\bm{\Pi}}_{h-1}$ using either Algorithm~\ref{alg:Choi} or~\ref{alg:Li}.
    \item Perform an eigen-decomposition for $Q_{\text{cov}, \tau} Q_{\text{cov}, \tau}^\top$ and retain the eigenvector $\bm{p}_{h}$ corresponding to the largest eigenvalue.
    \item Compute the component vector $\bm{t}_{h}$ for $\widetilde{\bm{\Pi}}$ as $\bm{t}_{h}=\widetilde{\bm{\Pi}}_{h-1} \bm{p}_{h}$.
    \item Calculate the loading vectors $\bm{\delta}_{h}$ for $\widetilde{\bm{\Pi}}$ and $\bm{q}_{h}$ for $\widetilde{\bm{\Lambda}}$ as $\bm{\delta}_{h}=\frac{\widetilde{\bm{\Pi}}_{h-1}^\top \bm{t}_{h}}{\bm{t}_{h}^\top \bm{t}_{h}}$ and $\bm{q}_{h} = \frac{\widetilde{\bm{\Lambda}}_{h-1}^\top \bm{t}_{h}}{\bm{t}_{h}^\top \bm{t}_{h}}$, respectively.
    \item Deflate the matrices $\widetilde{\bm{\Lambda}}_{h-1}$ and $\widetilde{\bm{\Pi}}_{h-1}$ to remove the information captured by component $\bm{t}_{h}$, obtaining $\widetilde{\bm{\Lambda}}_{h} = \widetilde{\bm{\Lambda}}_{h-1} - \bm{t}_{h} \bm{q}_{h}^\top$ and $\widetilde{\bm{\Pi}}_{h} = \widetilde{\bm{\Pi}}_{h-1} - \bm{t}_{h} \bm{\delta}_{h}^\top$, respectively.
\end{enumerate} 

\item[2)] \underline{Estimation of quantile regression parameters for $\bm{\Lambda}$ with retained FPQR components.} \\
Consider the matrix $\bm{T}$ of size $n \times h$, where each column represents the extracted FPQR component $\bm{t}_{h}$. Proceed with quantile regression to model $\bm{\Lambda}$ based on the retained FPQR components obtained in the previous step. This involves estimating:
\begin{equation*}
Q_{\tau}[\bm{\Lambda}_i \vert \bm{T}_i] = \bm{T}_i \bm{B}_{\tau},
\end{equation*} 
where $\bm{B}_{\tau}$ represents the parameter matrix. Utilize a standard loss function, such as the check loss, to estimate the model parameters, that is, solve:
\begin{equation*}
\widehat{\bm{B}}_{\tau} = \underset{\begin{subarray}{c}
  \bm{B}_{\tau}
  \end{subarray}}{\argmin}~ \sum_{i=1}^n \rho_{\tau} \left[ \bm{\Lambda}_i - \bm{T}_i \bm{B}_{\tau} \right].
\end{equation*}

\item[3)] \underline{Regression coefficient function estimation.} \\
The estimated parameter $\widehat{\bm{B}}_{\tau}$ is projected back into the original subspace defined by $\bm{\Pi}$ as $\widehat{\bm{\Omega}}_{\tau}=\bm{P}\left ( \bm{\delta}^\top \bm{P} \right )^{-1} \widehat{\bm{B}}_{\tau}$, where $\bm{P}$ and $\bm{\delta}$ are $K_{\X} \times h$ dimensional matrices, the columns of which consist of $\bm{p}_{h}$ and $\bm{\delta}_{h}$ obtained in the initial step, respectively. Subsequently, the FPQR estimate of the regression coefficient function is computed as follows:
\begin{equation*}
\widehat{\beta}_{\tau} \left (v, u\right) = \left[(\bm{\Psi}^{1/2})^{-1} \widehat{\bm{\Omega}}_{\tau} (\bm{\Phi}^{1/2})^{-1} \right ]\bm{\phi}(u) \bm{\psi}(v).
\end{equation*}
\end{enumerate}
Finally, the $\tau^\textsuperscript{th}$ quantile of the functional response conditioned on the functional predictor is derived as:
\begin{equation*}
\widehat{Q}_{\tau}\left [ \Y \vert \X \right ]= \int_{\mathcal{I}_{\X}} \X(v) \widehat{\beta}_{\tau}(v,u)dv. 
\end{equation*}

In contrast to the prevailing FPQR method introduced by \cite{BSMMA}, which necessitates resolving $K_{\Y} \times K_{\X}$ univariate quantile regression models to estimate the regression coefficient function $\beta_{\tau}(v,u)$, the proposed FPQR algorithm, utilizing the quantile covariance method outlined in \cite{Choi2018}, requires resolving $2 \times K_{\Y} \times K_{\X}$ univariate quantile regression models, leading to a substantial increase in computational time. Nevertheless, as evidenced by \cite{Alvaro2022}, and corroborated by our numerical findings in Section~\ref{sec:montecarlo}, this method yields superior parameter estimates compared to the FPQR approach by \cite{BSMMA}. Conversely, the proposed FPQR algorithm founded on the quantile covariance technique elucidated in \cite{Li2015} adopts a classical covariance-based strategy, facilitating seamless extension to multivariate response models. This characteristic renders it notably faster than its counterparts.

The computation of $\widehat{\beta}_{\tau} \left(v, u\right)$ requires optimizing three parameters: the number of B-spline basis functions for the response $K_{\Y}$, the number of B-spline basis functions for the predictor $K_{\X}$, and the number of FPQR components $h$. To determine the optimal values for these parameters, we employ a three-dimensional grid search algorithm combined with 5-fold cross-validation. Specifically, the grid search evaluates $K_{\Y} \in [4, 5, 8, 10, 20]$, $K_{\X} \in [4, 5, 8, 10, 20]$, and $h \in [1,2,3,4,5]$. For each combination, 5-fold cross-validation is performed, and the Bayesian information criterion (BIC) on the test sets is computed as follows:
\begin{equation*}
\text{BIC} = \ln \bigg \Vert \left[ \sum_{i=1}^n \rho_{\tau} \left( \Y_i(u) - \int_{\mathcal{I}_{\X}} \X_{i}(v) \widehat{\beta}_{\tau} \left ( v,u \right ) dv \right) \right] \bigg \Vert_{\mathcal{L}_2} + \omega \ln(n),
\end{equation*}
where $\omega = K_{\Y} + K_{\X} + h$ and $\Vert \cdot \Vert_{\mathcal{L}_2}$ is the $\mathcal{L}_2$ norm. The optimal tuning parameters are identified as those yielding the smallest mean BIC.

It is important to note that the computation of $\widehat{\beta}_{\tau} \left ( v,u \right )$ in the FPQR approach relies on FPQR components, which are updated iteratively. This iterative nature introduces complexities in deriving explicit theoretical results (for more details, see \cite{Delaigle2012}). Consequently, this study focuses on comparing the finite-sample performance and computation times of the proposed methods with existing ones rather than considering the asymptotic properties of the proposed estimators.

\section{Monte Carlo experiments} \label{sec:montecarlo}

Several Monte Carlo experiments are conducted to evaluate and compare the computational efficiency and numerical accuracy of the proposed FPQR methodologies against the established FPQR method by \cite{BSMMA}. Additionally, the finite-sample performances of the proposed methods are evaluated against the function-on-function linear regression model developed by \cite{Zhou2021} (abbreviated as ``FPLS''), which is based on the FPLS approach. The code for the FPLS method can be accessed at \url{https://github.com/ZhiyangGeeZhou/fAPLS}. All numerical computations are executed using \Rlogo\ version 4.3.1 on hardware equipped with an Intel Core i7 8550U 1.8 GHz Turbo processor. The FPQR algorithms, including those proposed, are implemented within the \Rlogo\ package \texttt{ffpqr}, available at \url{https://github.com/MugeMutis/ffpqr}.

\begin{figure}[!htb]
\centering
\includegraphics[width=8.5cm]{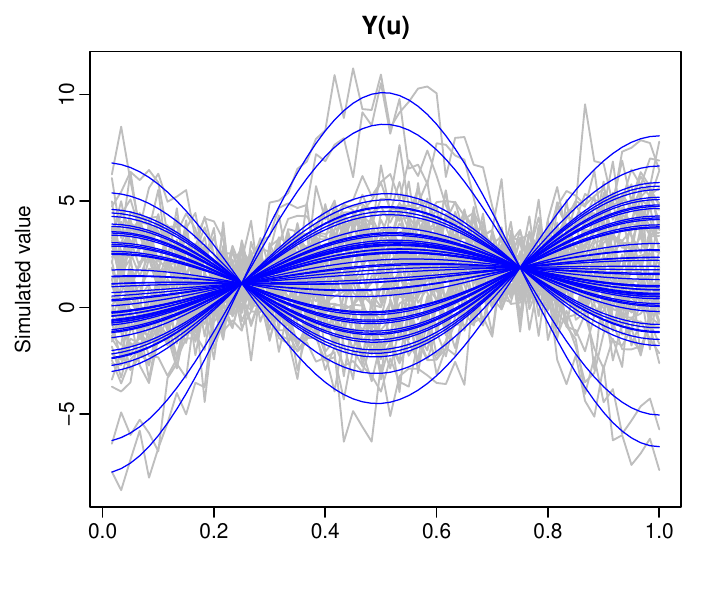}
\qquad
\includegraphics[width=8.5cm]{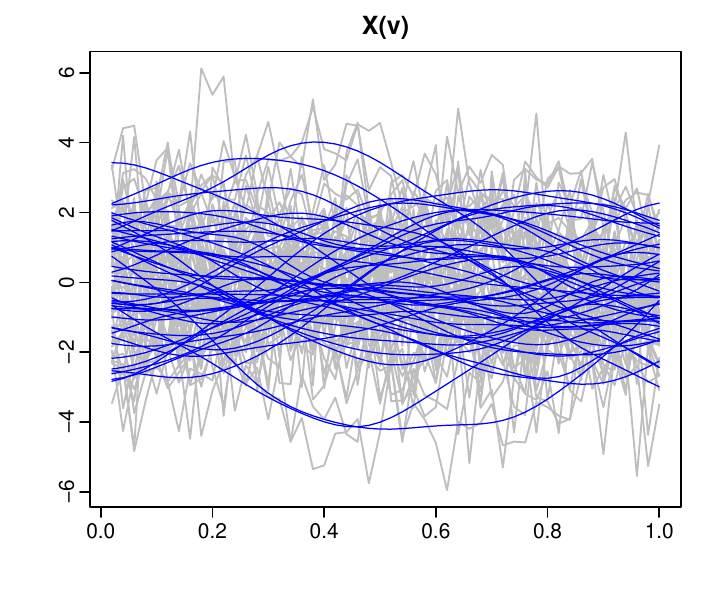}
\\  
\includegraphics[width=8.5cm]{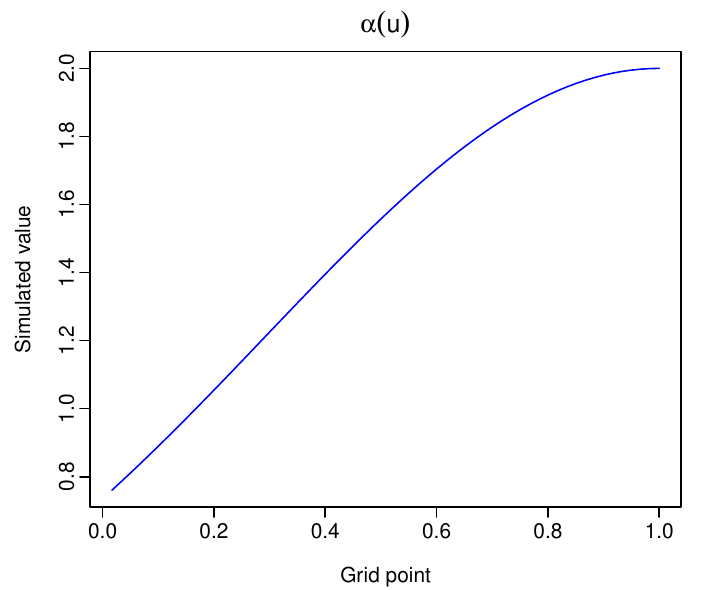}
\qquad
\includegraphics[width=8.5cm]{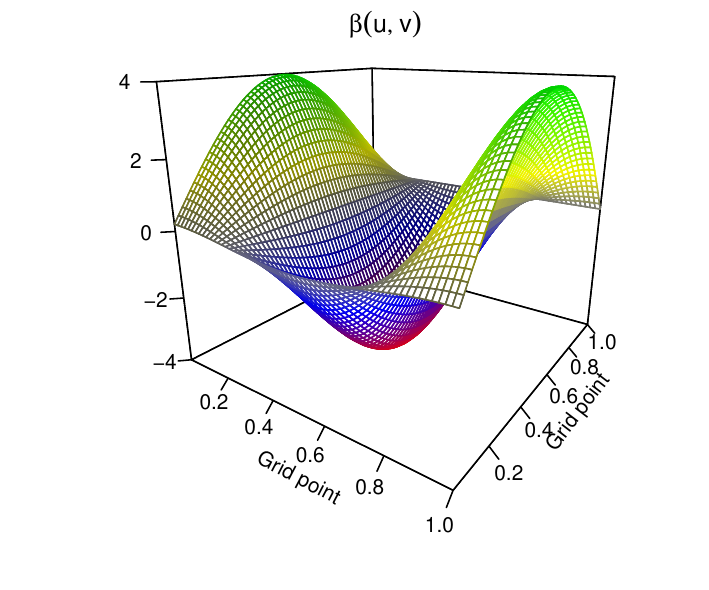}
\caption{\small{Graphs depicting 50 generated sample curves (true and noisy observations are shown as blue and gray curves, respectively) for functional variables are presented (top row; functional response on the left panel and functional predictor on the right panel), along with true regression coefficient functions (bottom row; intercept function on the left panel and bivariate regression coefficient function on the right panel). The data are generated  when $\varepsilon_{i} \sim \text{N}(0,1)$}.}
  \label{fig:Fig_1}
\end{figure}

In the experiments, we adopt the data generation process described by \cite{Cai2021}. The functional predictor $\X_i(v)$ is generated at 50 equidistant points within the interval $[0,1]$, that is $v \in \lbrace r/50: r = 1, \ldots, 50 \rbrace$, while the functional response $\Y_i(u)$ is generated at 60 equidistant points within $[0,1]$, that is $u \in \lbrace j/60: j = 1, \ldots, 50 \rbrace$. The functional predictor is generated using the process:
\begin{equation*}
\X_{i} \left ( v \right ) = \sum_{k=1}^{10} \frac{1}{k^{2}} \left\{ \zeta _{i1,k} \sqrt{2} \sin \left ( k \pi v \right ) + \zeta _{i2,k}\sqrt{2} \cos \left ( k \pi v \right ) \right\},
\end{equation*}
where the random variables $\zeta_{i1,k}$ and $\zeta_{i2,k}$ are generated from the standard normal distribution. The intercept and bivariate regression coefficient functions are specified as $\alpha \left ( u \right ) = 2e^{-\left ( u-1 \right )^{2}}$ and $ \beta \left ( v,u \right )= 4 \cos \left ( 2\pi u \right ) \sin \left ( \pi v \right )$, respectively. Subsequently, the functional response is generated according to the function-on-function linear regression model:
\begin{equation*}
\Y_{i} \left ( u \right ) = \alpha \left(u \right) +  \int_0^1 \X_{i} \left ( v \right ) \beta \left ( v,u \right ) dv + \varepsilon_{i}\left ( u \right ),
\end{equation*}
where $\alpha \left(u \right)$ is the intercep function and $\varepsilon_{i} \left(u\right)$ is the error function. Throughout the simulations, we consider three different distributions for the error function: standard normal distribution $\text{N}(0,1)$; student's t distribution with five degrees of freedom $t_{(5)}$; and chi-square distribution with one degree of freedom $\chi^2_{(1)}$. Note that the functional predictors are distorted by the Gaussian noise (generated from the standard normal distribution) before fitting the regression models. Figure~\ref{fig:Fig_1} illustrates a graphical representation of the generated random curves and parameter functions.

Throughout the experiments, six different sample sizes $n_{\text{train}} = [50, 100, 250, 500, 1000, 5000]$ are considered. Utilizing the generated training datasets, we construct models and assess the estimation performance of the methods through the calculation of root relative integrated squared percentage estimation errors (RRISPEE) for both the intercept and regression coefficient function consistent with \cite{BHS2024}:
\begin{equation*}
\text{RRISPEE}(\widehat{\alpha}_{\tau}) = 100 \times \sqrt{\frac{\Vert \alpha(u) - \widehat{\alpha}_{\tau}(u) \Vert_2^2}{\Vert \alpha(u) \Vert_2^2}}, \quad \text{RRISPEE}(\widehat{\beta}_{\tau}) = 100 \times \sqrt{\frac{\Vert \beta(v,u) - \widehat{\beta}_{\tau} \left ( v,u \right ) \Vert_2^2}{\Vert \beta(v,u) \Vert_2^2}},
\end{equation*}
where $\widehat{\alpha}_{\tau}(u)$ and $\widehat{\beta}_{\tau} \left ( v,u \right )$ respectively denote the estimates of $\alpha(u)$ and $\beta \left ( v,u \right )$. For each sample size, we generate separate test sets with a size of $n_{\text{test}}=1000$. Assessing the predictive capability of the methods involves employing the trained models derived from the training sets on the test samples. Subsequently, the root mean squared percentage error (RMSPE) is calculated using the following formula:
\begin{equation*}
\text{RMSPE} = 100 \times \sqrt{\frac{\Vert \Y(u) - \widehat{Q}_{\tau}(u) \Vert_2^2}{\Vert \Y(u) \Vert_2^2}},
\end{equation*}
where $\widehat{Q}_{\tau}(u)$ is the $\tau^\textsuperscript{th}$ quantile of the functional response $\Y$ in the test sample. It is worth noting that we conduct 100 Monte Carlo runs for different quantile levels $\tau \in[0.01, 0.1, 0.25, 0.5, 0.75, 0.90, 0.99]$ and the FPQR methods are compared with the FPLS for only $\tau = 0.5$. Note that FPLS assumes that both functional response and functional predictor variables are mean-zero processes and, thus, only the $\text{RRISPEE}(\widehat{\beta}_{\tau})$ is computed for this method. In the numerical analyses, the tuning parameters in the FPQR methods, that is, $K_{\Y}$, $K_{\X}$, and $h$, are determined using 5-fold cross-validation. We denote the FPQR methods based on the quantile covariances of \cite{Dodge2009}, \cite{Choi2018}, and \cite{Li2015} as Dodge, Choi, and Li, respectively.

\begin{figure}[!htb]
\centering
\includegraphics[width=5.6cm]{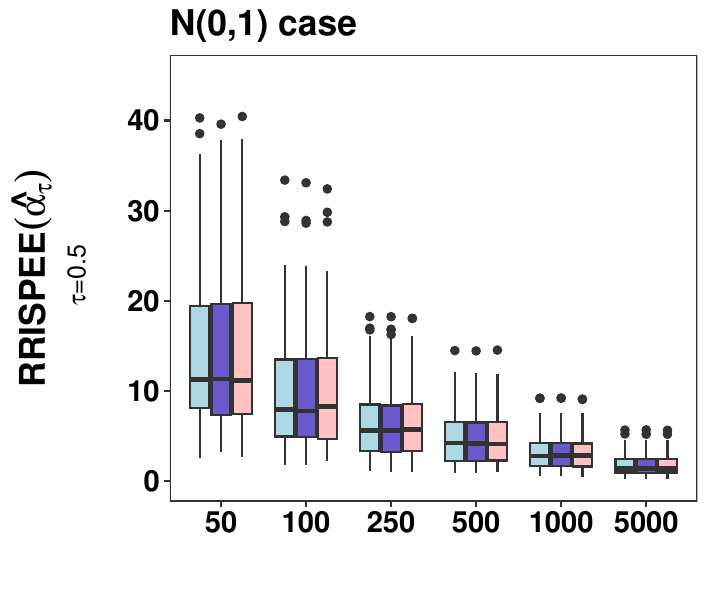}
\quad
\includegraphics[width=5.6cm]{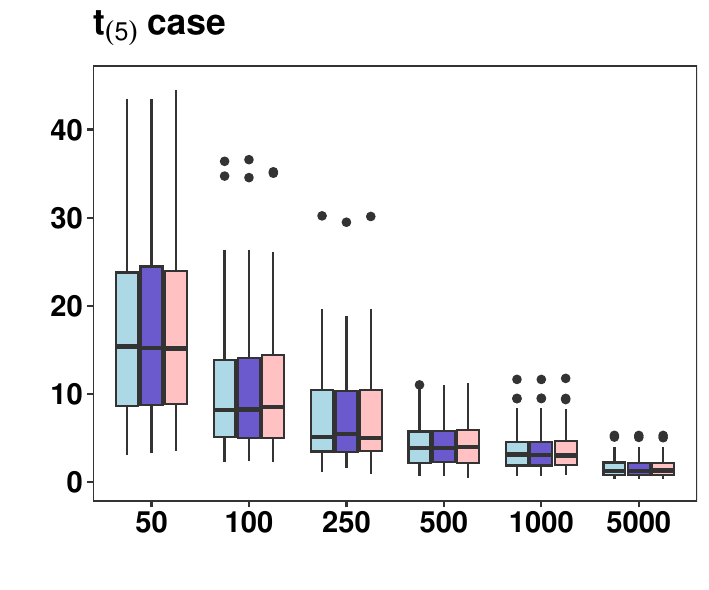}
\quad
\includegraphics[width=5.6cm]{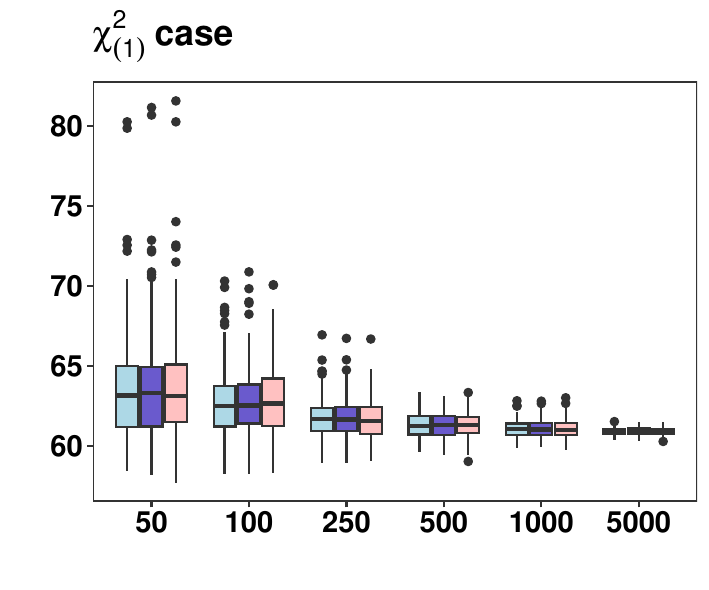}
\\
\includegraphics[width=5.6cm]{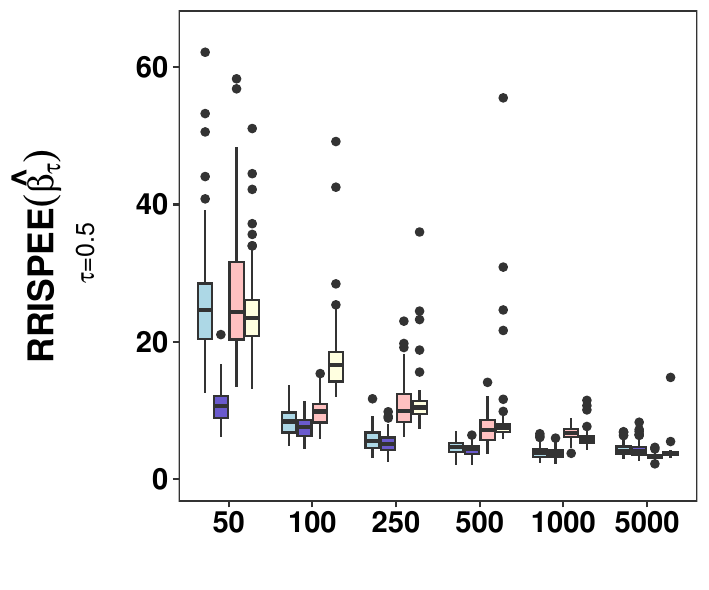}
\quad
\includegraphics[width=5.6cm]{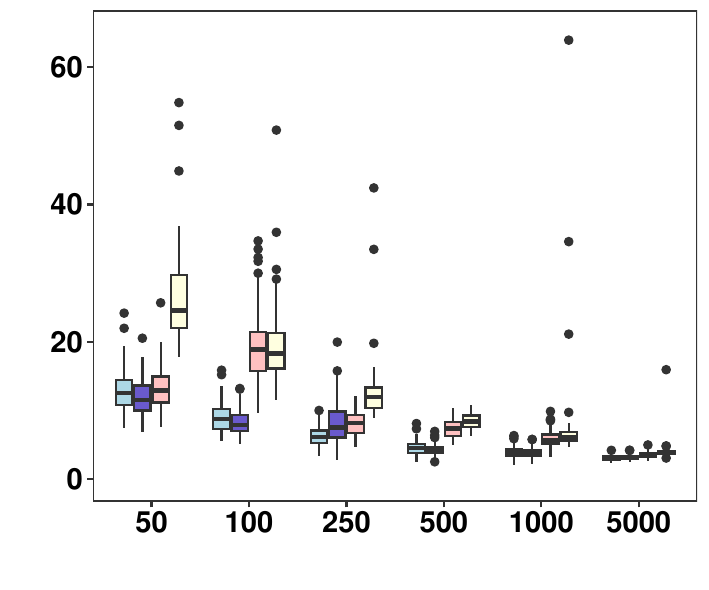}
\quad
\includegraphics[width=5.6cm]{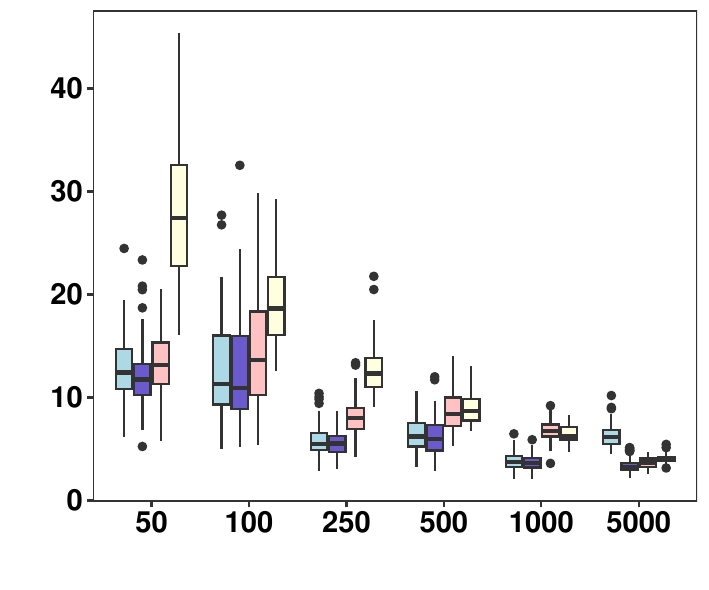}
\\
\includegraphics[width=5.6cm]{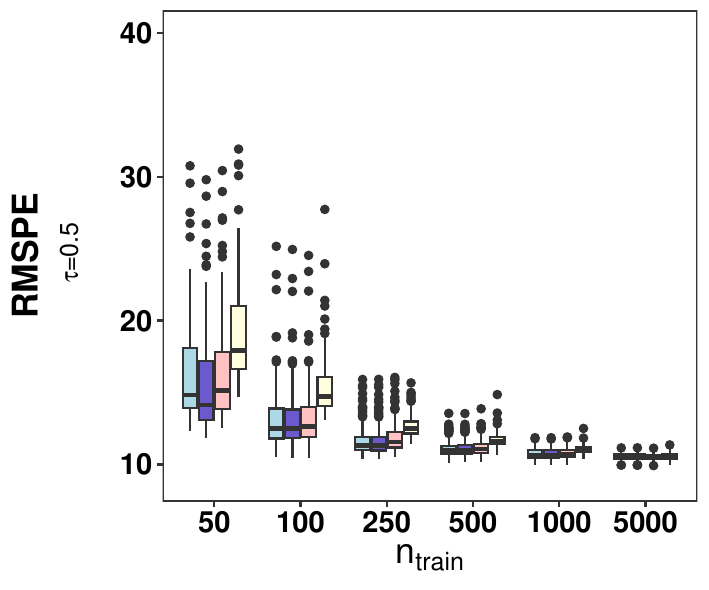}
\quad
\includegraphics[width=5.6cm]{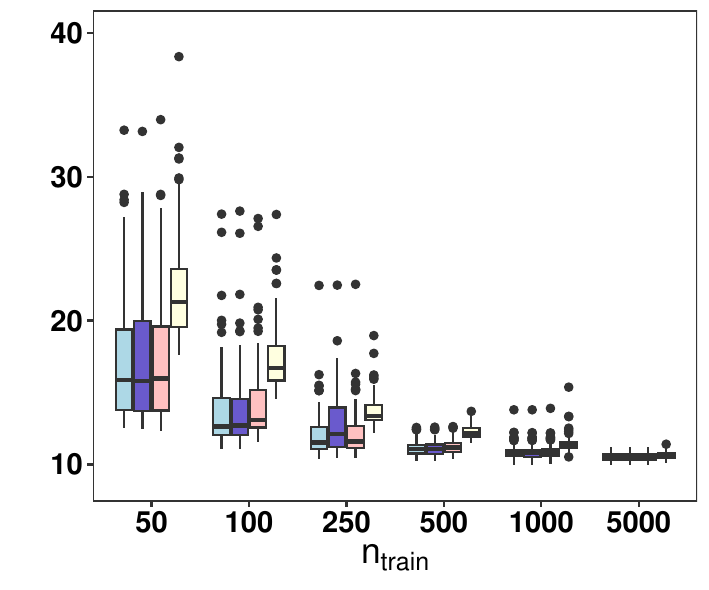}
\quad
\includegraphics[width=5.6cm]{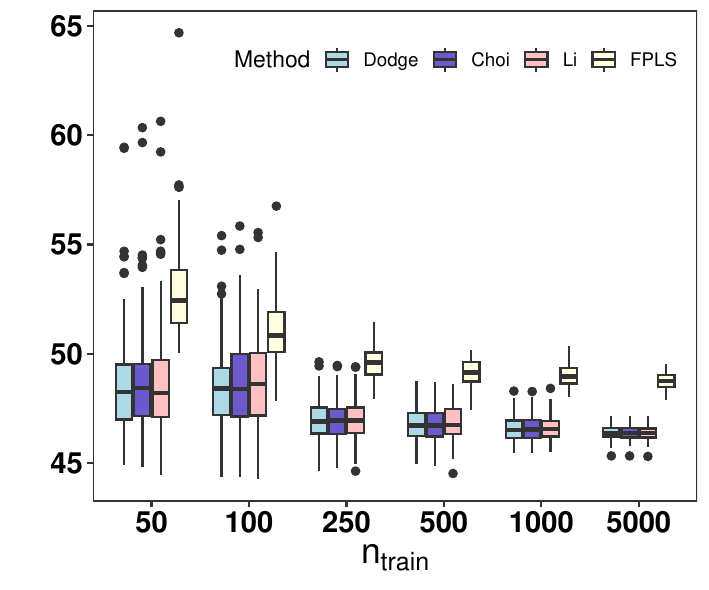}
\caption{\small{Boxplots of the calculated $\text{RREISPEE} \left ( \widehat{\alpha}_{\tau}  \right )$ (first row), $\text{RRISPEE} \left ( \widehat{\beta}_{\tau}  \right )$ (second row), and RMSPE (third row) values for the Dodge, Choi, Li, and FPLS methods under different sample sizes when $\tau=0.5$. Note that $\text{RREISPEE} \left ( \widehat{\alpha}_{\tau}  \right )$ cannot be calculated for the FPLS method. Rows and columns represent performance metrics and error distributions, respectively. $\text{RREISPEE} \left ( \widehat{\alpha}_{\tau}  \right )$, $\text{RRISPEE} \left ( \widehat{\beta}_{\tau}  \right )$, and RMSPE are computed under three error distributions; $\text{N}(0,1)$ (first column), $t_{(5)}$ (second column), and $\chi^2_{(1)}$ (third column)}.}
\label{fig:Fig_2}
\end{figure}

The computed $\text{RRISPEE}(\widehat{\alpha}_{\tau})$, $\text{RRISPEE}(\widehat{\beta}_{\tau})$, and RMSPE for $\tau = 0.5$ are presented in Figure~\ref{fig:Fig_2}. Analyzing Figure~\ref{fig:Fig_2} reveals that for all error distributions, the FPQR methods provide competitive $\text{RRISPEE}(\widehat{\alpha}_{\tau})$ values, which decrease as the training sample size increases. Specifically, when the errors follow $\text{N}(0,1)$ and $t_{(5)}$ distributions, all FPQR methods yield similar $\text{RRISPEE}(\widehat{\alpha}_{\tau})$ values, which are significantly smaller than those obtained under the $\chi^2_{(1)}$ distribution. For $\text{RRISPEE}(\widehat{\beta}_{\tau})$, the proposed Choi method demonstrates superior performance over other FPQR methods and FPLS for small to moderate sample sizes, irrespective of the error distribution. However, when the training sample size is large, all FPQR methods and FPLS yield comparable $\text{RRISPEE}(\widehat{\beta}_{\tau})$ values. Additionally, under the skewed $\chi^2_{(1)}$ error distribution, FPQR methods exhibit similar or improved $\text{RRISPEE}(\widehat{\beta}_{\tau})$ values compared to the $\text{N}(0,1)$ and $t_{(5)}$ distributions. Conversely, the FPLS method results in significantly larger $\text{RRISPEE}(\widehat{\beta}_{\tau})$ values under the $\chi^2_{(1)}$ distribution than under the $\text{N}(0,1)$ and $t_{(5)}$ distributions. Regarding RMSPE, Figure~\ref{fig:Fig_2} illustrates that all FPQR methods outperform FPLS across all error distributions. This outcome is attributed to the superior ability of FPQR methods to smooth noisy data compared to FPLS. While the proposed Choi method slightly improves RMSPE values under the $\text{N}(0,1)$ distribution, all methods generally produce similar RMSPE values under symmetric error distributions, which are considerably smaller than those obtained under $\chi^2_{(1)}$ distributed errors. Notably, the error metrics for $\chi^2_{(1)}$ distributed errors differ significantly from those for $\text{N}(0,1)$ and $t_{(5)}$ distributed errors. Consequently, the y-axes of the boxplots for $\chi^2_{(1)}$ distributed errors differ from those of the other two cases.

\begin{figure}[!htb]
\centering
\includegraphics[width=4.7cm]{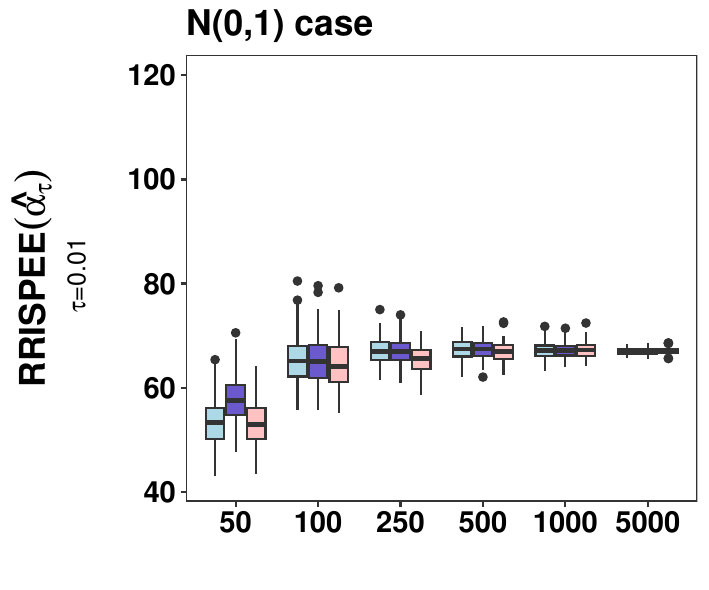}
\quad
\includegraphics[width=4.7cm]{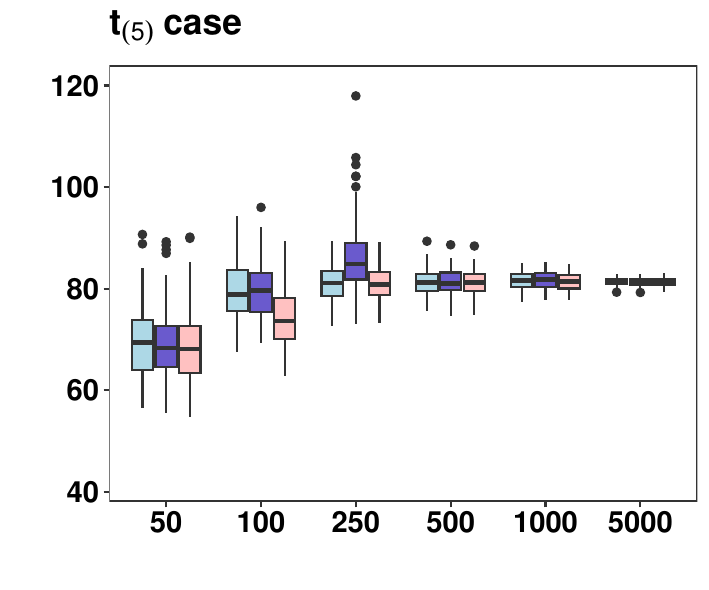}
\quad
\includegraphics[width=4.7cm]{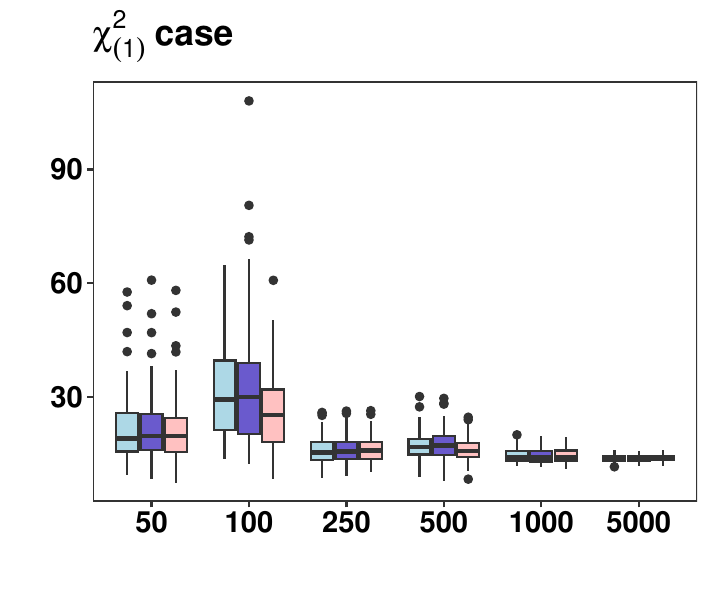}
\\
\includegraphics[width=4.7cm]{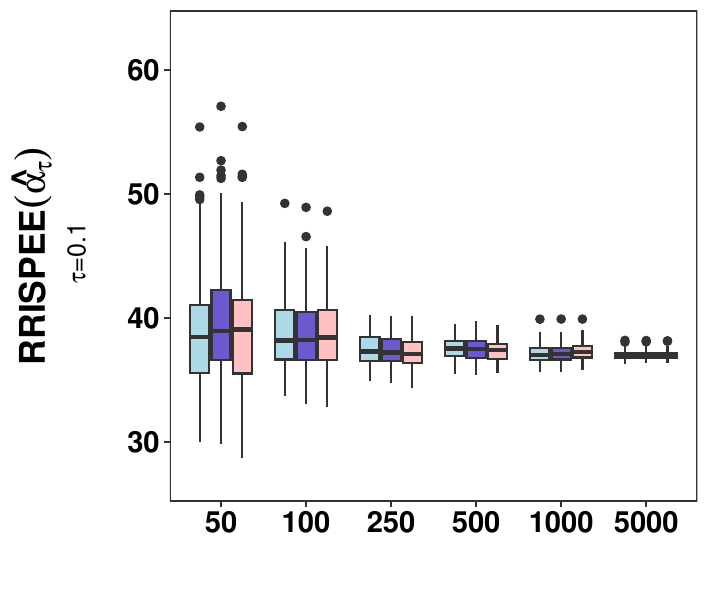}
\quad
\includegraphics[width=4.7cm]{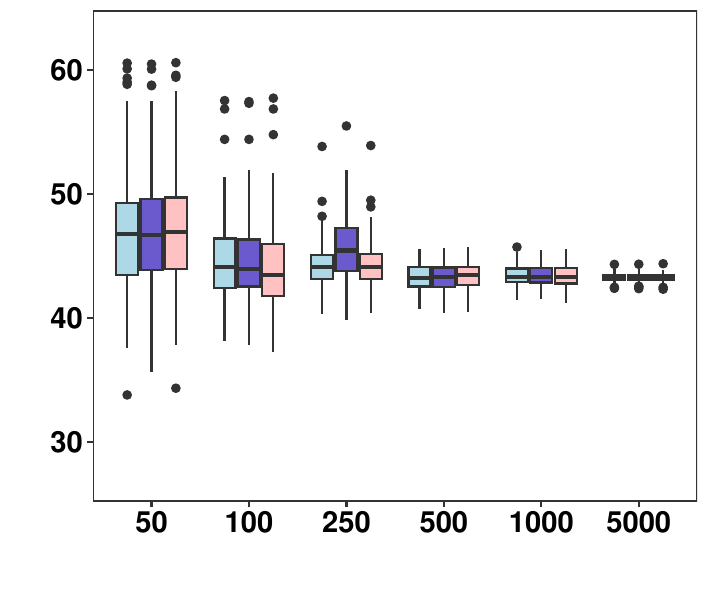}
\quad
\includegraphics[width=4.7cm]{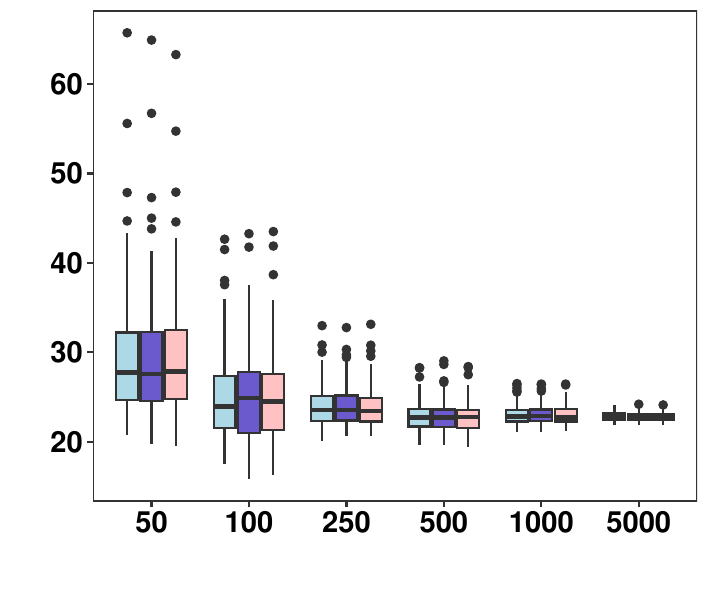}
\\
\includegraphics[width=4.7cm]{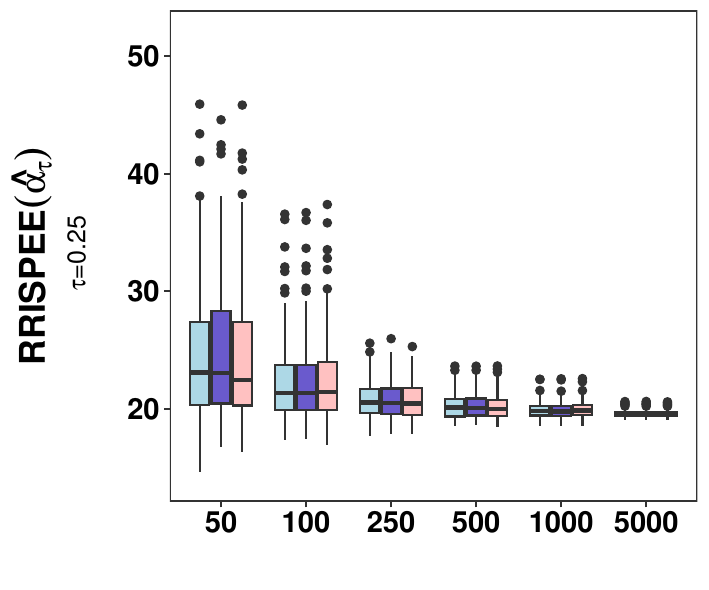}
\quad
\includegraphics[width=4.7cm]{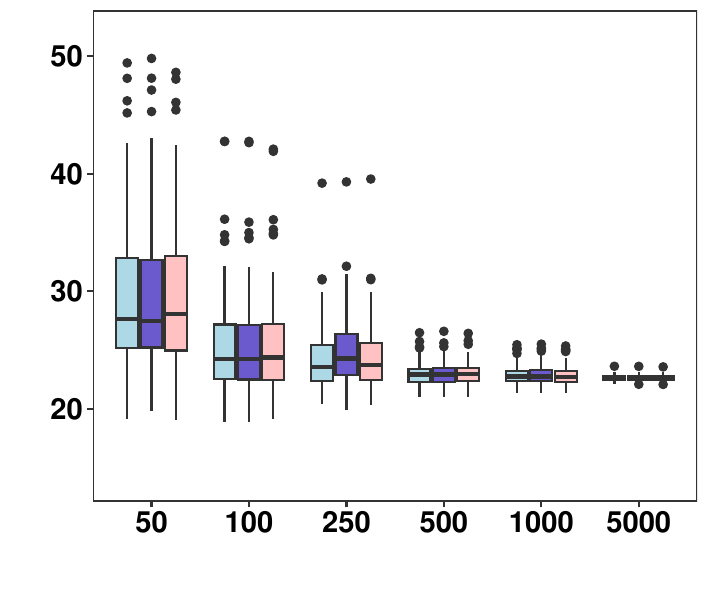}
\quad
\includegraphics[width=4.7cm]{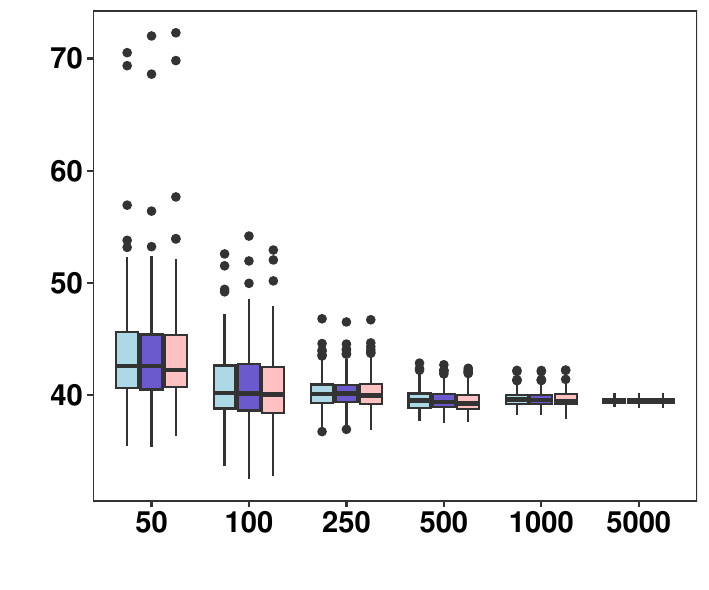}
\\
\includegraphics[width=4.7cm]{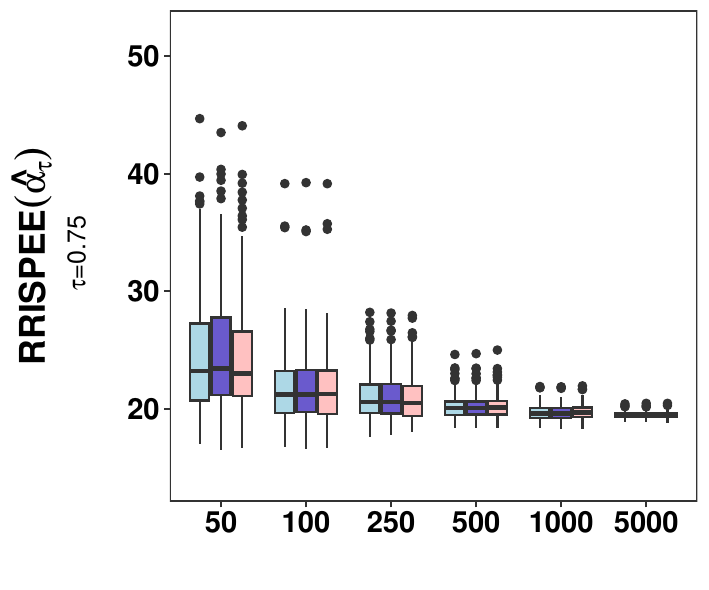}
\quad
\includegraphics[width=4.7cm]{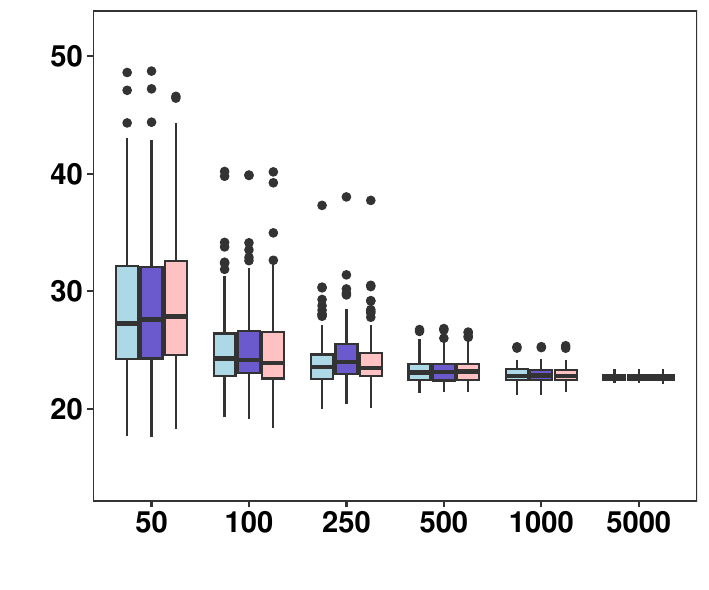}
\quad
\includegraphics[width=4.7cm]{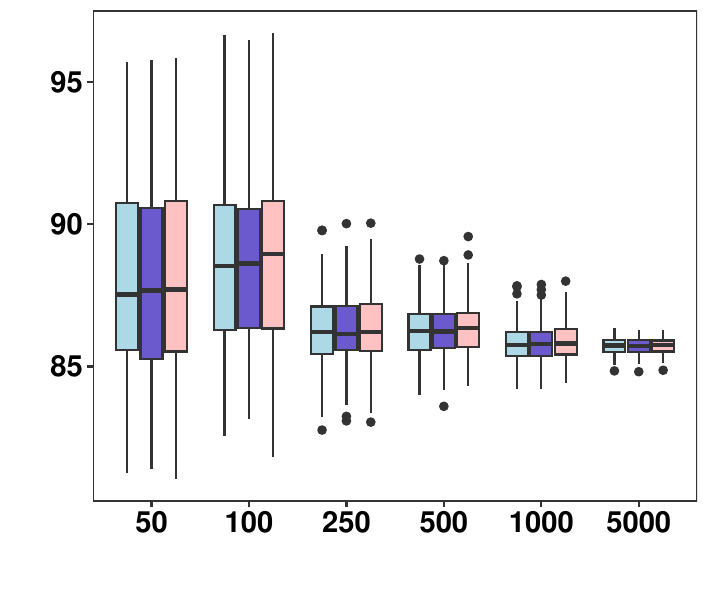}
\\
\includegraphics[width=4.7cm]{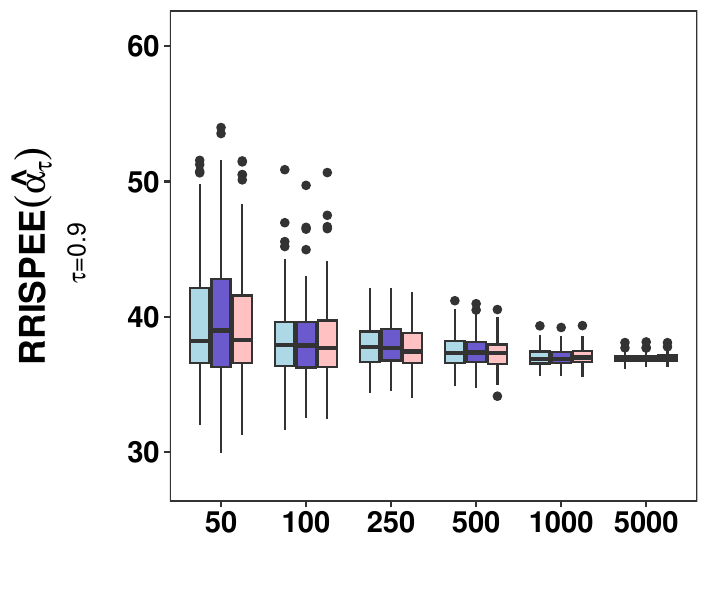}
\quad
\includegraphics[width=4.7cm]{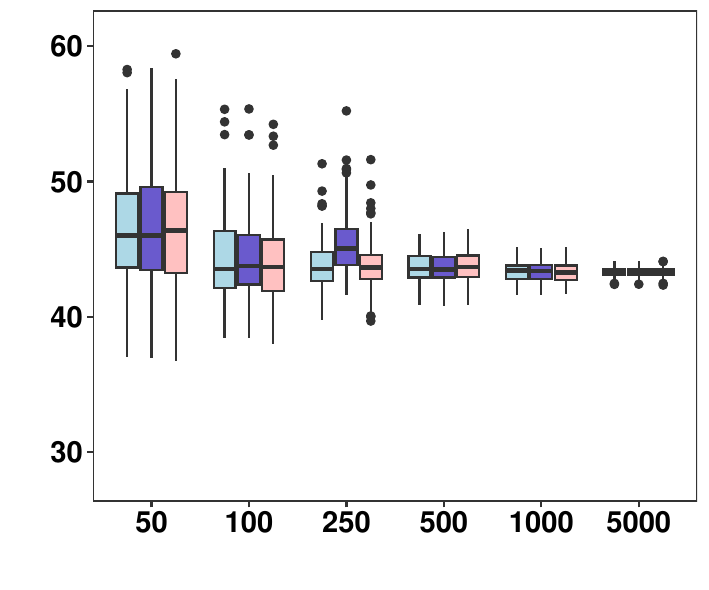}
\quad
\includegraphics[width=4.7cm]{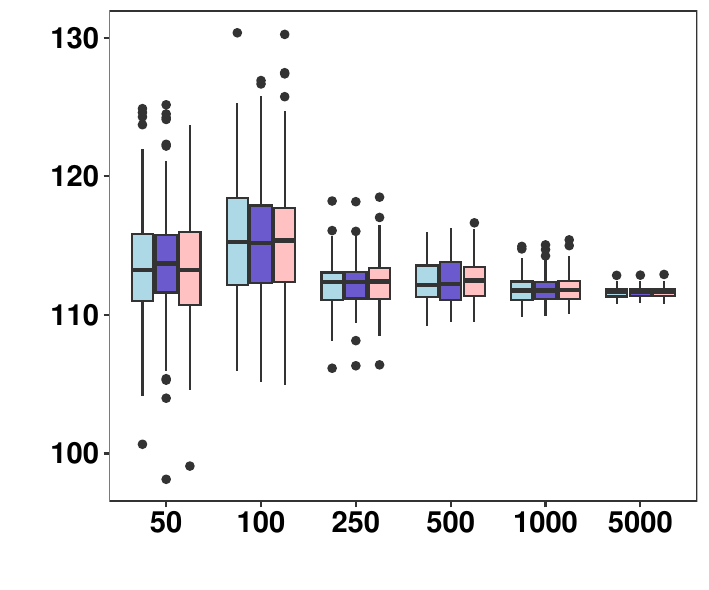}
\\
\includegraphics[width=4.7cm]{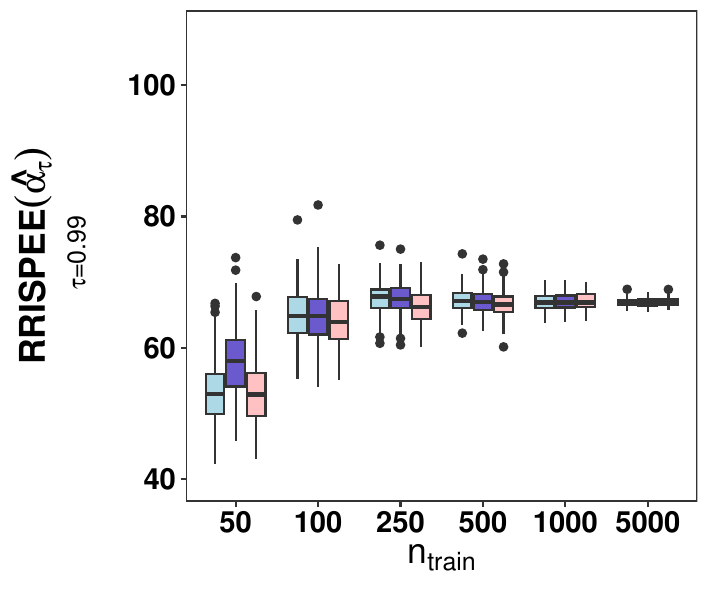}
\quad
\includegraphics[width=4.7cm]{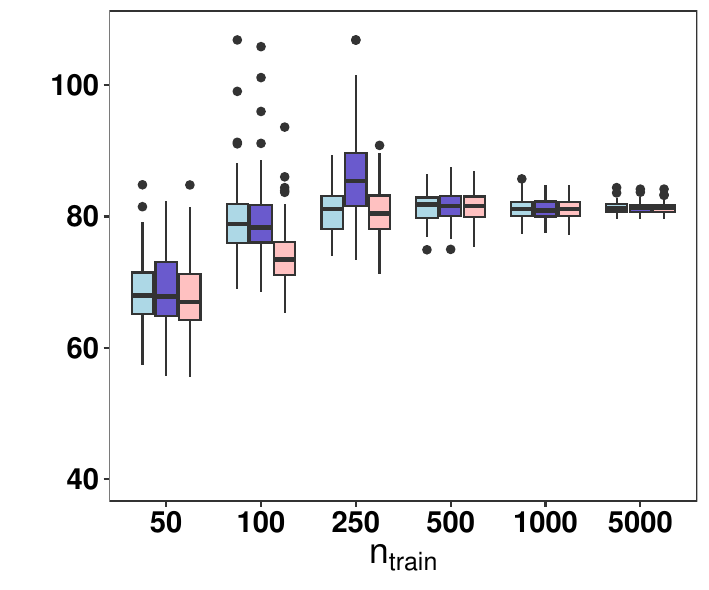}
\quad
\includegraphics[width=4.7cm]{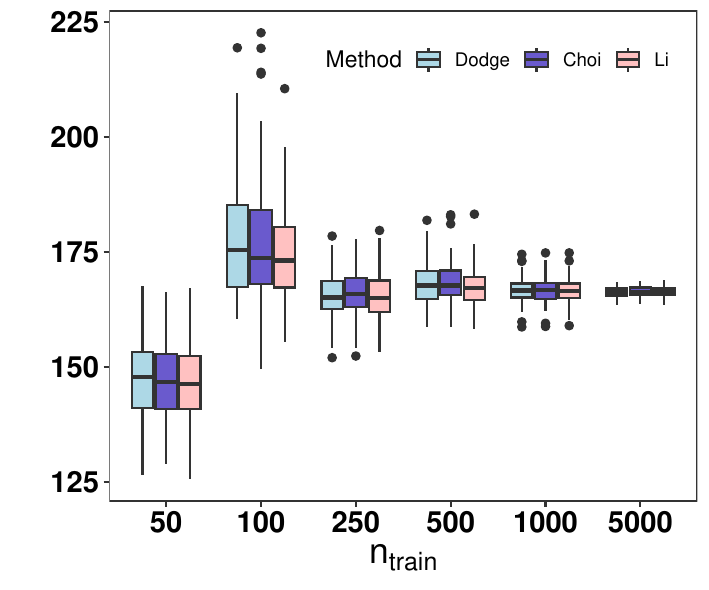}
\caption{\small{Boxplots of the calculated $\text{REISPE} \left ( \widehat{\alpha}_{\tau}  \right )$ values for the Dodge, Choi, and Li methods under different sample sizes. Error metrics are computed for different quantile levels (rows) and error distributions (columns)}.}\label{fig:Fig_3}
\end{figure}

\begin{figure}[!htb]
\centering
\includegraphics[width=4.6cm]{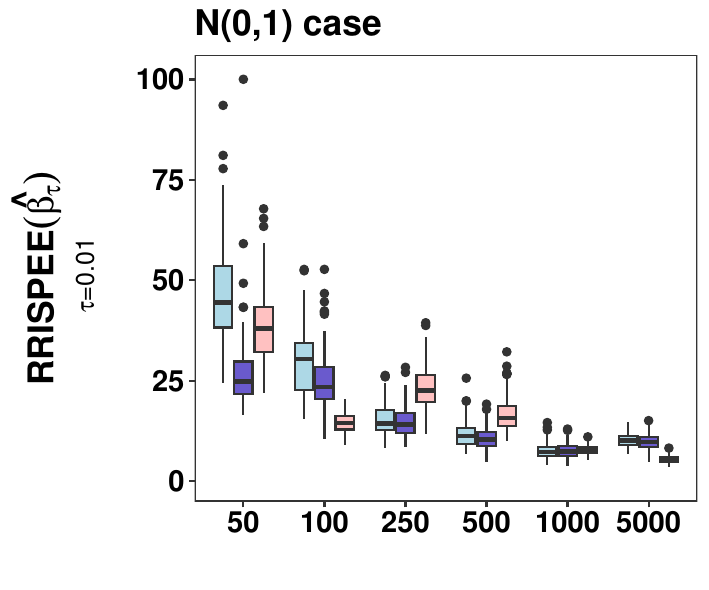}
\quad
\includegraphics[width=4.6cm]{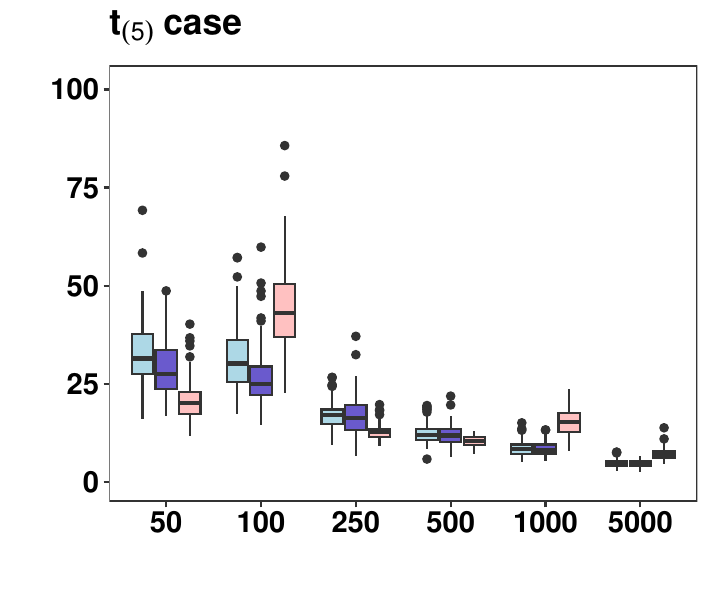}
\quad
\includegraphics[width=4.6cm]{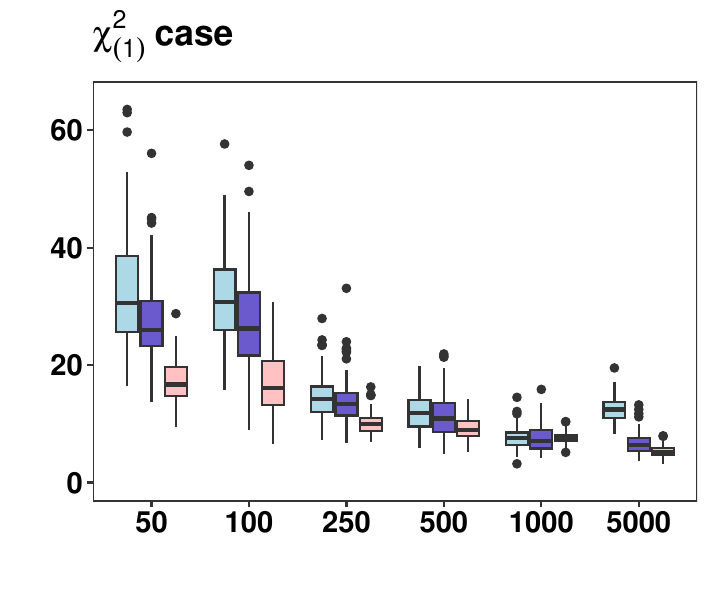}
\\
\includegraphics[width=4.6cm]{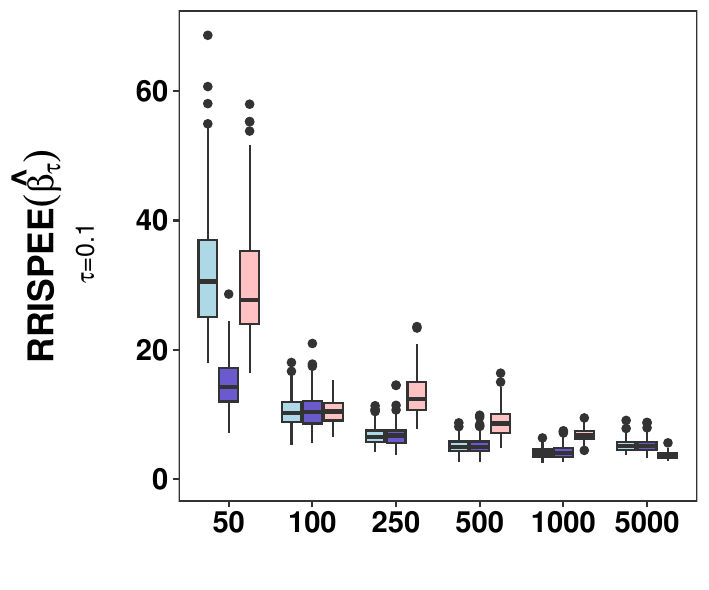}
\quad
\includegraphics[width=4.6cm]{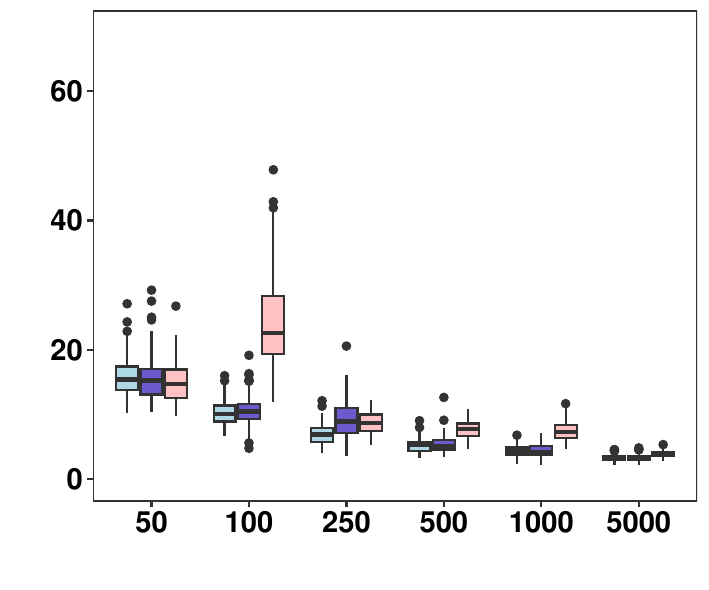}
\quad
\includegraphics[width=4.6cm]{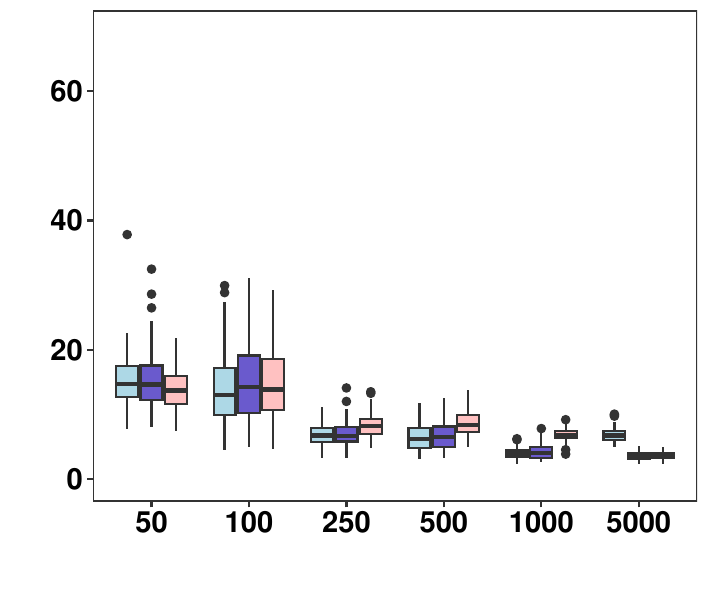}
\\
\includegraphics[width=4.6cm]{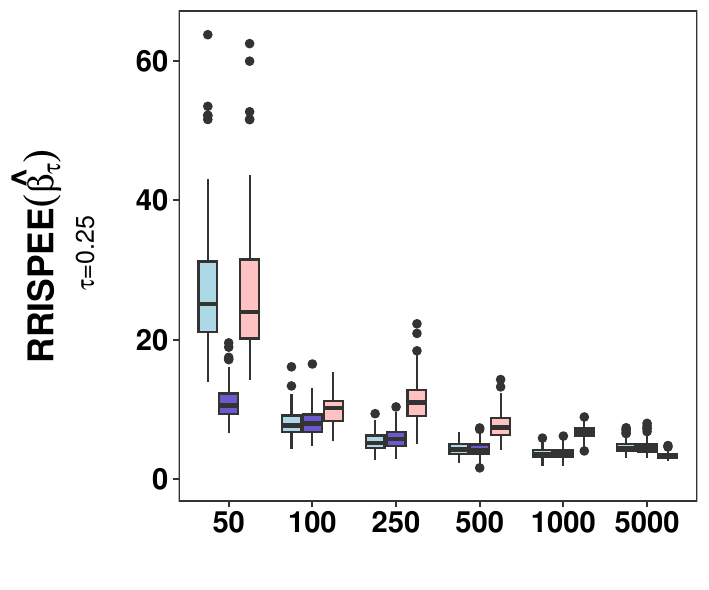}
\quad
\includegraphics[width=4.6cm]{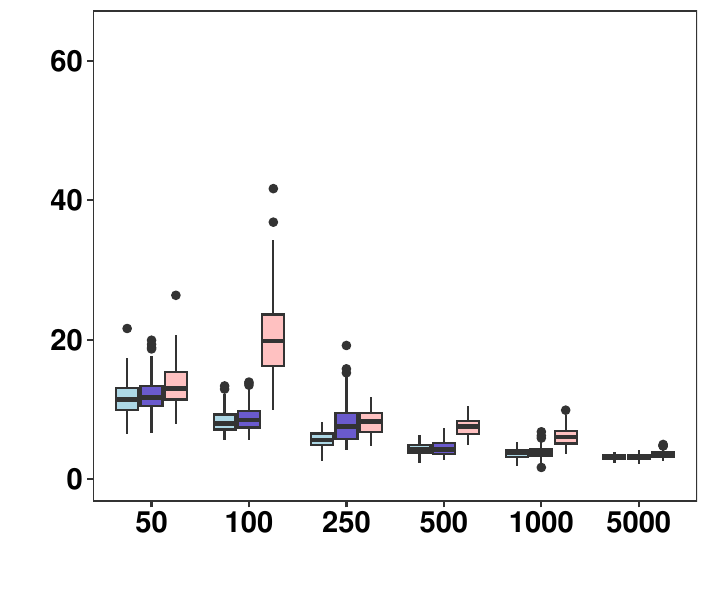}
\quad
\includegraphics[width=4.6cm]{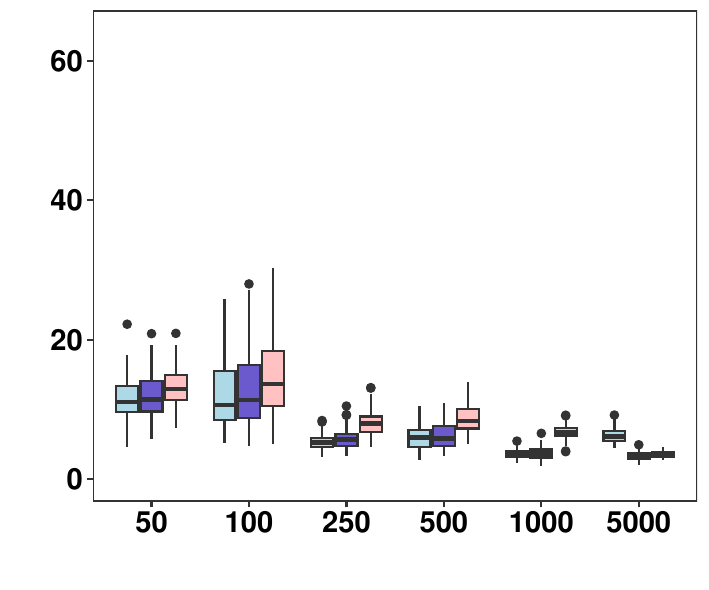}
\\
\includegraphics[width=4.6cm]{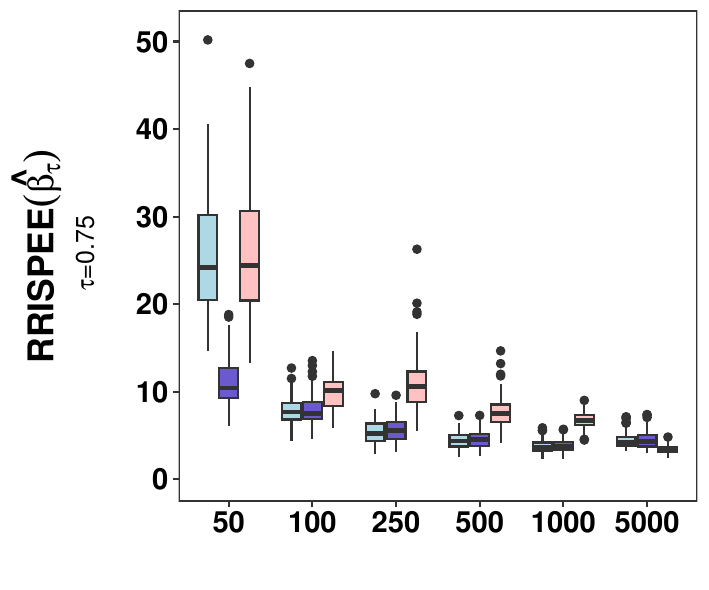}
\quad
\includegraphics[width=4.6cm]{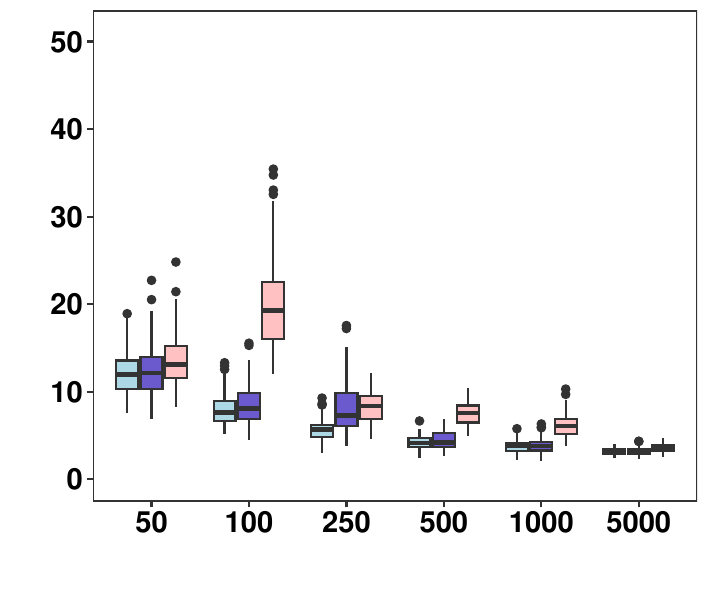}
\quad
\includegraphics[width=4.6cm]{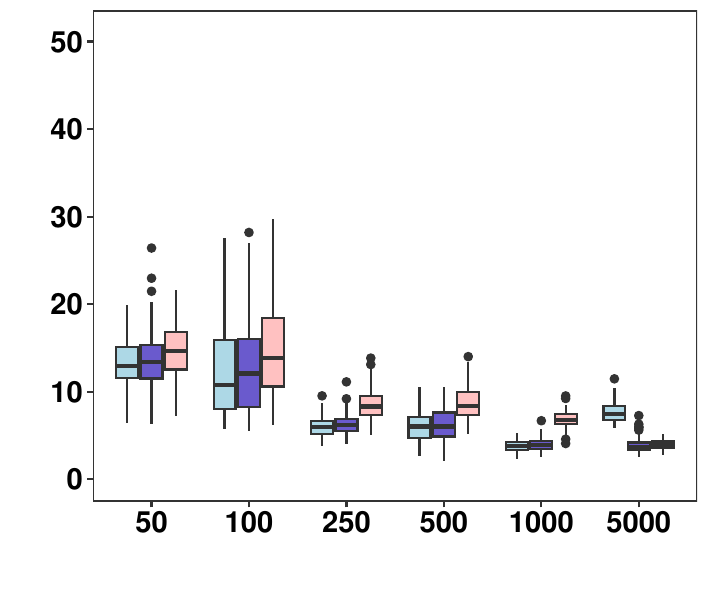}
\\
\includegraphics[width=4.6cm]{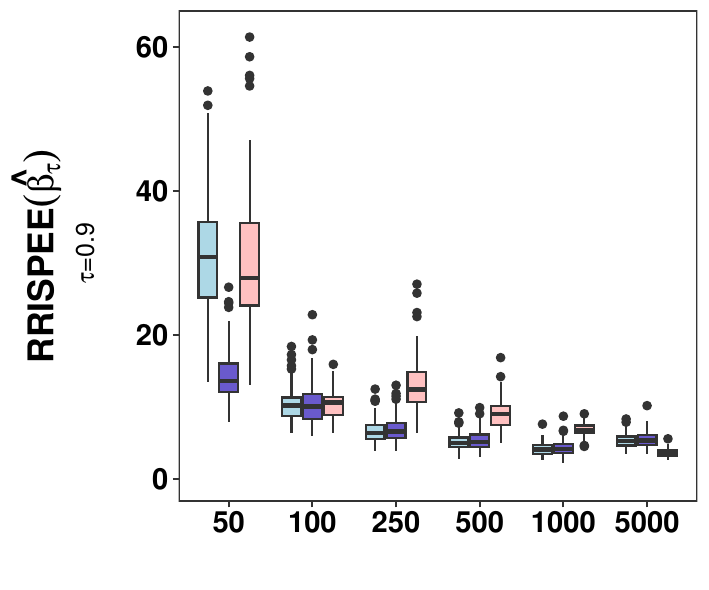}
\quad
\includegraphics[width=4.6cm]{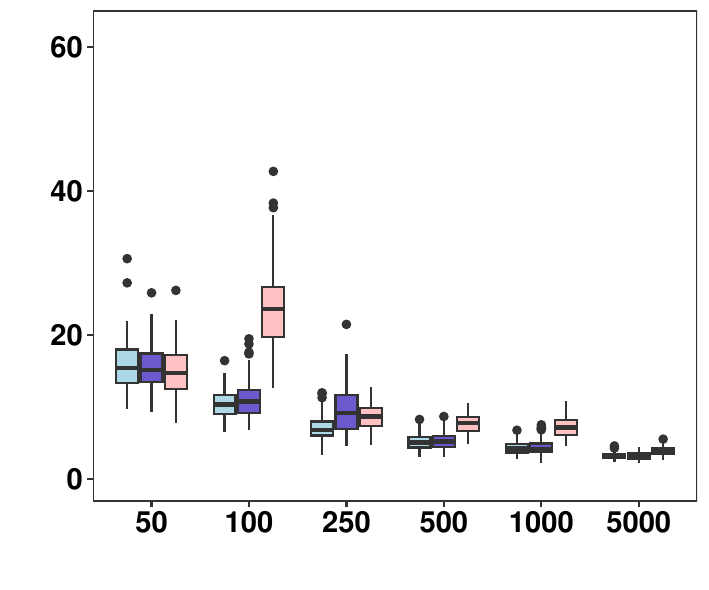}
\quad
\includegraphics[width=4.6cm]{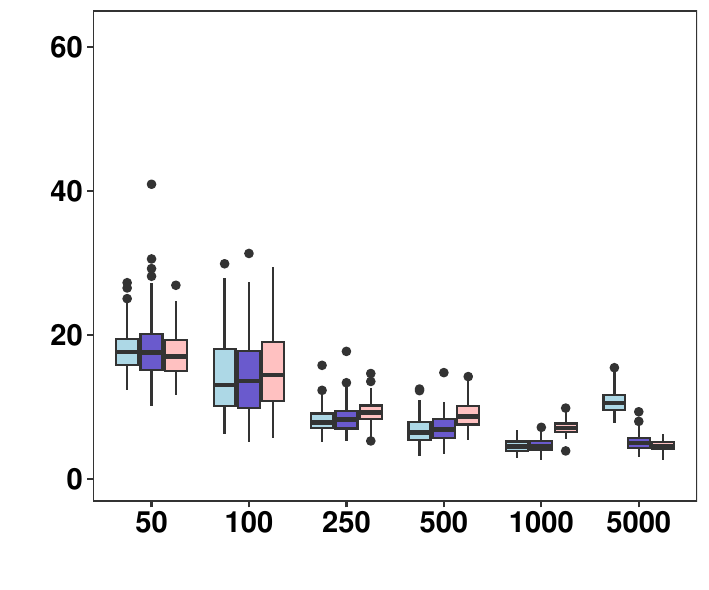}
\\
\includegraphics[width=4.6cm]{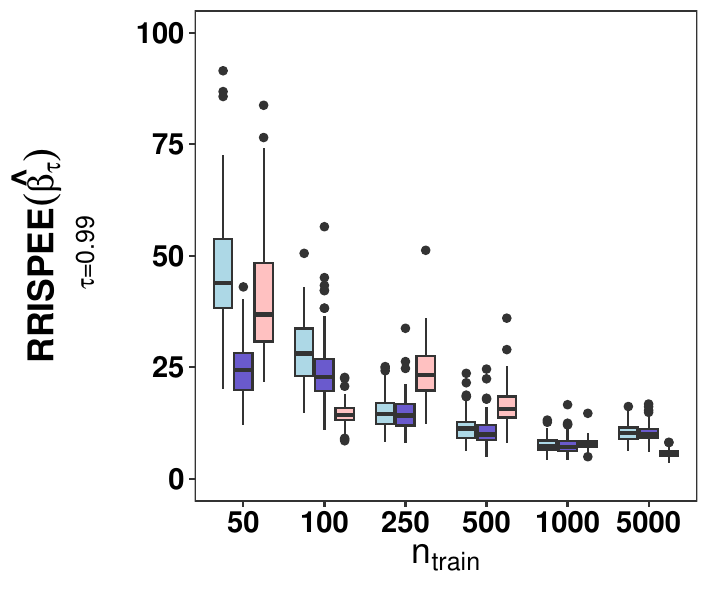}
\quad
\includegraphics[width=4.6cm]{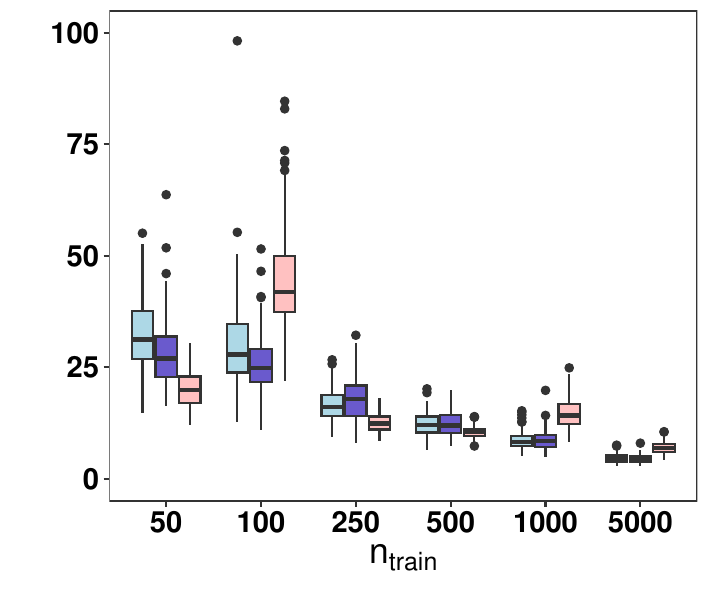}
\quad
\includegraphics[width=4.6cm]{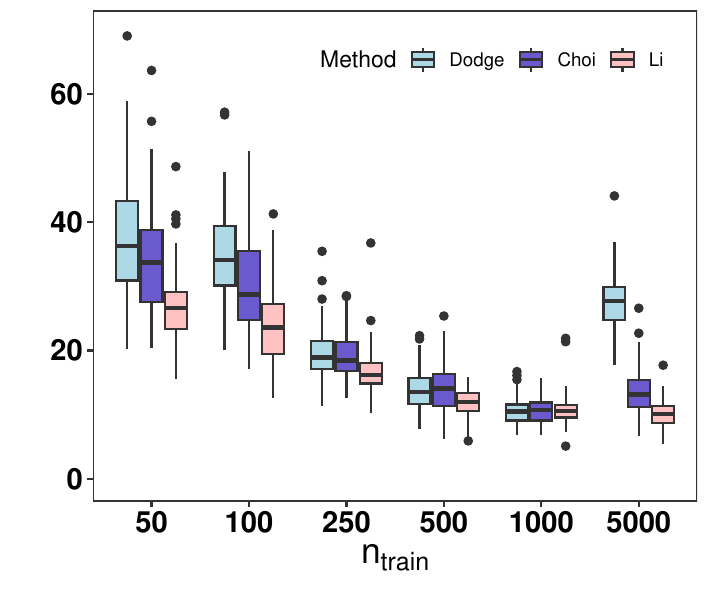}
\caption{\small{Boxplots of the calculated $\text{RISPE} \left ( \widehat{\beta}_{\tau}  \right )$ values for the Dodge, Choi, and Li methods under different sample sizes. Error metrics are computed for different quantile levels (rows) and error distributions (columns)}.}\label{fig:Fig_4}
\end{figure}

\begin{figure}[!htb]
\centering
\includegraphics[width=4.7cm]{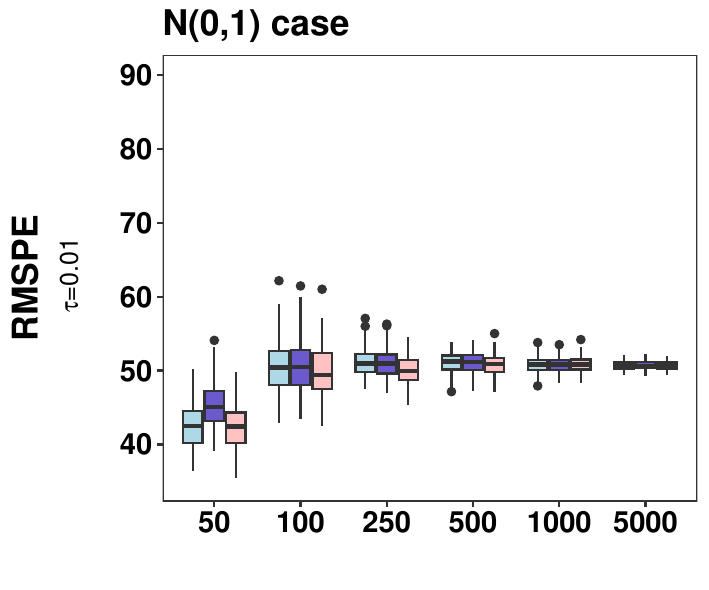}
\quad
\includegraphics[width=4.7cm]{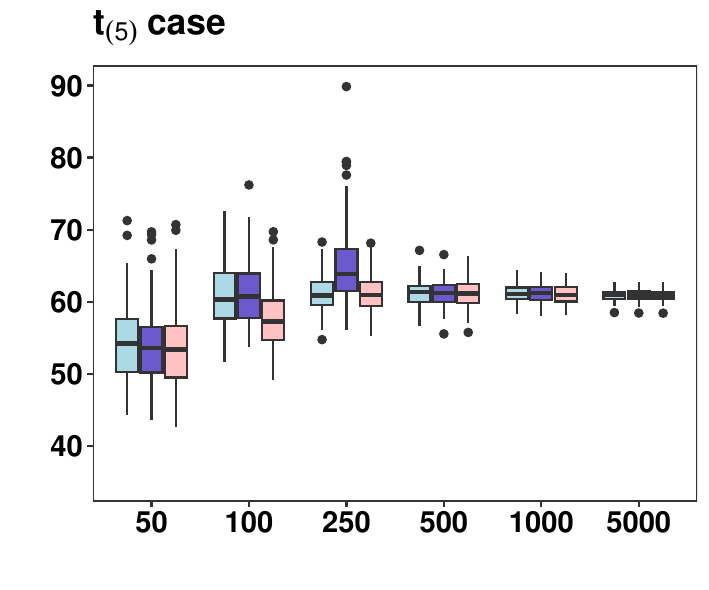}
\quad
\includegraphics[width=4.7cm]{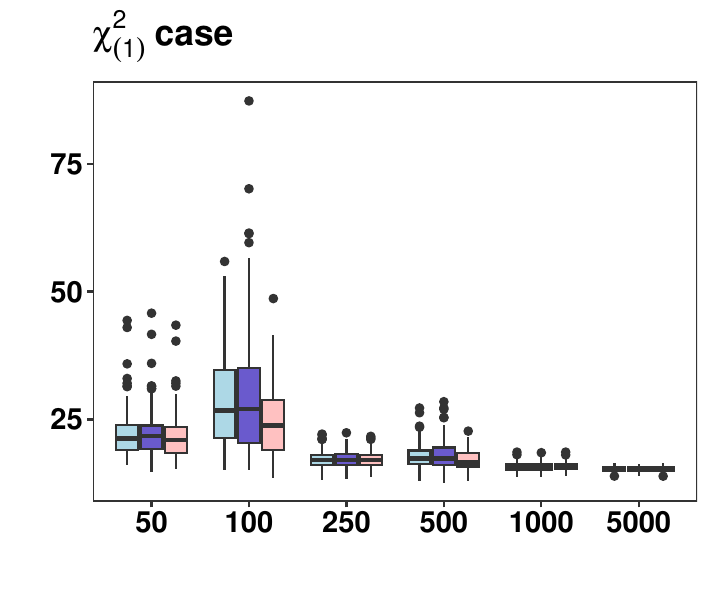}
\\
\includegraphics[width=4.7cm]{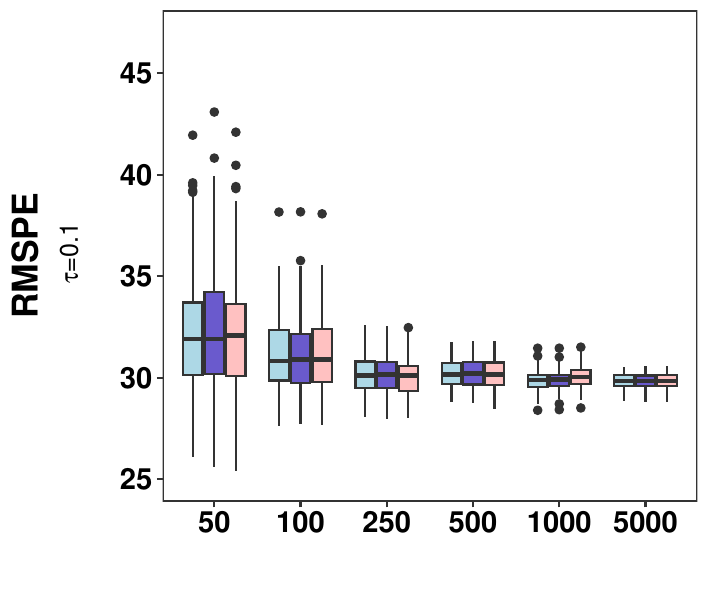}
\quad
\includegraphics[width=4.7cm]{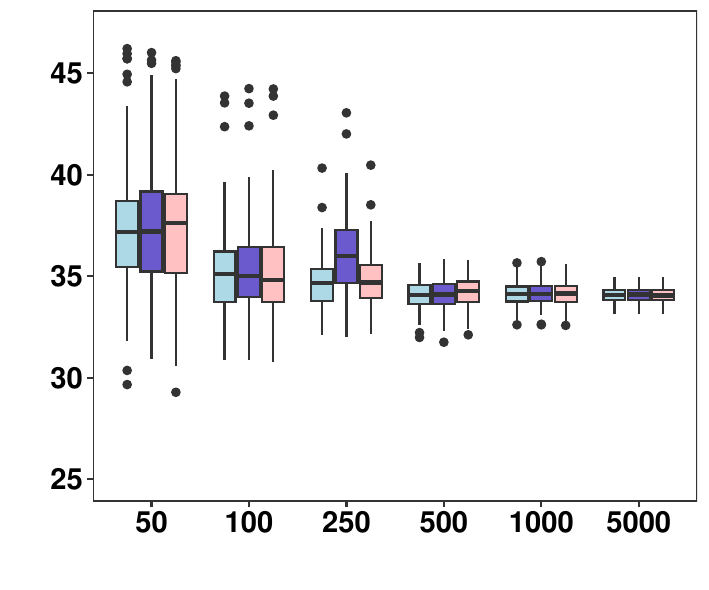}
\quad
\includegraphics[width=4.7cm]{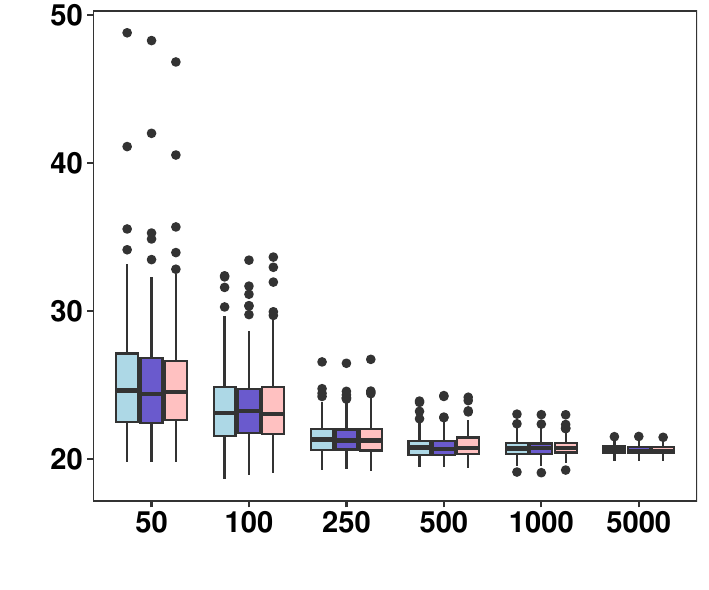}
\\
\includegraphics[width=4.7cm]{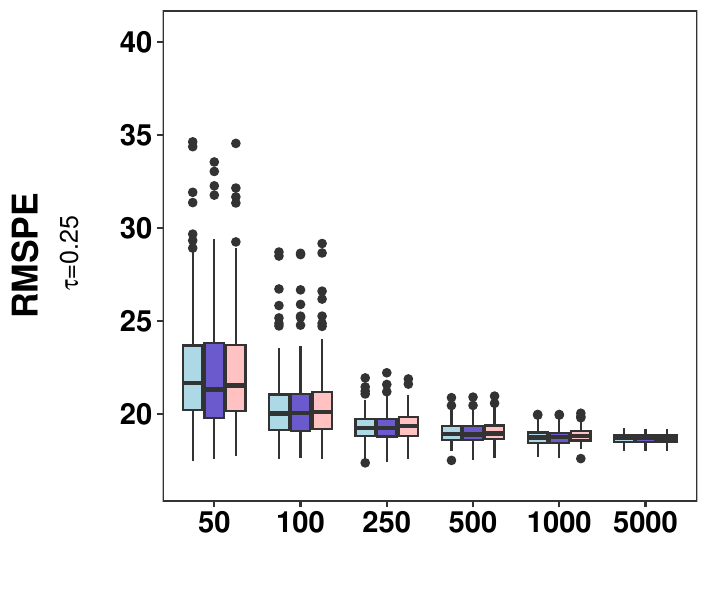}
\quad
\includegraphics[width=4.7cm]{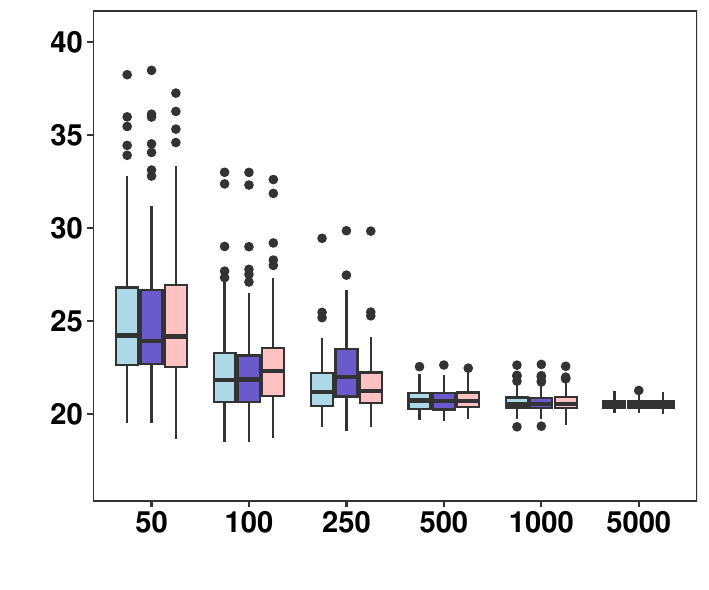}
\quad
\includegraphics[width=4.7cm]{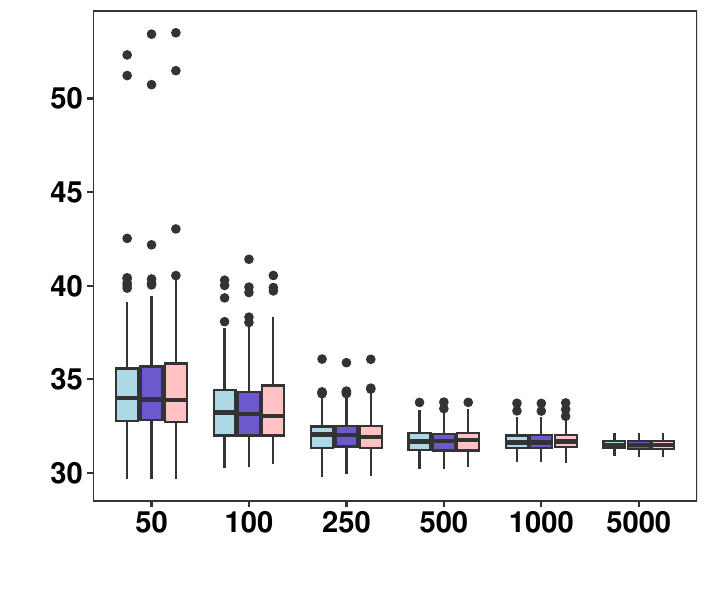}
\\
\includegraphics[width=4.7cm]{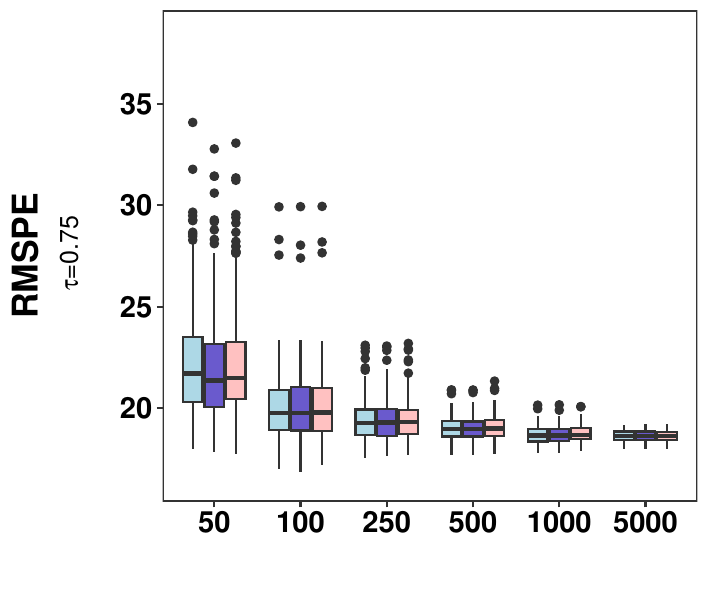}
\quad
\includegraphics[width=4.7cm]{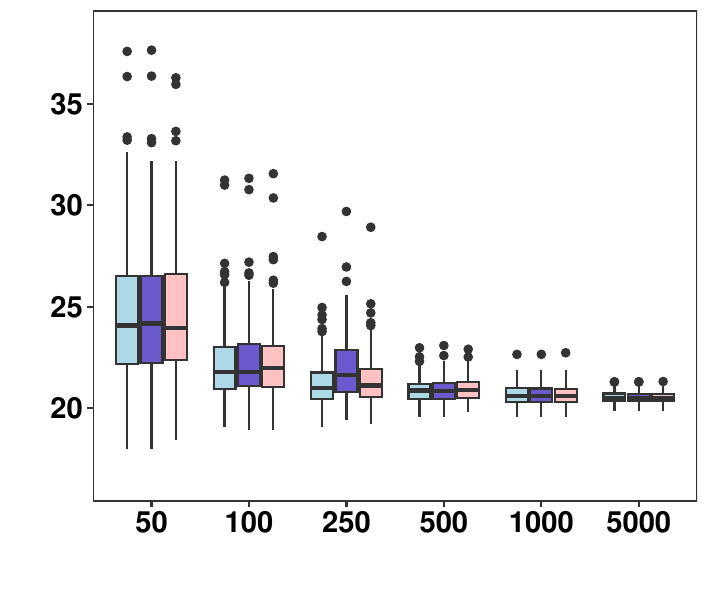}
\quad
\includegraphics[width=4.7cm]{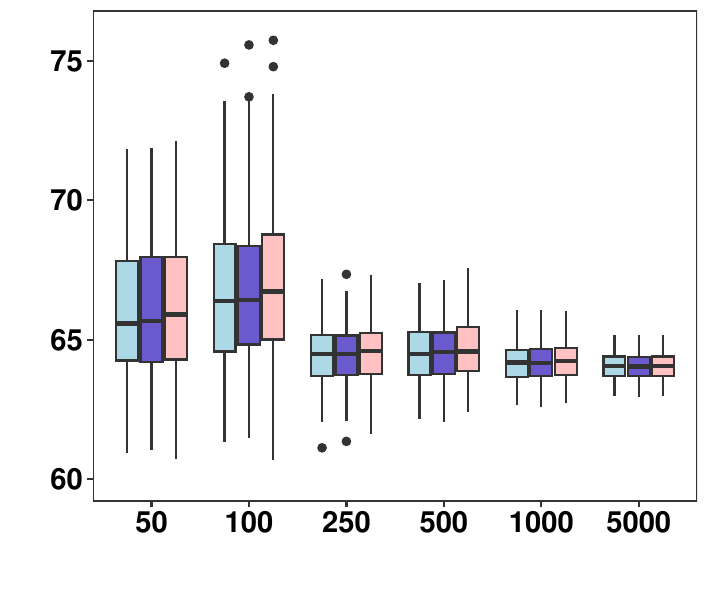}
\\
\includegraphics[width=4.7cm]{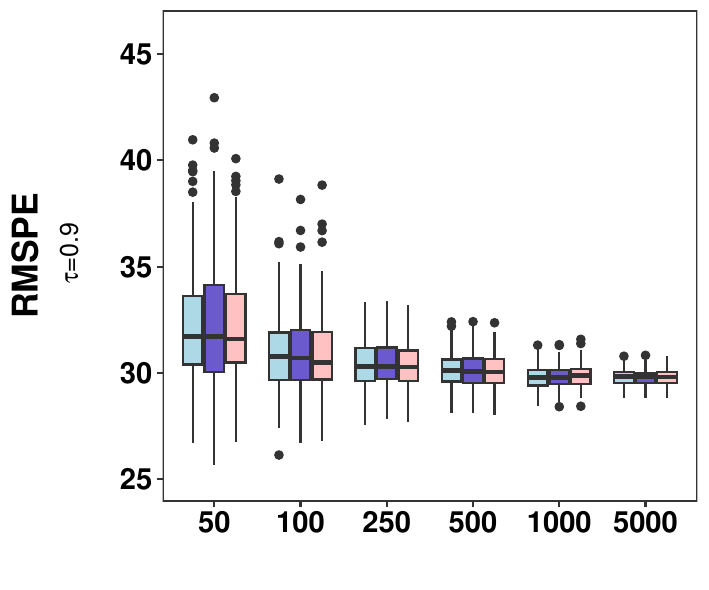}
\quad
\includegraphics[width=4.7cm]{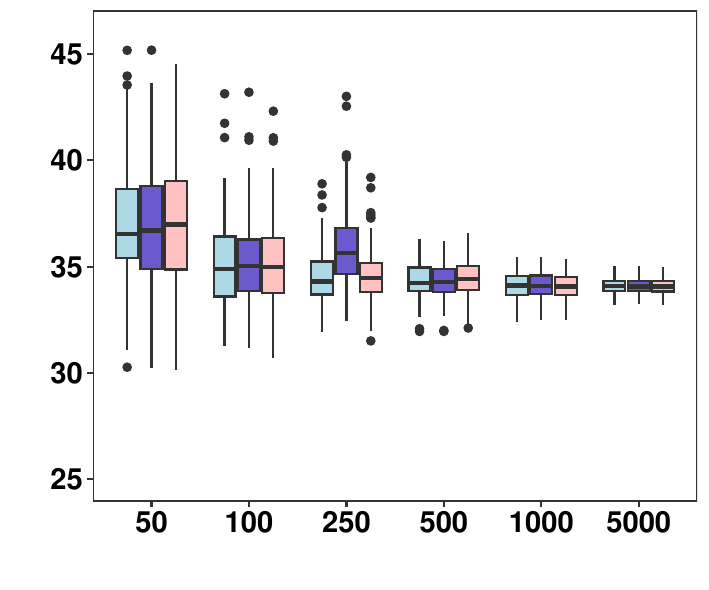}
\quad
\includegraphics[width=4.7cm]{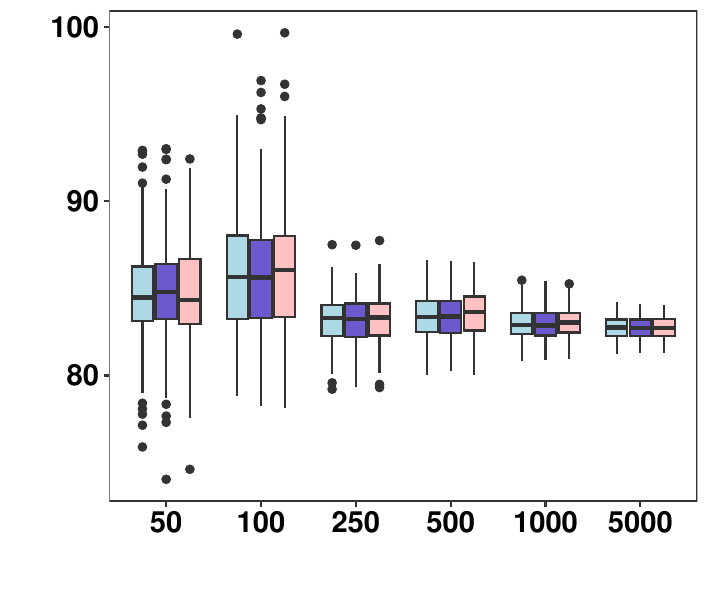}
\\
\includegraphics[width=4.7cm]{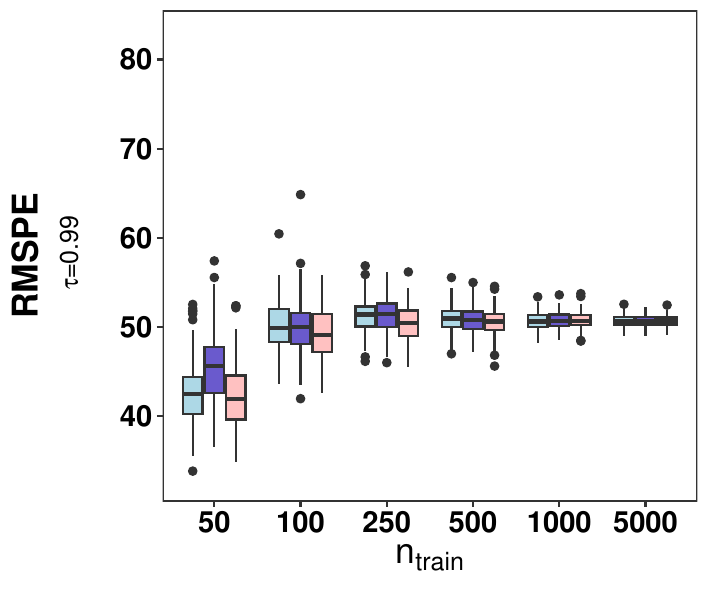}
\quad
\includegraphics[width=4.7cm]{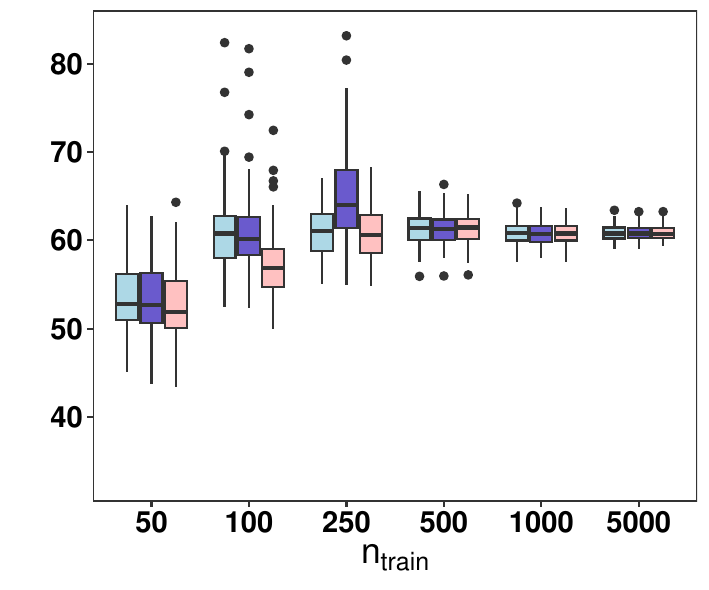}
\quad
\includegraphics[width=4.7cm]{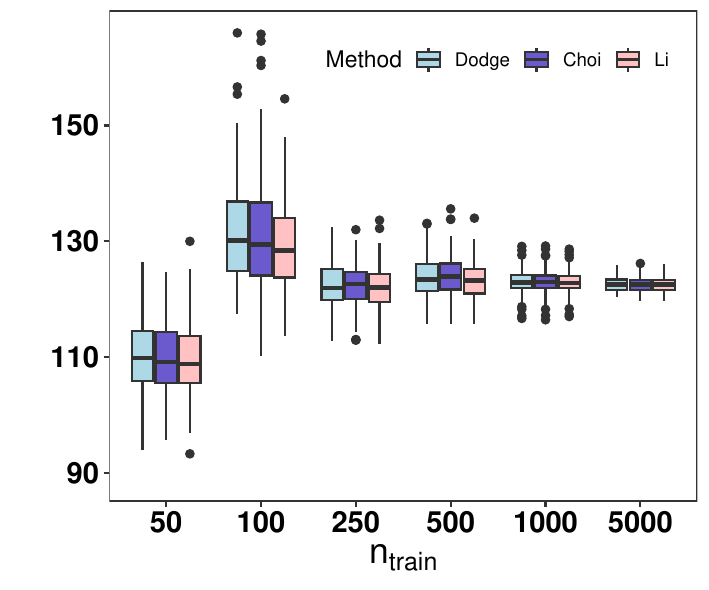}
\caption{\small{Boxplots of the calculated RMSPE values for the Dodge, Choi, and Li methods under different sample sizes. Error metrics are computed for different quantile levels (rows) and error distributions (columns)}.}\label{fig:Fig_5}
\end{figure}

The computed $\text{RRISPEE}(\widehat{\alpha}_{\tau})$, $\text{RRISPEE}(\widehat{\beta}_{\tau})$, and RMSPE for other quantile levels, specifically $\tau \in[0.01, 0.1, 0.25, 0.75, 0.90, 0.99]$, are presented in Figures~\ref{fig:Fig_3}, \ref{fig:Fig_4}, and \ref{fig:Fig_5}, respectively. These figures indicate that the error metrics from all FPQR methods for quantile levels around 0.5, namely $\tau \in [0.25, 0.75]$, are similar to those obtained when $\tau = 0.5$. However, the FPQR methods produce significantly larger error metrics for extreme quantile levels compared to $\tau = 0.5$. This is expected since the datasets in Monte Carlo experiments are generated from a conditional mean function-on-function regression model. Among the methods, the proposed Choi method yields worse results for $\text{RRISPEE}(\widehat{\alpha}_{\tau})$ (Figure~\ref{fig:Fig_3}) and RMSPE (Figure~\ref{fig:Fig_5}) than Dodge and Li at extreme quantile levels when the errors follow the $\text{N}(0,1)$ distribution. However, for other quantile levels and error distributions, all FPQR methods tend to produce similar $\text{RRISPEE}(\widehat{\alpha}_{\tau})$ and RMSPE values. Regarding $\text{RRISPEE}(\widehat{\beta}_{\tau})$, the proposed Choi method generally produces smaller values than Dodge and Li for all quantile levels when the errors follow the $\text{N}(0,1)$ distribution. When the errors follow the $t_{(5)}$ and/or $\chi^2_{(1)}$ distributions, the Li method generally produces better $\text{RRISPEE}(\widehat{\beta}_{\tau})$ values than Dodge and Choi for the extreme quantile levels ($\tau \in [0.01, 0.99]$). Conversely, Dodge and Choi outperform Li for the other quantile levels.

Furthermore, we assess the robustness of the FPQR methods in the presence of outliers. To achieve this, we conduct an additional set of 100 Monte Carlo simulations. In these simulations, the datasets are generated as previously described, but the error terms follow a contaminated normal distribution. Specifically, for each simulation, $\gamma \%$ of the training errors are drawn from a standard normal distribution, while the remaining $(1 - \gamma) \%$ of the training errors, chosen at random, are sampled from a normal distribution with a mean of 8 and a standard deviation of 1. Two $\gamma$ values as $\gamma \in[0.05, 0.1]$ are considered; the generated datasets are contaminated at 5\% and 10\% contamination levels. Note that to assess the robustness property of the methods, only the quantile level $\tau = 0.5$ is considered.

\begin{figure}[!htb]
\centering
\includegraphics[width=5.5cm]{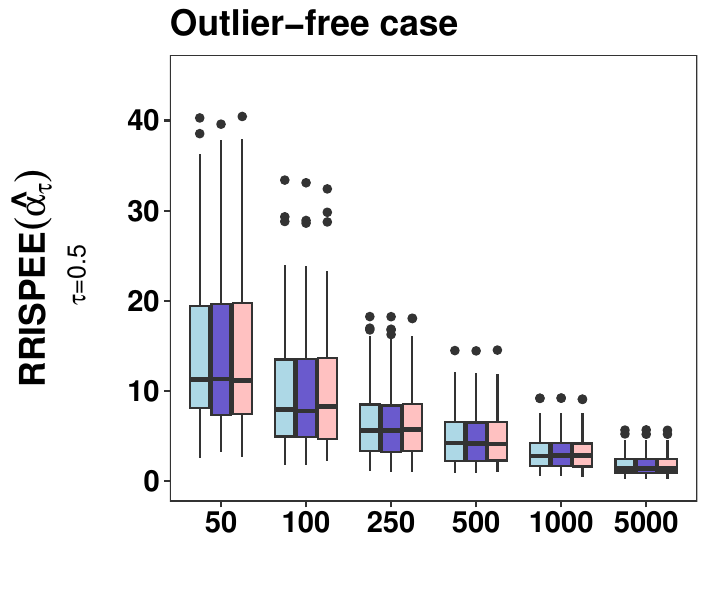}
\quad
\includegraphics[width=5.5cm]{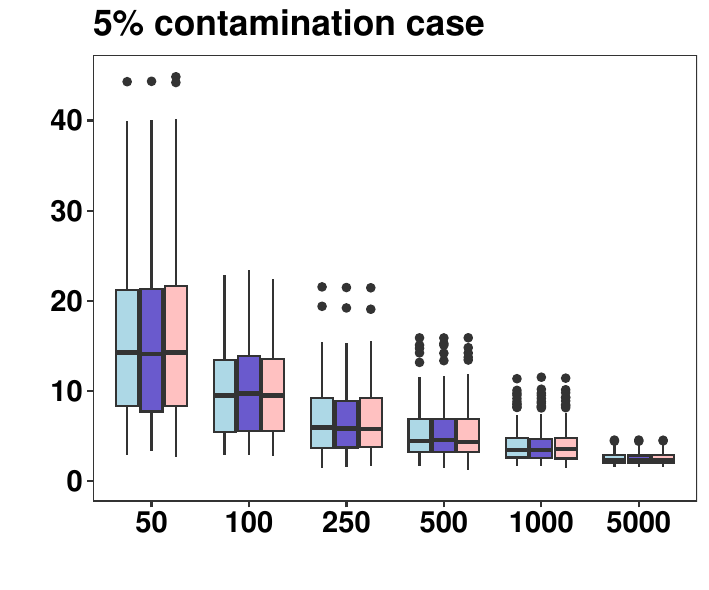}
\quad
\includegraphics[width=5.5cm]{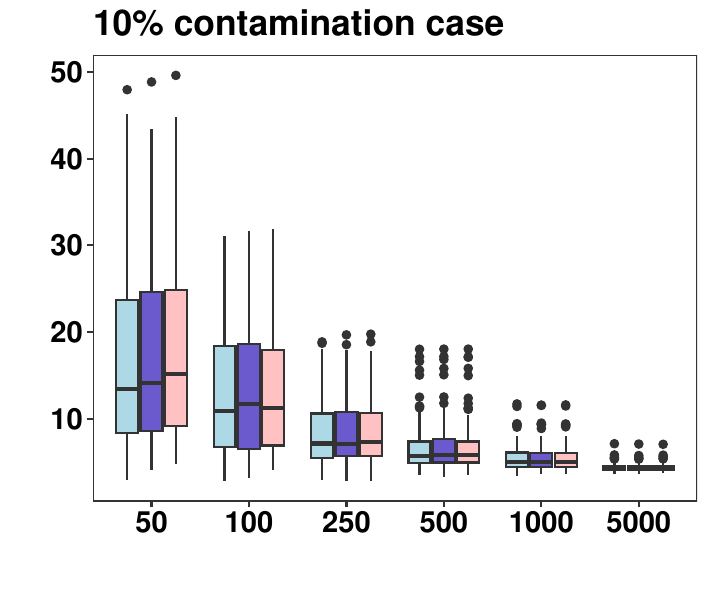}
\\
\includegraphics[width=5.5cm]{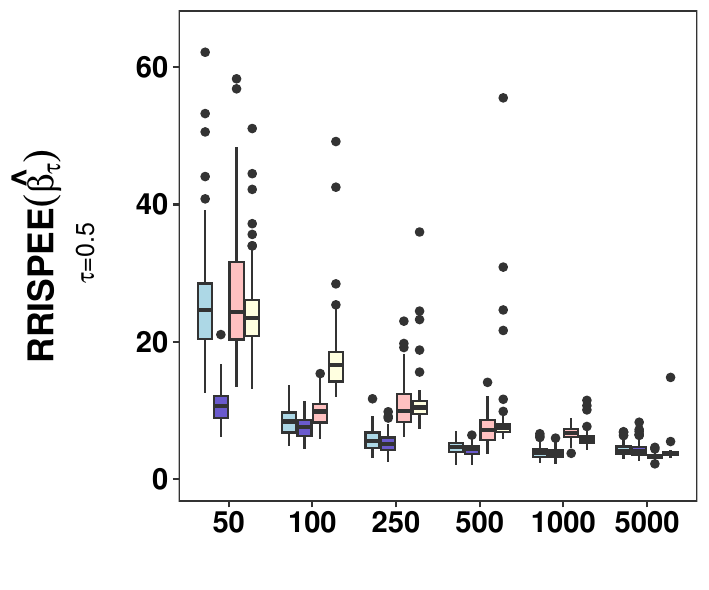}
\quad
\includegraphics[width=5.5cm]{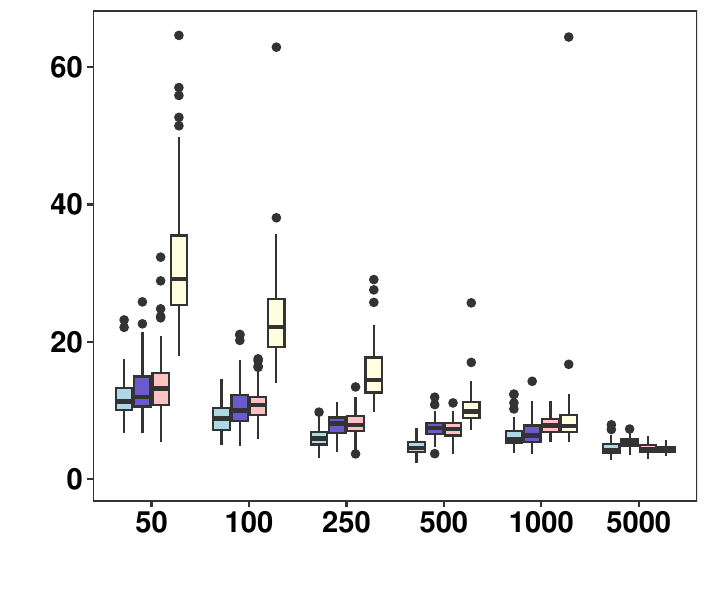}
\quad
\includegraphics[width=5.5cm]{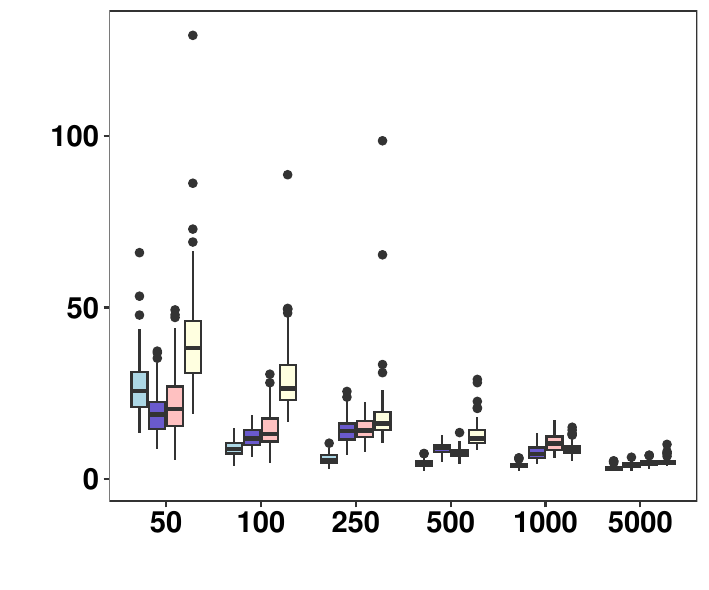}
\\
\includegraphics[width=5.5cm]{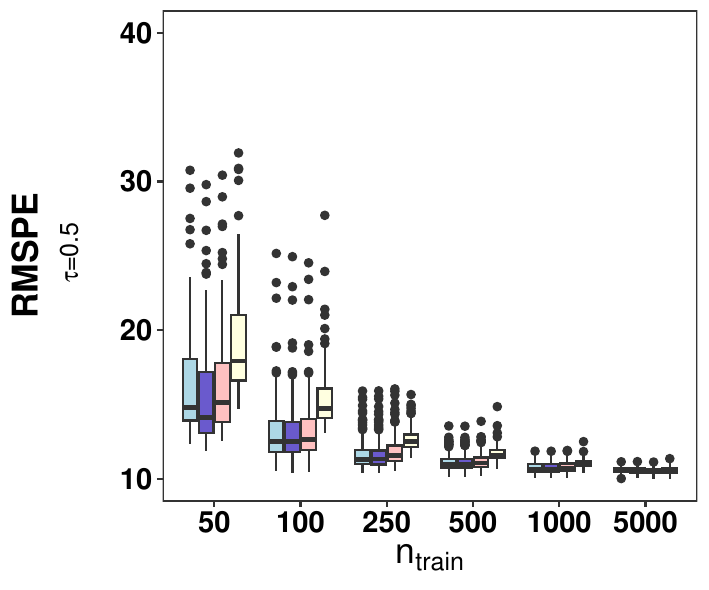}
\quad
\includegraphics[width=5.5cm]{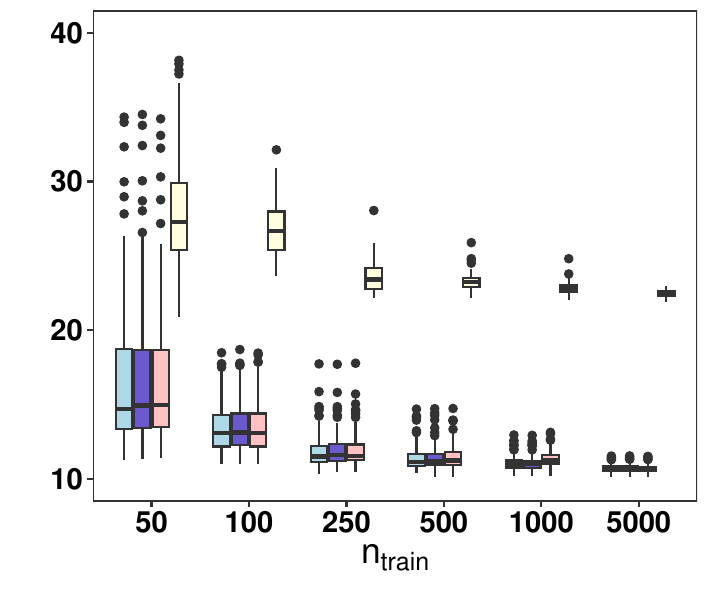}
\quad
\includegraphics[width=5.5cm]{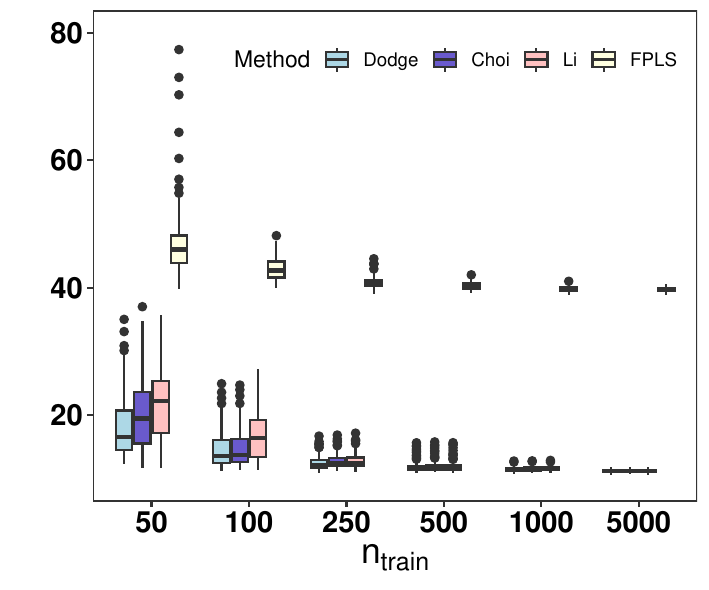}
\caption{\small{Boxplots of the calculated $\text{REISPE} \left ( \widehat{\alpha}_{\tau}  \right )$ (first row), $\text{RISPE} \left ( \widehat{\beta}_{\tau}  \right )$ (second row), and RMSPE (third row) values for the Dodge, Choi, Li, and FPLS methods under different sample sizes when $\tau=0.5$. Note that $\text{REISPE} \left ( \widehat{\alpha}_{\tau}  \right )$ cannot be calculated for the FPLS method. $\text{RREISPEE} \left ( \widehat{\alpha}_{\tau}  \right )$, $\text{RRISPEE} \left ( \widehat{\beta}_{\tau}  \right )$, and RMSPE are computed under three outlier contamination cases; outlier-free case (first column), 5\% contamination case (second column), and 10\% contamination case (third column)}.}
  \label{fig:Fig_6}
\end{figure}

The computed $\text{RRISPEE}(\widehat{\alpha}_{\tau})$, $\text{RRISPEE}(\widehat{\beta}_{\tau})$, and RMSPE when the data are contaminated by outliers are presented in Figure~\ref{fig:Fig_6}. The findings reveal that all FPQR methods yield comparable $\text{RRISPEE}(\widehat{\alpha}_{\tau})$ values across all contamination levels, indicating their ability to mitigate the influence of outliers and provide consistent estimates for the intercept function. In terms of $\text{RRISPEE}(\widehat{\beta}_{\tau})$ and RMSPE, all FPQR methods demonstrate similar performance to their performance when outliers are absent. At the same time, FPLS is notably impacted by outlier presence, leading to inflated error metrics. Among the FPQR methods, Dodge exhibits marginally smaller error values than the proposed Choi and Li methods for smaller sample sizes. Yet, the performance of all FPQR methods converges with increasing sample size.

We delve deeper into the computational efficiencies of the FPQR techniques via a sequence of Monte Carlo experiments. In this analysis, a single Monte Carlo simulation is conducted, considering the same six sample sizes mentioned earlier and varying the number of FPQR components from $h = 1$ to $h = 10$ when $K_{\Y} = K_{\X} = 20$. The resulting computation times for all sample sizes and numbers of FPQR components are documented in Table~\ref{tab:Tab1}. The findings reveal that the Li method exhibits significantly shorter computation times than other methods. For instance, the Li method requires between 0.47 and 24 times less computing time than the FPQR Dodge method and between 1 and 48 times less than the Choi method. Notably, the Choi method is the most time-consuming among the examined methods.

In summary, the outcomes of our Monte Carlo experiments indicate that, among the considered methods, the Choi method generally yields improved estimates for the regression coefficient functions but requires the longest computation time. Conversely, the Li method delivers comparable parameter estimates while demanding less computation time than other methods.
\begin{table}[!htbp] 
\tabcolsep 0.13in
\caption{\small{Elapsed computing times (in seconds) for the FPQR methods based on different sample sizes $n_{\text{train}}$ and number of FPQR components $h$}.}
\centering
\begin{tabular}{@{}llcccccccccc@{}}
\toprule
& & \multicolumn{10}{c}{$h$} \\
\cmidrule{3-12}
$n_{train}$ & Method & 1 & 2 & 3 & 4 & 5 & 6 & 7 & 8 & 9 & 10 \\ [0.5ex]
\midrule
& Dodge & $0.44$ & $0.66$ & $0.94$ & $1.23$ & $1.55$ & $1.82$ & $2.14$ & $2.36$ & $2.69$ & $2.99$ \\ [0.5ex]
50 & Choi & $0.66$ & $1.28$ & $1.86$ & $2.65$ & $3.08$ & $3.68$ & $4.24$ & $4.99$ & $5.75$ & $6.14$ \\ [0.5ex]
& Li & $0.03$ & $0.06$ & $0.06$ & $0.07$ & $0.09$ & $0.09$ & $0.11$ & $0.11$ & $0.12$ & $0.13$ \\ [0.5ex]
\cmidrule{2-12}
 & Dodge & $0.49$ & $0.68$ & $0.95$ & $1.27$ & $1.57$ & $1.83$ & $2.21$ & $2.48$ & $2.93$ & $3.13$ \\ [0.5ex]
100 & Choi & $0.64$ & $1.31$ & $1.89$ & $2.54$ & $3.23$ & $3.75$ & $4.31$ & $4.78$ & $5.64$ & $6.18$ \\ [0.5ex]
& Li & $0.05$ & $0.06$ & $0.06$ & $0.08$ & $0.09$ & $0.10$ & $0.10$ & $0.11$ & $0.12$ & $0.15$ \\ [0.5ex]
\cmidrule{2-12}
 & Dodge & $0.47$ & $0.67$ & $1.00$ & $1.34$ & $1.66$ & $2.10$ & $2.22$ & $2.63$ & $2.94$ & $3.33$ \\ [0.5ex]
250 & Choi & $0.72$ & $1.39$ & $2.01$ & $2.66$ & $3.30$ & $4.03$ & $4.69$ & $5.33$ & $6.08$ & $6.61$ \\ [0.5ex]
& Li & $0.06$ & $0.07$ & $0.07$ & $0.11$ & $0.13$ & $0.13$ & $0.14$ & $0.14$ & $0.17$ & $0.19$ \\ [0.5ex]
\cmidrule{2-12}
& Dodge & $0.50$ & $0.79$ & $1.12$ & $1.58$ & $1.64$ & $2.21$ & $2.50$ & $3.12$ & $3.44$ & $3.67$ \\ [0.5ex]
500 & Choi & $0.82$ & $1.53$ & $2.53$ & $2.83$ & $3.82$ & $4.69$ & $5.24$ & $5.86$ & $6.75$ & $7.56$ \\ [0.5ex]
& Li & $0.05$ & $0.09$ & $0.13$ & $0.13$ & $0.16$ & $0.17$ & $0.19$ & $0.19$ & $0.24$ & $0.25$ \\ [0.5ex]
\cmidrule{2-12}
 & Dodge & $0.69$ & $1.10$ & $1.56$ & $2.00$ & $2.50$ & $3.12$ & $3.35$ & $3.92$ & $4.39$ & $4.86$ \\ [0.5ex]
1000 & Choi & $1.09$ & $2.29$ & $3.06$ & $4.23$ & $5.02$ & $5.87$ & $7.06$ & $7.96$ & $8.84$ & $9.88$ \\ [0.5ex]
& Li & $0.09$ & $0.12$ & $0.16$ & $0.17$ & $0.22$ & $0.25$ & $0.28$ & $0.33$ & $0.36$ & $0.37$ \\ [0.5ex]
\cmidrule{2-12}
 & Dodge & $2.37$ & $5.15$ & $7.11$ & $8.79$ & $10.17$ & $11.76$ & $13.17$ & $15.60$ & $18.09$ & $19.08$ \\ [0.5ex]
5000 & Choi & $4.77$ & $10.22$ & $14.28$ & $17.35$ & $20.55$ & $23.60$ & $26.84$ & $31.22$ & $35.20$ & $38.83$ \\ [0.5ex]
& Li & $0.48$ & $0.74$ & $1.06$ & $1.20$ & $1.23$ & $1.43$ & $1.53$ & $1.86$ & $2.12$ & $2.21$ \\ [0.5ex]
\bottomrule
\end{tabular}
\label{tab:Tab1}
\end{table}

\section{North Dakota weather data} \label{sec:empiricaldata}

In this section, we conduct two empirical data analyses, namely Case-I and Case-II, to assess the empirical performance of the proposed methodologies. For both cases, we utilize North Dakota weather data, accessible at the following link: \url{https://ndawn.ndsu.nodak.edu/}. In Case I, our focus is on evaluating the forecasting accuracy of the methodologies. Conversely, in Case II, we concentrate on assessing the accuracies of quantile predictions. In the empirical data analyses, the performance of the FPQR methods is compared with that of the FLQR method proposed by \cite{BSA2022}, whose code is available at \url{https://github.com/UfukBeyaztas/FFLQR}. For both methods, B-spline basis expansion functions are utilized to reconstruct the functional form of the discretely observed data. Consistent with the Monte Carlo experiments, the optimal number of basis functions and FPQR components is determined using 5-fold cross-validation.

\subsection{Case-I}\label{sec:6.1}

\begin{figure}[!htb]
  \centering
  \includegraphics[width=5.5cm]{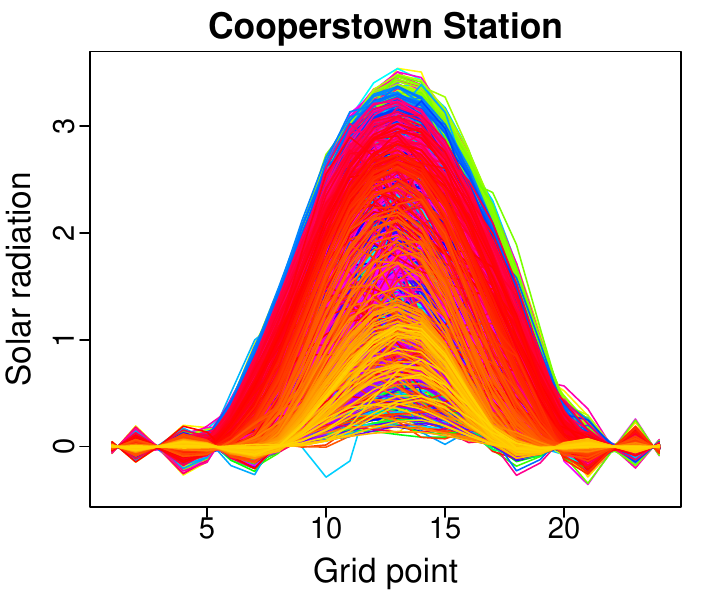}
\quad
  \includegraphics[width=5.5cm]{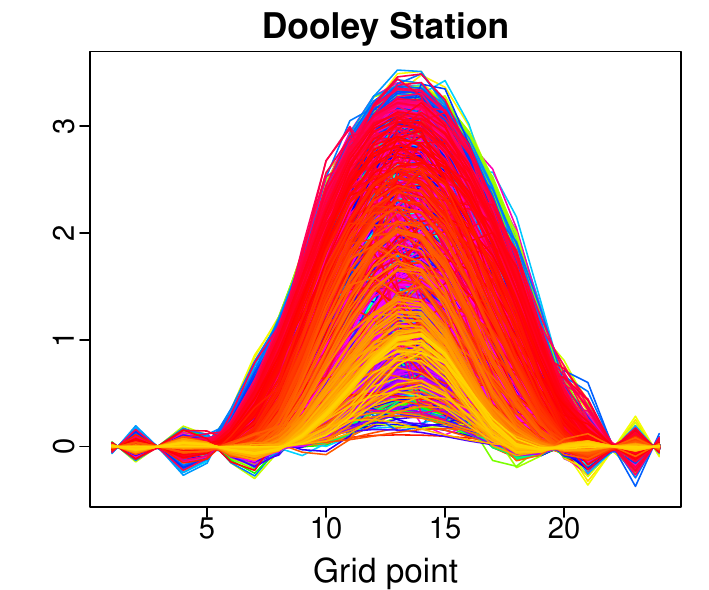}
  \quad
\includegraphics[width=5.5cm]{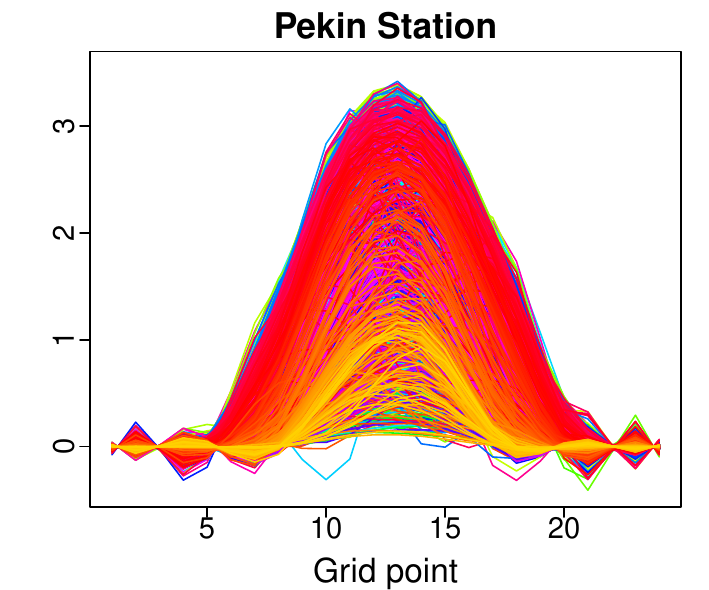}
\caption{\small{Plots of the solar radiation for three different stations: Cooperstown (left panel), Dooley (middle panel), and Pekin (right panel). Different colors correspond to different hours}.}
  \label{fig:Fig_7}
\end{figure}

For Case I, we analyze hourly solar radiation measurements collected from three distinct stations in the US state of North Dakota, (Cooperstown, Dooley, and Pekin) between 01/01/2018 and 31/12/2023. In each dataset, the solar radiation measurements are treated as a function of hours, resulting in a total of $n = 2191$ functional observations, denoted as $\lbrace \Y_{i}(u), 1 \leq u \leq 24, i=1, 2, \ldots, 2191 \rbrace$. Figure~\ref{fig:Fig_7} visually represents the datasets corresponding to each station.

Our objective is to assess the influence of previous solar radiation curve time series $\Y_{i-1}(u)$ on the current solar radiation curve time series $\Y_i(u)$. To achieve this, we consider the autoregressive Hilbertian process of order one (ARH(1)), the functional counterpart to the well-established autoregressive process of the order one model. The ARH(1) model is defined as follows:
\begin{equation*}
\Y_i(u) = \alpha(u) + \int_1^{24} \Y_{i-1}(v) \beta(v,u) dv + \varepsilon_{i}\left ( u \right ), \quad \forall u,v \in [1,24], \quad i = 1, \ldots, n-1.
\end{equation*}
Let $\widehat{\alpha}(u)$ and $\widehat{\beta}(v,u)$ denote the estimates of $\alpha(u)$ and $\beta(v,u)$, respectively. Consequently, the forecast of the $(i+1)$\textsuperscript{th} solar radiation curve time series is obtained as follows:
\begin{equation*}
\widehat{\Y}_{i+1}(u) = \widehat{\alpha}(u) + \int_1^{24} \Y_{i}(v) \widehat{\beta}(v,u)dv.
\end{equation*}

We adopt an expanding-window strategy to assess the predictive capabilities of the methods. Initially, the datasets are partitioned into two segments: training samples comprising days from 01/01/2018 to 10/05/2022 ($1591$ days) and test samples spanning days from 11/05/2022 to 31/12/2023 ($600$ days). Models are then trained using the entire training dataset to forecast solar radiation on 11/05/2022. Subsequently, predictions for solar radiation measurements on 12/05/2022 are made by gradually expanding the training dataset by one day. This iterative process continues until all observations in the test sample are predicted, ensuring the training dataset encompasses the entire dataset. During each iteration, the RMSPE is computed to evaluate the predictive accuracy of the methods. Notably, predictions are obtained using a quantile level of $\tau=0.5$ for a fair comparison. Additionally, leveraging the advantages of quantile regression-based methodologies, we derive pointwise 95\% prediction intervals for the functional response variable in the test set, denoted by $ \left [ \widehat{Q}_{0.025},\widehat{Q}_{0.975} \right ]$, by fitting two quantile regression models (employing quantile levels $\tau_1 = 0.025$ and $\tau_2 = 0.975$) to the training data. Subsequently, to assess the accuracy of the pointwise prediction intervals, we compute the coverage probability deviance (CPD) and interval score (IS) metrics as follows:
\begin{align*}
\begin{split}
\text{CPD} &= 0.95 - \frac{1}{J} \sum_{j=1}^J \mathds{1}\Big\{ \widehat{Q}_{0.025}(u) \leq  \Y(u_j) \leq \widehat{Q}_{0.975}(u) \Big\}, \\
\text{IS} &= \frac{1}{J} \Big\{ \left[\widehat{Q}_{0.975}(u)-\widehat{Q}_{0.025}(u) \right]+ \frac{2}{0.05} \sum_{j=1}^J \left[ \widehat{Q}_{0.025}(u)-\Y(u_j) \right]\mathds{1}\left[ \Y(u_j)<\widehat{Q}_{0.025}(u) \right] \\
&\hspace{1.74in} + \frac{2}{0.05} \sum_{j=1}^J \left[\Y(u_j)-\widehat{Q}_{0.975}(u) \right]\mathds{1}\left[\Y(u_j) > \widehat{Q}_{0.975}(u)\right] \Big\},
 \end{split}
\end{align*}
where $\{j = 1, \ldots, J \}$ are the discrete time points at which the response functions are observed.

\begin{figure}[!htb]
\centering
\includegraphics[width=5.6cm]{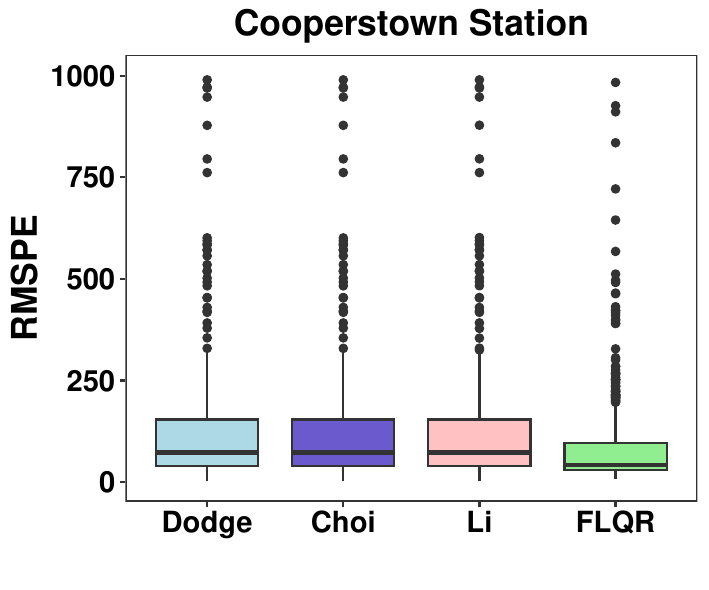}
\quad
\includegraphics[width=5.6cm]{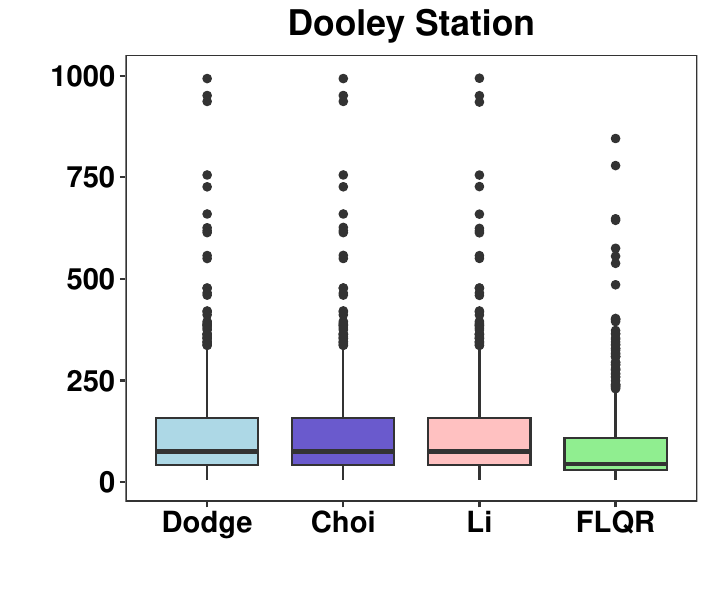}
\quad
\includegraphics[width=5.6cm]{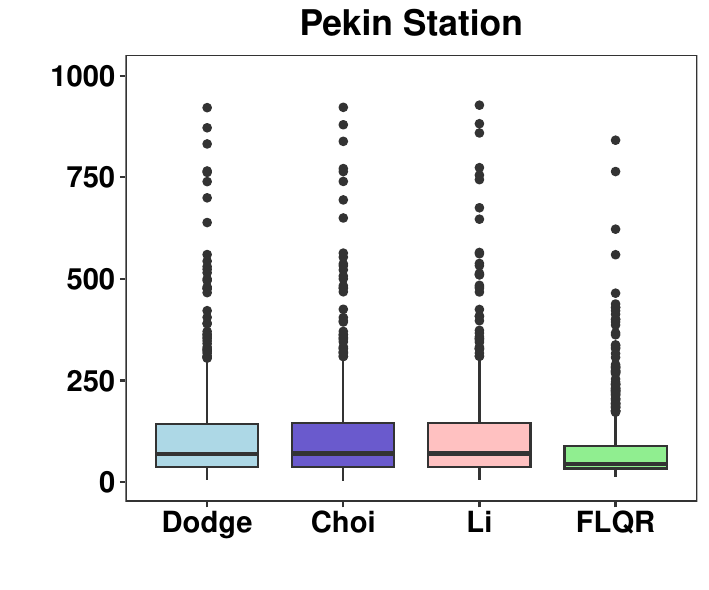}
\\
\includegraphics[width=5.6cm]{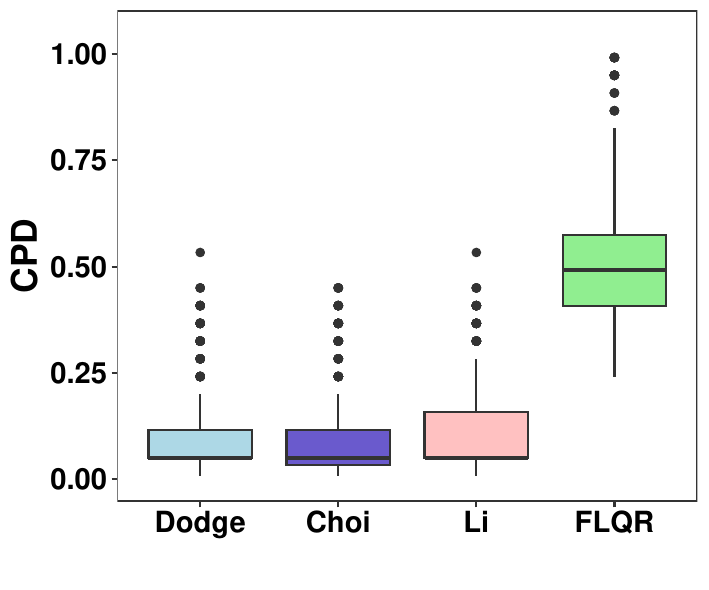}
\quad
\includegraphics[width=5.6cm]{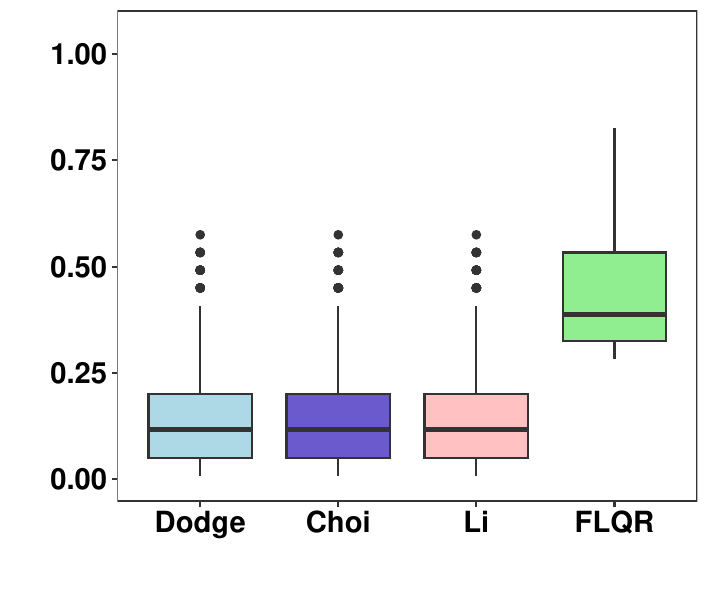}
\quad
\includegraphics[width=5.6cm]{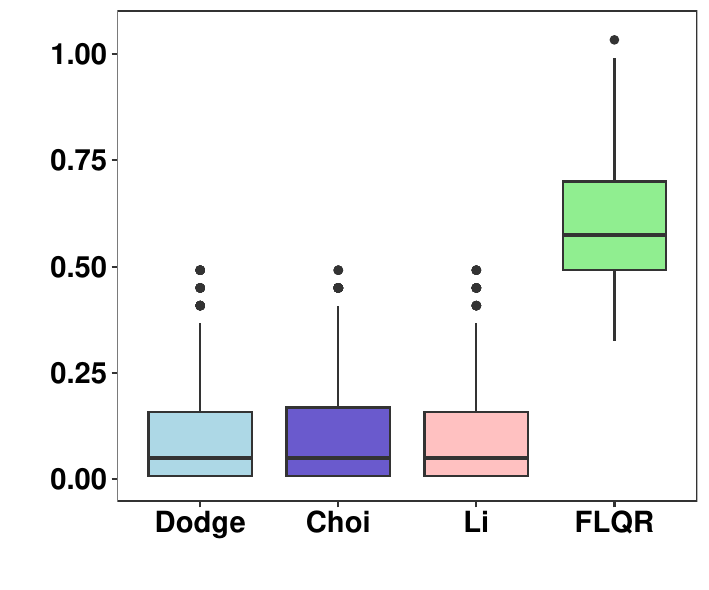}
\\  
\includegraphics[width=5.6cm]{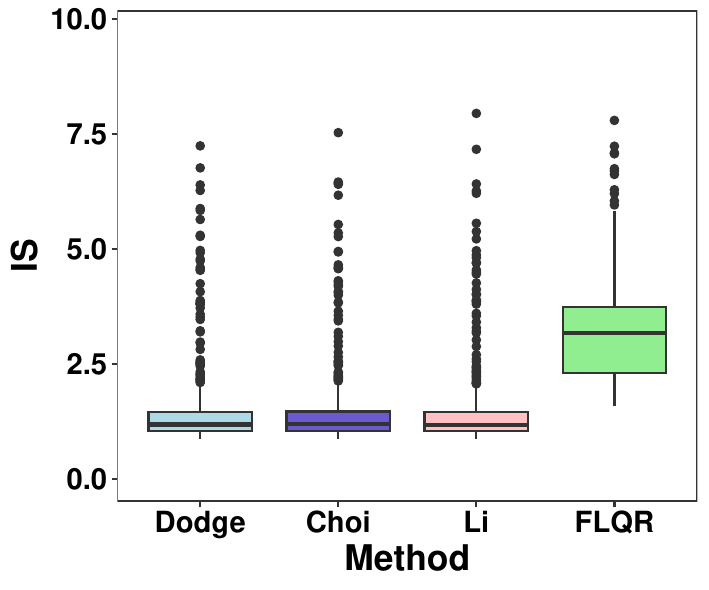}
\quad
\includegraphics[width=5.6cm]{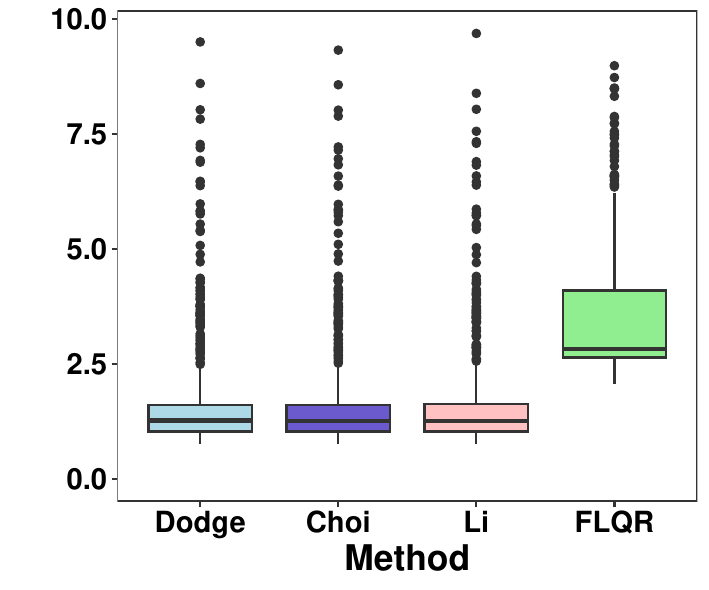}
\quad
\includegraphics[width=5.6cm]{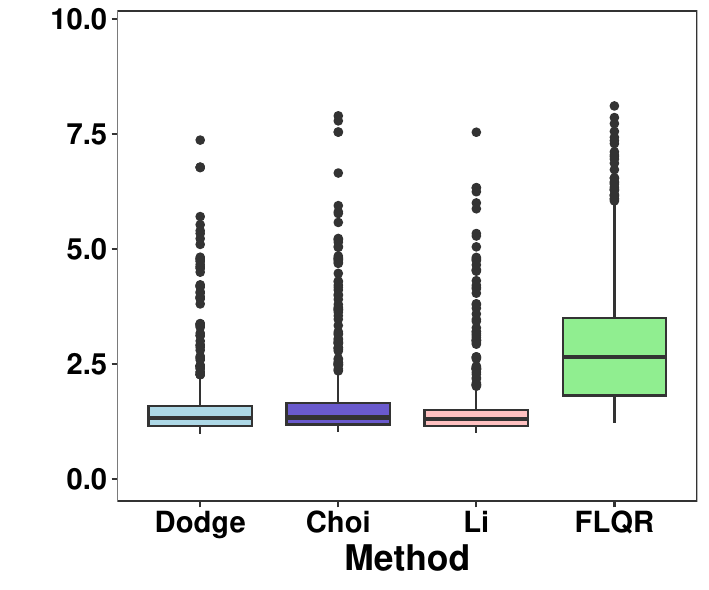}
\caption{\small{Boxplots of the calculated RMSPE, CPD and IS values for the FLQR, Dodge, Choi and Li methods. Rows and columns represent performance metrics and stations, respectively.}}\label{fig:Fig_8}
\end{figure}

The obtained RMSPE, CPD, and IS values for all stations are depicted in Figure~\ref{fig:Fig_8}. The figure reveals that, across all stations, the FLQR method produces slightly improved RMSPE values compared to all FPQR approaches, indicating lower RMSPE. All FPQR approaches yield similar RMSPE values across all stations, demonstrating negligible variance in their predictive performances. Furthermore, as illustrated in Figure~\ref{fig:Fig_8}, all FPQR techniques produce significantly improved CPD values, meaning smaller CPD values, than the FLQR method. The CPD values from all FPQR techniques are comparable and predominantly near zero across all stations. This suggests that the prediction intervals constructed by the FPQR methods effectively encompass a significant portion of the observations in the test sample. Conversely, the large CPD values obtained by the FLQR method indicate that its prediction intervals fail to cover a significant portion of the observations in the test sample. Similarly, the IS values, which concurrently assess the coverage probability and the width of the pointwise prediction intervals, approach zero for all FPQR methods across all stations. It is evident from Figure~\ref{fig:Fig_8} that the FPQR methods produce smaller IS values than the FLQR method. In other words, considering both CPD and IS values, the FPQR methods tend to produce narrower and more accurate prediction intervals than the FLQR method. The computational efficiency of all methods for this dataset is summarized in Table~\ref{tab:Tab2}. The results indicate that FLQR requires considerably more computing time than all FPQR methods. Among the FPQR methods, the proposed Li method demands 8 to 12 times less computational time than the Choi method and 6 to 10 times less than the Dodge method. Our findings from Case I suggest that the Li method presents a compelling alternative to the existing FPQR Dodge method, especially considering its computational efficiency.

\begin{table}[!htbp] 
\tabcolsep 0.48in
\caption{\small{Elapsed computing time (in seconds) of the FLQR and FPQR methods for Case-I. The computing times are recorded for two different training sample sizes}.}
\centering
\begin{tabular}{@{}lccccc@{}}
\toprule
& & \multicolumn{3}{c}{Station} \\
\cmidrule{3-5}
$n_{\text{train}}$ & Method & Cooperstown & Dooley & Pekin \\ [0.5ex]
\midrule
\multirow{4}{*}{1591} & FLQR & 21.10 & 21.60 & 4.04 \\ [0.5ex]
& Dodge & 1.13 & 1.46 & 0.78 \\ [0.5ex]
& Choi & 1.69 & 1.46 & 0.91 \\ [0.5ex]
& Li & 0.14 & 0.14 & 0.11 \\ [0.5ex]
\midrule
\multirow{4}{*}{2191} & FLQR & 28.60 & 28.30 & 5.12 \\ [0.5ex]
& Dodge & 1.30 & 1.45 & 0.82 \\ [0.5ex]
& Choi & 1.94 & 2.04 & 1.20 \\ [0.5ex]
& Li & 0.19 & 0.18 & 0.13 \\ [0.5ex]
\bottomrule
\end{tabular}
\label{tab:Tab2}
\end{table}

\subsection{Case II}

In Case II, our objective shifts towards assessing the predictive accuracy of the methods across different quantiles beyond the conventional level of $\tau=0.5$. In this scenario, we investigate the relationship between solar radiation, which is treated as the functional response, and temperature, which is considered the functional predictor. The data comprises solar radiation and temperature readings from 141 weather stations (i.e., $n = 141$) in North Dakota, US, from January 1, 2022, to December 31, 2022. Each observation consists of hourly measurements, denoted as $\lbrace (\Y_{i}(u), \X_i(v)), 1 \leq u,v \leq 24, i=1, 2, \ldots, 141 \rbrace$.

\begin{figure}[!htb]
\centering
\includegraphics[width=8.5cm]{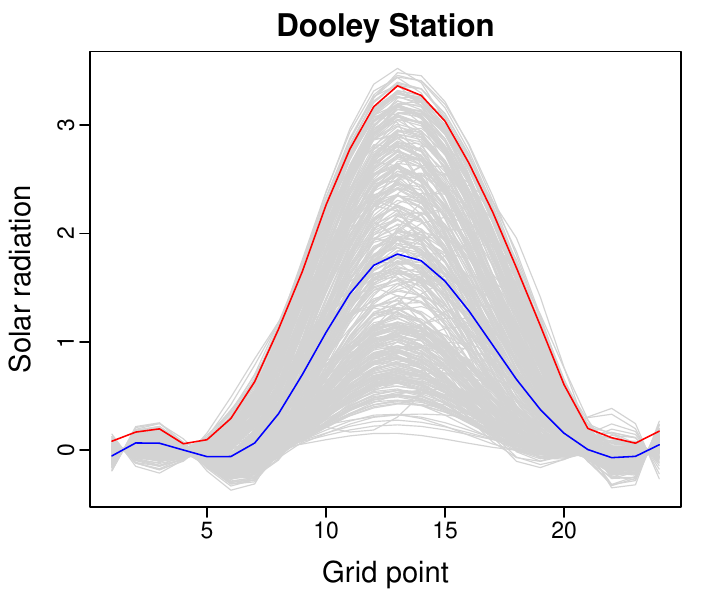}
\quad
\includegraphics[width=8.5cm]{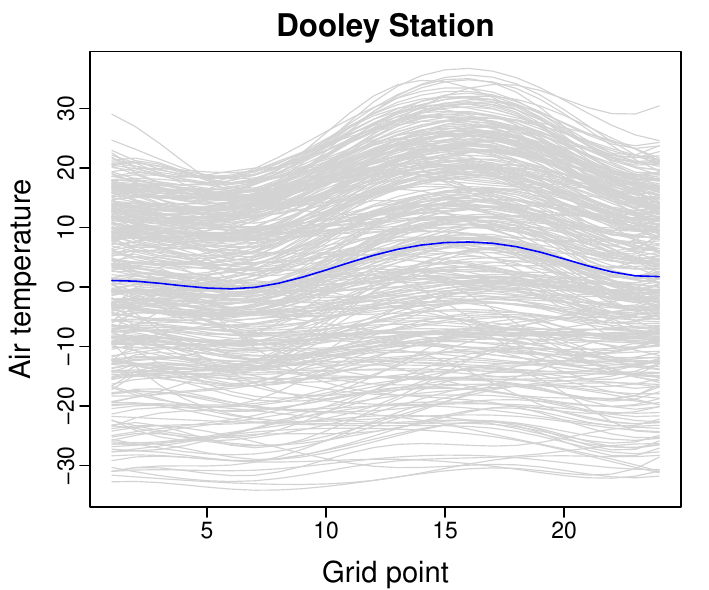} 
\caption{\small{Plots of the annual observations for solar radiation (left panel) and air temperature (right panel) with annual average (blue line) and the $97.5$\textsuperscript{th} percentile (red line) for any station (Dooley).}}
  \label{fig:Fig_9}
\end{figure}

To establish the true quantiles of the solar radiation curves, we aggregate hourly average solar radiation curves for the year 2022 for each station, resulting in 365 hourly averaged solar radiation curves per station. Subsequently, we compute the functional mean and the pointwise $0.975^\textsuperscript{th}$ quantile of these hourly averaged curves to derive the true mean and true $0.975^\textsuperscript{th}$ quantile of the solar radiation for each station. We aim to assess the methods' predictive capabilities in estimating these actual percentiles. Our interest lies solely in predicting the upper tail (specifically for $\tau=0.975$) of the distribution, given the proximity of the lower tail of solar radiation measurements to zero. Figure~\ref{fig:Fig_9} provides a visual representation of the hourly averaged solar radiation and temperature measurements, their means, and the $0.975$\textsuperscript{th} quantile of the solar radiation measurements for the Dooley station.

\begin{figure}[!htb]
\centering
\includegraphics[width=8.5cm]{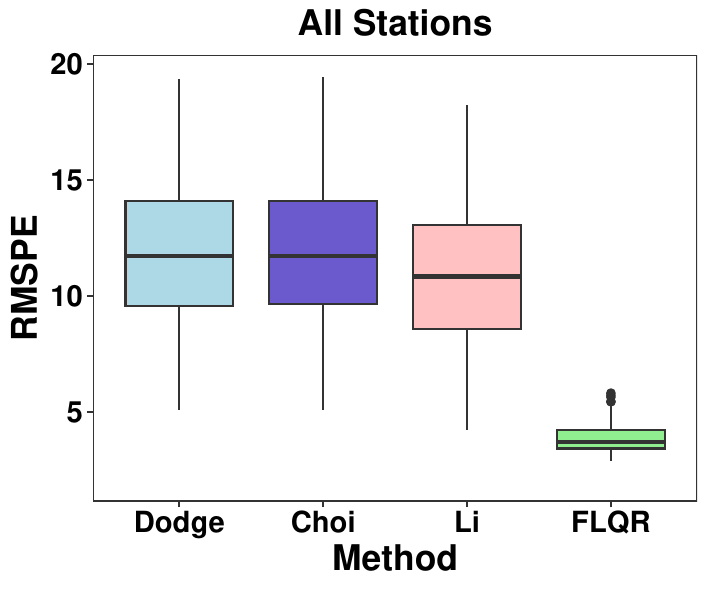}
\quad
\includegraphics[width=8.5cm]{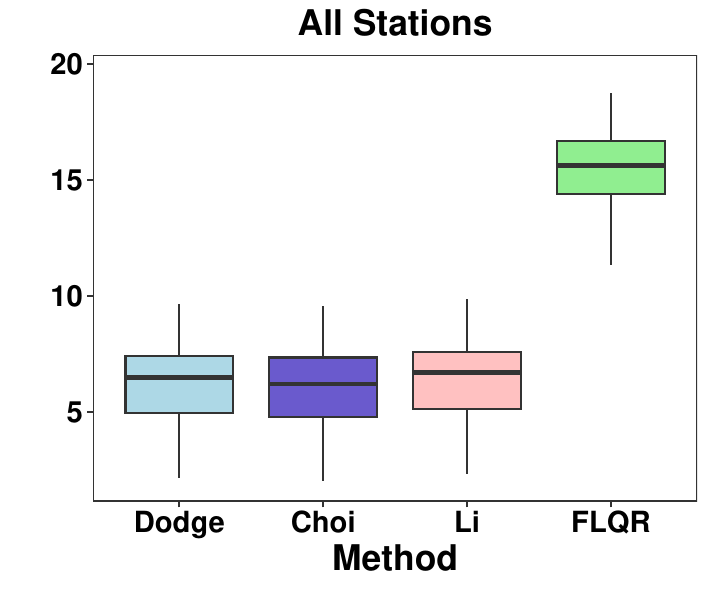}
\caption{\small{Boxplot of the computed RMSPE values for $\tau=0.5$ (left panel) and $\tau=0.975$ (right panel ) quantile levels under Case~II}.}\label{fig:Fig_10}
\end{figure}

To assess the predictive efficacy of the methods, we adopt the following procedure for each station. The FLQR and FPQR techniques are employed on the complete set of functional observations spanning the entire year, considering quantile levels of $\tau=0.5$ and $\tau=0.975$. Utilizing these models, we predict the annual distribution of solar radiation at both quantile levels based on the annual average of temperature measurements. Subsequently, we calculate the RMSPE values for each station corresponding to both $\tau=0.5$ and $\tau=0.975$ quantile levels, facilitating a comparison of the methods' predictive capabilities. The obtained results are depicted in Figure~\ref{fig:Fig_10}.

\begin{figure}[!htb]
\centering
\includegraphics[width=7cm]{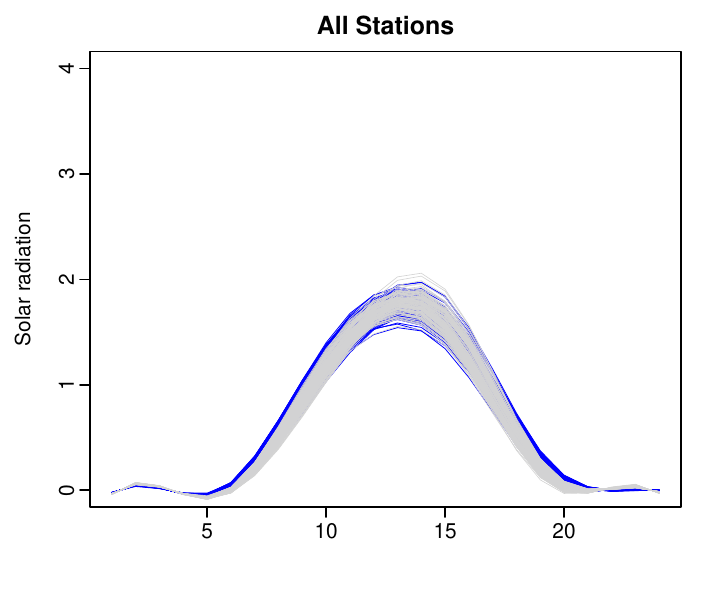}
\quad
\includegraphics[width=7cm]{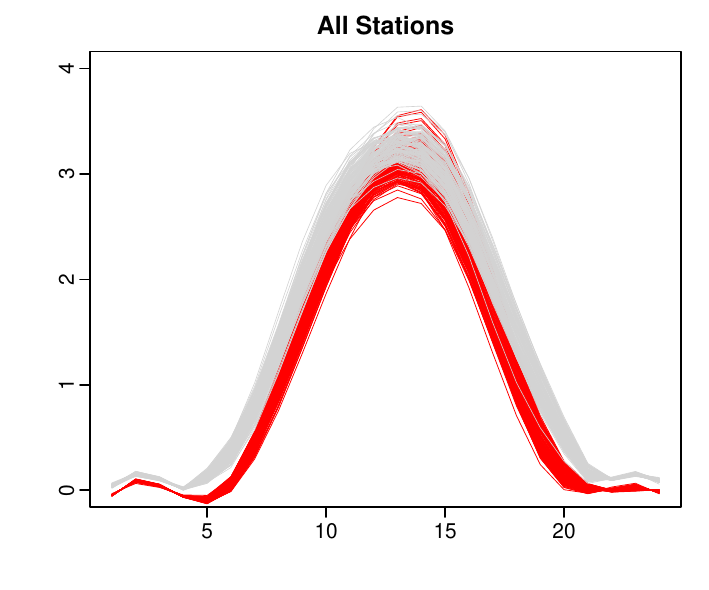}
\\
\includegraphics[width=7cm]{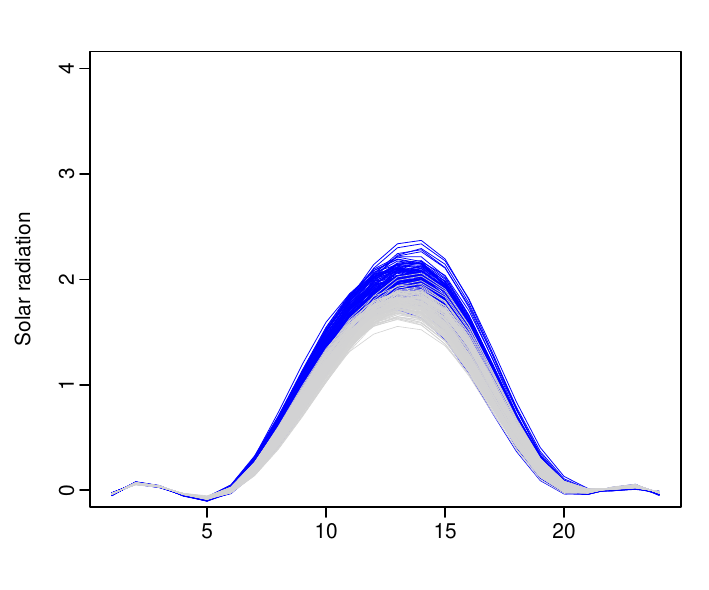}
\quad
\includegraphics[width=7cm]{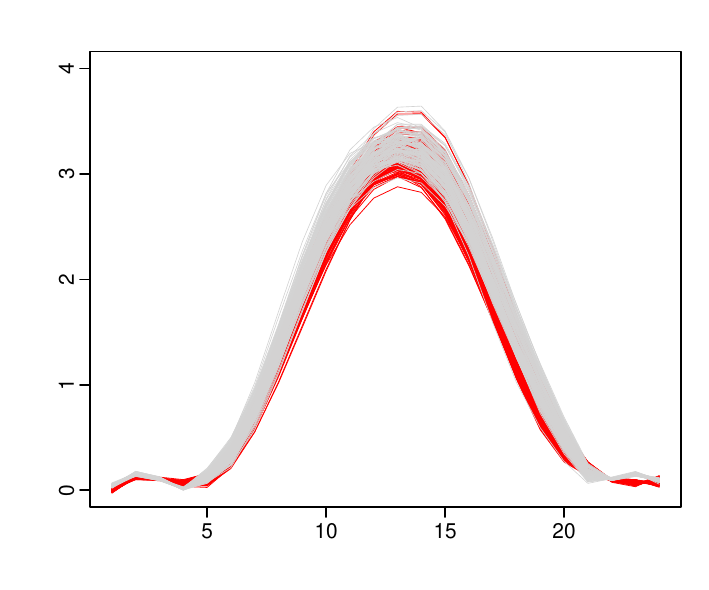}
\\
\includegraphics[width=7cm]{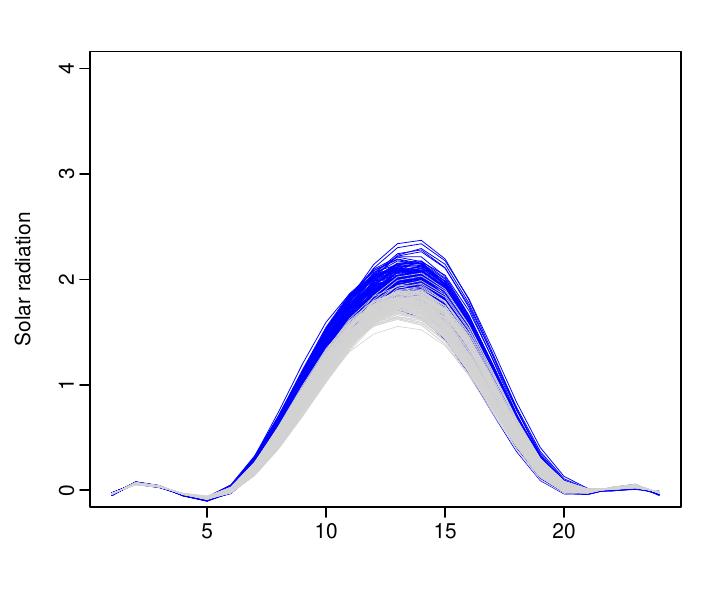}
\quad
\includegraphics[width=7cm]{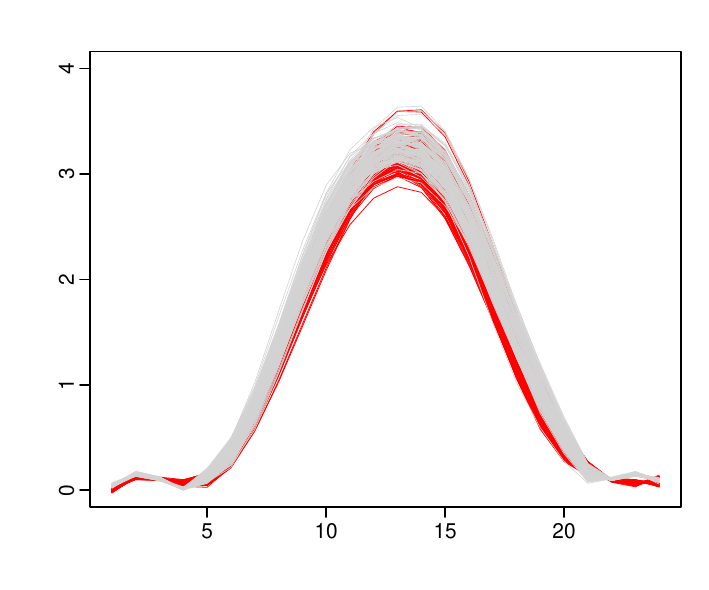}
\\
\includegraphics[width=7cm]{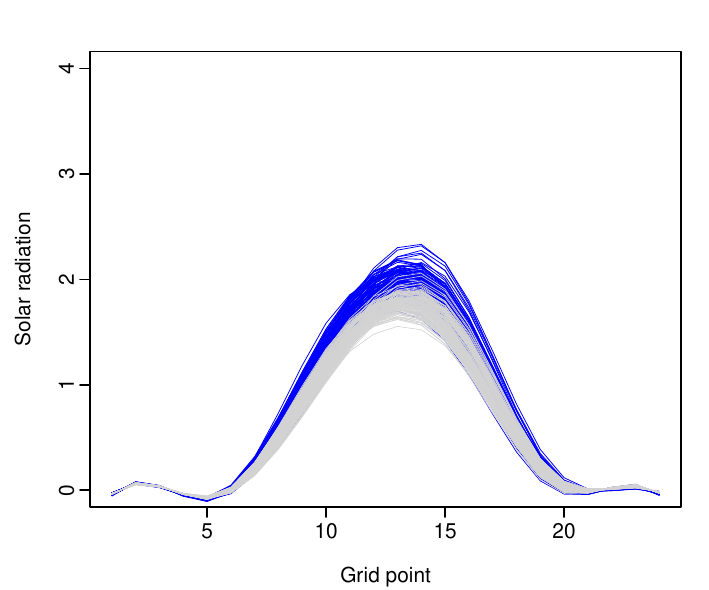}
\quad
\includegraphics[width=7cm]{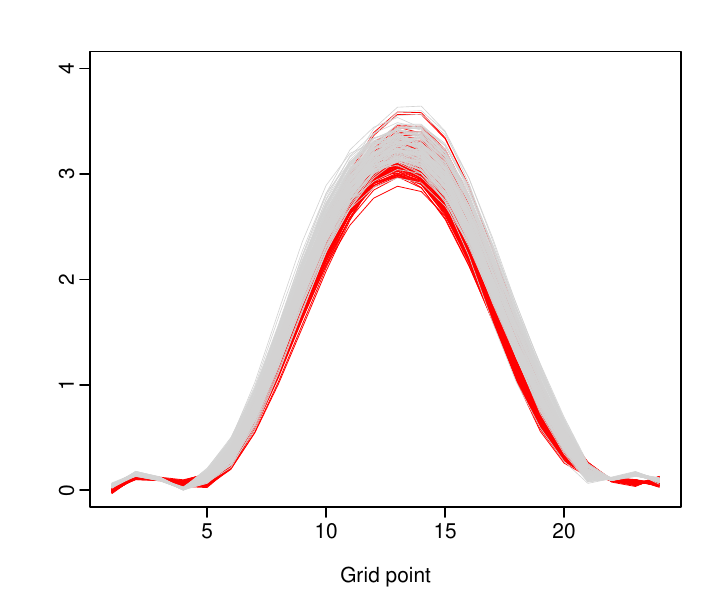}
\caption{\small{Graphical display of the true (gray lines) and predicted (blue and red lines) solar radiation measurements for all 141 stations. The predictions are obtained when $\tau=0.5$ (left panels) and $\tau=0.975$ (right panels) for the methods FLQR (first row), Dodge (second row), Choi (third row), and Li (fourth row)}.}\label{fig:Fig_11}
\end{figure}

Observing Figure~\ref{fig:Fig_10}, it becomes evident that FLQR produces improved RMSPE values compared to all FPQR methods for the $\tau=0.5$ quantile level. For this quantile level, all FPQR methods yield comparable outcomes. Conversely, for the upper tail of solar radiation measurements ($\tau=0.975$), the FPQR methods significantly outperform FLQR, producing considerably smaller RMSPE values. Among the FPQR methods, the proposed Choi method exhibits slightly superior predictions compared to Li and Dodge. The empirical analyses conducted under Case~II indicate that FLQR tends to produce better predictions for the conditional mean (quantile level $\tau=0.5$). In comparison, the FPQR methods produce better predictions for the upper tail of the response variable (quantile level $\tau=0.975$). Visual representations of both true and predicted solar radiation measurements for both quantile levels are illustrated in Figure~\ref{fig:Fig_11}.

\begin{figure}[!htb]
\centering
\includegraphics[width=7cm]{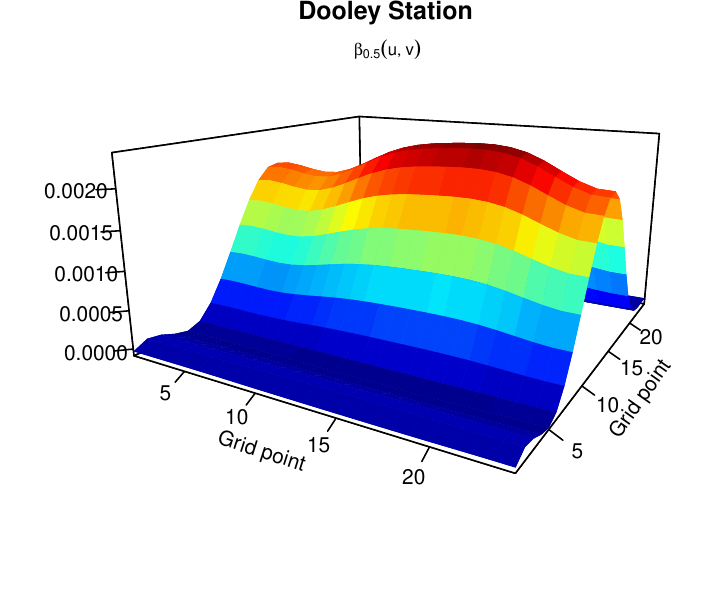}
\quad
\includegraphics[width=7cm]{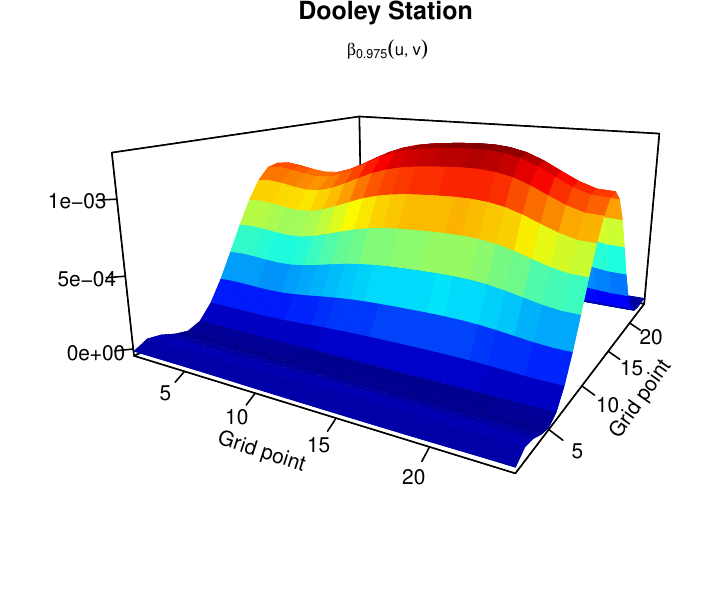}
\\
\includegraphics[width=7cm]{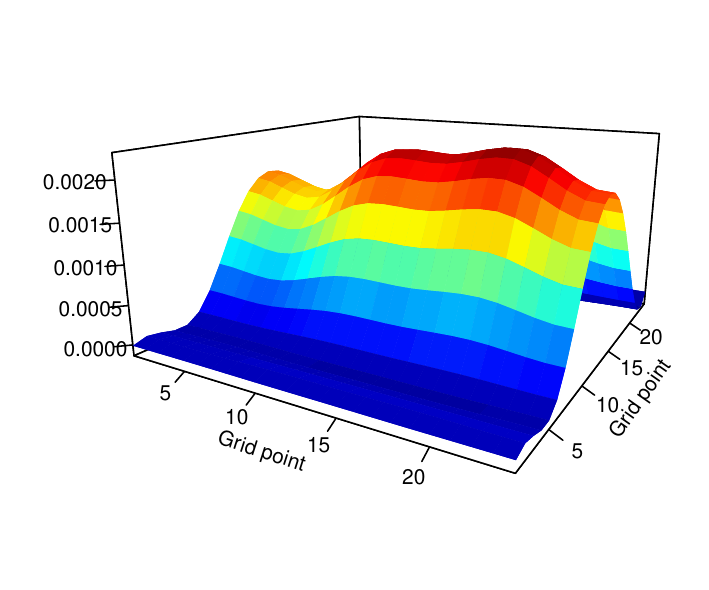}
\quad
\includegraphics[width=7cm]{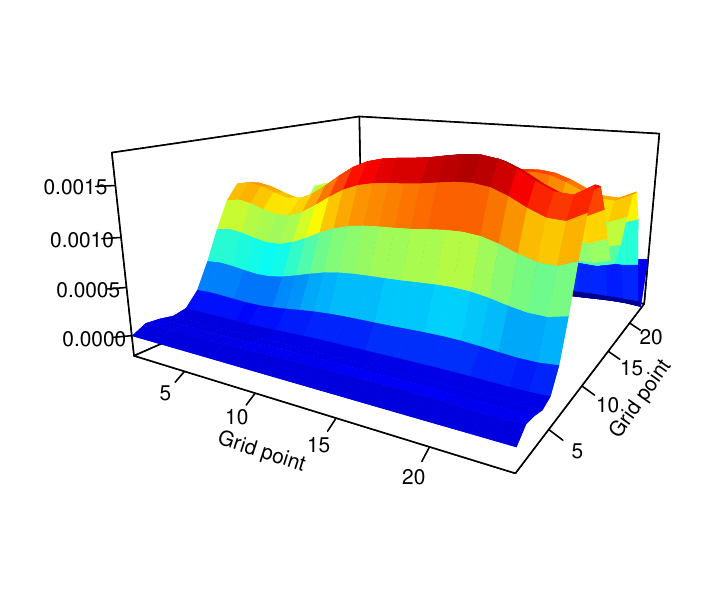}
\\
\includegraphics[width=7cm]{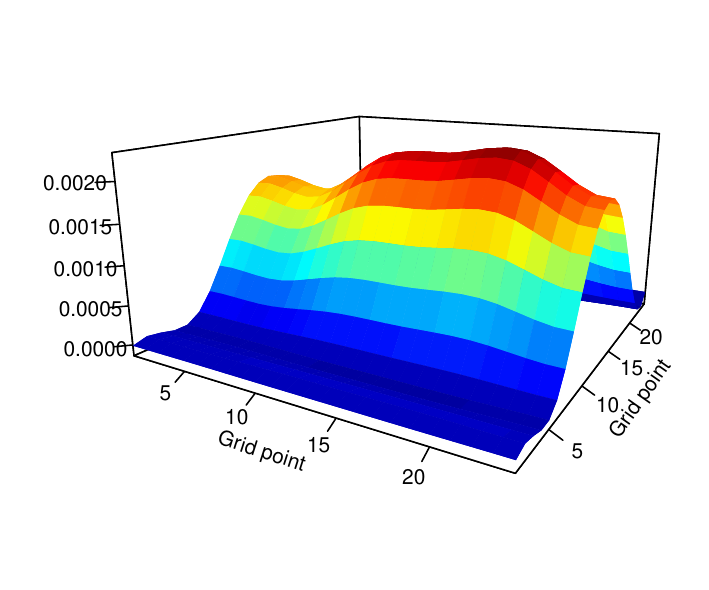}
\quad
\includegraphics[width=7cm]{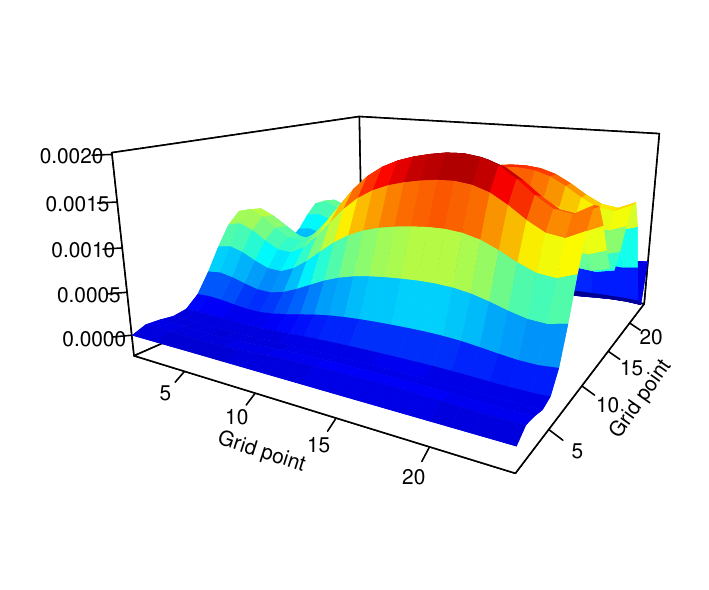}
\\
\includegraphics[width=7cm]{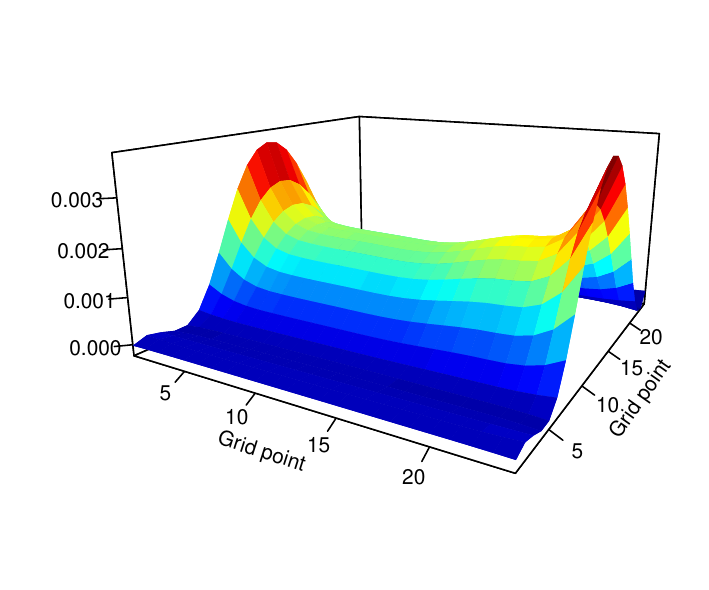}
\quad
\includegraphics[width=7cm]{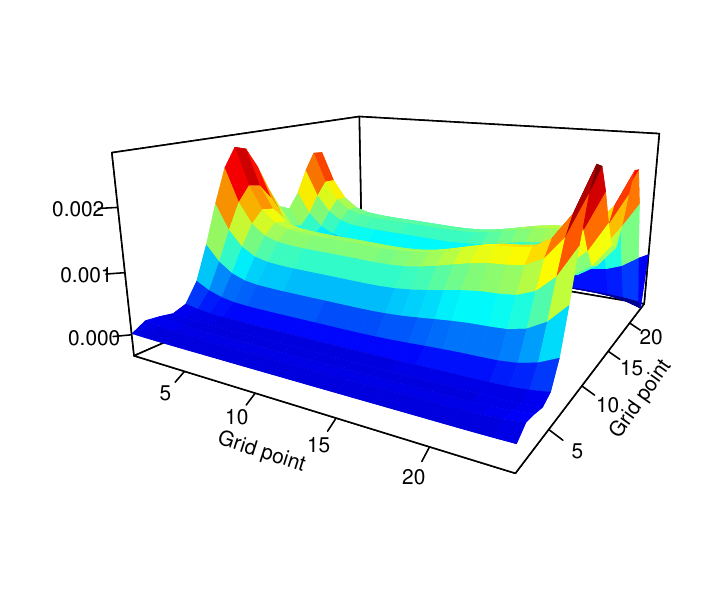}
\caption{\small{Estimated bivariate regression parameter functions under $\tau=0.5$ (left panels) and $\tau=0.975$ (right panels) for any station (Dooley) obtained by the methods FLQR (first row), Dodge (second row), Choi (third row), and Li (fourth row)}.}\label{fig:Fig_12}
\end{figure}

The estimated bivariate regression parameter functions for all methods are presented in Figure~\ref{fig:Fig_12}. This figure demonstrates that temperature (functional predictor) has minimal to no effect on solar radiation during the morning and night, while it has a substantial effect around noon. This pattern is consistently observed in all the estimated regression coefficient functions. Additionally, Figure~\ref{fig:Fig_12} indicates that for the upper quantile level ($\tau=0.975$), the effect of temperature on solar radiation is more pronounced in the estimated regression coefficient functions of the FPQR methods compared to those obtained by the FLQR method. This result clearly explains the superior performance of the FPQR methods over FLQR for the quantile level $\tau=0.975$, as shown in Figure~\ref{fig:Fig_11}.

From an environmental science perspective, solar radiation is the earth's primary energy source and plays a crucial role in surface radiation balance, hydrological cycles, vegetation photosynthesis, and weather and climate extremes. Accurate solar radiation prediction is essential for effective planning, reserve management, and avoiding penalties due to the intermittent nature of solar energy (see, for example, \cite{Huang2021, Krishnan2023}). Beyond predicting the conditional mean of solar radiation, forecasting its quantiles, particularly the upper quantiles, is critical for optimizing the design and management of solar energy systems. Accurate predictions of extreme solar radiation levels aid in maximizing energy capture and ensuring the stability and efficiency of power grids by facilitating better planning and resource allocation during peak solar conditions. Our empirical data analysis demonstrates that FLQR provides superior predictions for solar radiation at the conditional mean (quantile level $\tau=0.5$) compared to FPQR methods. Conversely, FPQR methods yield more accurate predictions at the upper quantile level ($\tau=0.975$) than FLQR. Notably, the proposed Choi method offers improved predictions for solar radiation at the upper quantile level, enhancing decision making for extreme solar conditions.

\section{Conclusion} \label{sec:conc}

In this study, we introduce two novel algorithms aimed at robustly and effectively estimating the regression coefficient function within the FFLQR framework by extending the methodology proposed by \cite{Alvaro2022} to accommodate functional data. Our approach involves initially projecting functional observations onto a finite-dimensional space using FPQR decomposition, where FPQR bases are approximated through B-spline basis expansion. Subsequently, the infinite-dimensional FFLQR model is approximated by constructing a multivariate quantile regression model using FPQR components derived from the functional response and predictor. Unlike conventional FPQR methods that rely on the quantile covariance proposed by \cite{Dodge2009}, our algorithms utilize the quantile covariances introduced by \cite{Choi2018} and \cite{Li2015} in the FPQR decomposition. We assess our proposed methods' numerical accuracy and computational efficiency through Monte Carlo simulations and empirical analyses using North Dakota weather data. Our findings suggest that FLQR produces better predictions than FPQR methods, including the proposed methods, for the conditional mean of the response variable. However, the proposed and existing FPQR methods provide superior predictions for other quantile levels of the response variable. Among the FPQR methods, our approaches yield predictive performance comparable to existing FPQR techniques. The Choi method generally provides enhanced estimation of the regression coefficient function within the FFLQR framework compared to alternative methods, albeit at the cost of increased computational overhead. Conversely, the Li method is the most computationally efficient among the proposed methods while delivering competitive regression coefficient function estimates.

The proposed method has the potential to be extended in several research directions. Here, we briefly list two.
First, the current framework considers a single functional predictor in the model but can readily accommodate multiple functional predictors, akin to the approach proposed by \cite{BSMMA}.
Second, our methods can be straightforwardly extended to handle function-on-scalar and scalar-on-function regression models, broadening their applicability and utility in various statistical contexts.

\section*{Acknowledgment}

The authors would like to acknowledge the insightful comments from two reviewers that led to a much-improved paper.

\section*{Disclosure statement}

The authors declare no conflict of interest to any party.

\newpage
\bibliographystyle{agsm}
\bibliography{ffpqr.bib}

\end{document}